\documentclass[aps,prb,onecolumn,twoside,floatfix,superscriptaddress,showpacs]{revtex4-2}
\usepackage{graphicx}
\usepackage{subfig}
\usepackage{amsmath}
\usepackage{rotating}
\usepackage{color}
\usepackage{multirow}
\usepackage{xcolor}
\usepackage{ulem}
\usepackage{appendix}
\usepackage{dsfont}

\DeclareGraphicsExtensions{.eps,.pdf,.png}
\graphicspath{{figures/}}
\usepackage{epstopdf}

\begin{document}
\title{First principles model for voids decorated by transmutation solutes: Short-range order effects and application to neutron irradiated tungsten}
\author{Duc Nguyen-Manh}
\email[Corresponding author: ]{duc.nguyen@ukaea.uk}
\affiliation{CCFE, United Kingdom Atomic Energy Authority, Abingdon, Oxfordshire OX14~3DB, United Kingdom}
\affiliation{Department of Materials, University of Oxford, Parks Road, OX1 3PH, United Kingdom}
\author{Jan S. Wr\'obel}
\affiliation{Faculty of Materials Science and Engineering, Warsaw University of Technology, ul. Wo\l{}oska 141, 02-507 Warsaw, Poland}
\author{Michael Klimenkov}
\affiliation{Karsruhe Institute of Technology, Institute for Applied Materials, 76021 Karsrule, Germany}
\author{Matthew J. Lloyd}
\affiliation{Department of Materials, University of Oxford, Parks Road, OX1 3PH, United Kingdom}
\author{Luca Messina} 
\affiliation{EdF-R$\&$D, D\'epartement Mat\'eriaux et M\'ecanique des Composants, Les Renardi\`eres, F-77250 Moret sur Loing, France}
\author{Sergei L. Dudarev}
\affiliation{CCFE, United Kingdom Atomic Energy Authority, Abingdon, Oxfordshire OX14~3DB, United Kingdom}
\affiliation{Department of Materials, University of Oxford, Parks Road, OX1 3PH, United Kingdom}

\date{\today}

\begin{abstract}
Understanding how properties of materials change due to nuclear transmutations is a major challenge for the design of structural components for a fusion power plant. In this study, by combining a first-principles matrix Hamiltonian approach with thermodynamic integration we investigate quasi-steady state configurations of multi-component alloys, containing defects, over a broad range of temperature and composition. The model enables simulating transmutation-induced segregation effects in materials, including tungsten where the phenomenon is strongly pronounced. Finite-temperature analysis shows that voids are decorated by Re and Os, but there is no decoration by tantalum (Ta). The difference between the elements is correlated with the sign of the short range order (SRO) parameter between impurity and vacancy species, in agreement with Atom Probe Tomography (APT) observations of irradiated W-Re, W-Os, W-Ta alloys in the solid solution limit. Statistical analyses of Re and Os impurities in vacancy-rich tungsten show that the SRO effects involving the two solutes are highly sensitive to the background concentration the species. In quaternary W-Re-Os-Vac alloys containing 1.5$\%$ Re and 0.1$\%$ Os, the SRO Re-Os parameter is negative at 1200K, driving the formation of concentrated Re and Os precipitates. Comparison with experimental Transmission Electron Microscopy (TEM) and APT data on W samples irradiated at the High Flux Reactor (HFR) shows that the model explains the origin of anomalous segregation of transmutation products (Re,Os) to vacancy clusters and voids in the high temperature limit pertinent to the operating conditions of a fusion power plant.                     

\end{abstract}


\maketitle

\section{Introduction}
Fusion energy generation requires materials with extraordinary properties, able to withstand exceptionally high fluxes of alpha particles and high-energy 14.1 MeV neutrons \cite{stork2017}.
As well as producing structural distortions at the atomic scale, called radiation defects, the neutrons initiate transmutation nuclear reactions leading to changes in the chemical composition of materials. The accumulation of transmutation elements results in  segregation and precipitation effects that contribute to swelling and embrittlement of materials. The development of reliable plasma facing components for the divertor and first wall of a prototype fusion reactor, such as DEMO, poses many challenges that need to be addressed for fusion energy to become a viable power source \cite{Bloom1988,Knaster2016,Rowcliffe2018}.  Design constraints based on the activation of reactor components limits material selection and necessitates the development of novel solutions for the high heat flux components. Tungsten is currently one of the leading candidate materials for these applications, however, there is a lack of data on its performance under realistic reactor conditions \cite{Reith2013,gonzalez2017,Abernethy2017,Jaime2017}. The combined effects of radiation damage and transmutation result in an increase in the inherent brittleness of W, a significant increase in its ductile to brittle transition temperature and radiation induced hardening \cite{abernethy2019}.  The pertinent transmutation pathways in a material depends on its composition, the distribution of isotopes, and on the incident neutron energy spectrum. In a DEMO, an average Re production rate of 0.1at$\%$Re/dpa is expected from the available estimates \cite{Gilbert2011,gilbert2013,Gilbert2015,gilbert2017}, whereas in a typical fission test reactor this rate, involving resonance reactions triggered by relatively low energy neutrons \cite{Gilbert2011,gilbert2017}, is much higher, typically between 3 and 10 of at$\%$Re/dpa \cite{Klopp1975,hasegawa2016,katoh2019}.  Reliable prediction of component performance in a fusion reactor requires investigating samples with realistic post-irradiation composition. 

So far, neutron irradiation experiments on microstructural evolution  of tungsten and tungsten alloys were carried out in fission reactors, where transmutation rates are high, including the Experimental Breeder Reactor II (EBR-II) \cite{herschitz1984,herschitz1984_2}, Joyo \cite{Tanno2009,Tanno2011,Hasegawa2011,fukuda2012,Fukuda2013,hasegawa2016} and the Japan Materials Test Reactor (JMTR) \cite{Fujitsuka2000,Hasegawa2011,hasegawa2016}, the High Flux Isotope Reactor (HFIR) \cite{Hasegawa2011,hasegawa2016,hwang2018,Edmondson2020} and  the High Flux Reactor (HFR) \cite{Klimenkov2016,Abernethy2017,Lloyd2019}. 
Transmutation in reactor components occurs through a series of neutron capture or neutron loss reactions, followed by either $\beta$ or $\alpha$ decay. $\beta$ decays results in a change of the material composition, and in W results primarily in the production of Re and to a lesser extent Ta \cite{Gilbert2015}. Os is also produced during the transmutation of W by neutron capture and the subsequent beta decay of Re atoms \cite{Gilbert2015}. 
Transmutation elements, formed in W under neutron irradiation over the expected period of fusion reactor operation can be either detrimental or beneficial to mechanical and engineering properties of the original material. It has long been recognized that room-temperature brittleness of tungsten can be alleviated by alloying it with rhenium \cite{raffo1969}. Early studies investigated W and W-Re alloys as candidate materials for high temperature thermo-couples, and showed that the presence of Re suppressed void formation \cite{matolich1974}, something that was also demonstrated later in less concentrated alloys\cite{hasegawa2016}. Transmission Electron Microscopy (TEM) analysis by Hu and collaborators \cite{Hu2016a,Hu2019} showed that at high neutron doses, under a neutron energy spectrum in which transmutations were significant, irradiation-induced hardening in W was caused primarily by voids and Re-rich precipitates. Atom Probe Tomography (APT) analyses of neutron irradiated W have found Re and Os rich precipitates \cite{hwang2018,Lloyd2019,katoh2019,Edmondson2020} and Re/Os decorated voids \cite{Klimenkov2016,hwang2018,Lloyd2019}. It was found in Refs. \cite{Hasegawa2014,He2006,Tanno2011,Hasegawa2011,Fukuda2013} that hardening is caused by radiation-induced precipitation of $\sigma$ phase (WRe) and $\chi$ phase (WRe{$_3$}), which form in W-5Re and W-10Re alloys after irradiation to 0.5-0.7 dpa at 600-1500$^{\circ}$C. This happens despite the fact that the solubility limit of Re in W is high, close to 30 at.$\%$ \cite{Ekman2020}. Precipitation of $\alpha$-Mn phase (similar to $\sigma$ phase) in W-25Re alloy, resulting from exposure to neutron doses of several dpa, was earlier reported in \cite{Sikka1974}. Platelet-like precipitates in neuron-irradiated W-10Re and W-25Re alloys were found using a combination of Field-Ion Microscopy (FIM) and APT in \cite{herschitz1984,herschitz1984_2}. Most recently, an APT study of neutron irradiated single W crystal in the HFIR at 770 $^{\circ}$C exposed to damage level of 1.8 dpa,  concludes that the composition of precipitates progresses towards that of the $\sigma$ phase with average 19.62$\pm$0.41  13.03$\pm$0.46 at.$\%$ for Re and Os, respectively \cite{Edmondson2020}.      

Ion irradiation also provides the means for investigating radiation damage effects, allowing for the accumulation of high damage doses, and at the same time avoiding sample activation and transmutation induced changes in sample composition \cite{armstrong2011,Armstrong2013,Yi2013,Xu2015,Xu2017,Yi2018}. Heavy-ion irradiation experiments in W-(Re,Os,Ta) alloys \cite{Xu2015,Xu2017}, have measured an increase in hardness using nanoindentation. High dose experiments  have linked this increase to the formation of Re and Os rich clusters but there was no indication of intermetallic formation in these samples. APT analysis of W-1Re-1Os irradiated to 33dpa at 773K showed that Os had acted to suppress Re clustering, when compared with W-2Re, implanted using equivalent conditions \cite{Xu2015}. Interestingly, in the W-4.5Ta alloy, no evidence of irradiation-induced clustering was found. In the W-2Re-1Ta alloy, at both irradiation temperatures studied,the presence of Ta reduced the W-Re cluster number density and volume fraction compared to that in the W-Re alloy \cite{Xu2017}. However, differences in the dose, dose rate and irradiation temperature between ion and neutron irradiation experiments make comparison difficult. A combined microstructural characterisation of ion and neutron irradiated materials with equivalent dose (1.7dpa) and irradiation temperature (1173K) was recently performed using a combination of Atom Probe Tomography (APT), Transmission Electron Microscopy (TEM) and nanoindentation techniques \cite{Lloyd2019,Lloyd2021}. For the neutron irradiated W samples at the High Flux Reactor (HFR) in Petten, with an average production rate of 0.8at$\%$Re/dpa (1.67dpa, 1173K), Lloyd et al. discovered strong evidence for void decoration with both Re and Os in APT analyses \cite{Lloyd2019}. Large Re clusters with a core of Os were observed in neutron irradiated samples, but in the ion irradiated material, clusters were smaller and comprised of mainly Os atoms. Nanoindentation showed greater hardening in the neutron irradiated material compared to the ion irradiated, despite the same composition and dose. More importantly, in all of the APT analyses of neutron irradiated W of low transmutation rate there was no indication that either $\sigma$ or $\chi$ phase formation had occurred \cite{Lloyd2021}.

In the absence of a dedicated high-intensity 14.1 MeV neutron source, multi-scale materials modelling based on first-principles calculations is an indispensable tool for exploring the irradiation conditions similar to those to be experienced by the first wall in a fusion reactor \cite{gonzalez2017,nguyen2006,Derlet2007,Nguyen2007,NguyenManh_JMS2012,fitzgerald2008,Muzyk2011,Dudarev_ARMR2013,NguyenManh2015,Hofmann_Acta2015,Wrobel2017,Jaime2017,Mason2017,Dudarev_NF2018,Mason2019,nordlund2019,Derlet2020}. Precipitation of non-equilibrium phases in irradiated material is a commonly observed phenomenon and is a result of radiation induced segregation (RIS), whereas radiation enhanced diffusion (RED) can lead to a faster rate of formation of equilibrium phases \cite{Russell1984,Barbu1977,Okamoto1979,Cauvin1979,Martin1980,cauvin1981,Shu2013}. For example, the RIS of solute Cr was found, using APT, in Fe-5Cr alloy, following self-ion irradiation at 400$^{\circ}$C\cite{Hardie2013}. This finding is particularly striking and significant since binary Fe-Cr alloys are a well known model alloy family, the thermodynamic properties of which are linked to those of ferritic/martensitic steels, which are the low-activation structural materials for fusion as power plants \cite{Nguyen2008,Boutard2008}. Interpretation of experimental observations of solute segregation effects in the context of theoretical models, linking solute segregation to binding between the defects and solute atoms, and/or different rates of diffusion of solute and solvent atoms, have so far focused primarily on kinetic approaches to the treatment of phase stability \cite{Martin1980,Okamoto1979,cauvin1981,Krasnochtchekov2007}. In a combination with Density Functional Theory (DFT),  such models, treating solute fluctuations in solid solution under irradiation, provide insight into the highly complex phenomena of solute-defect trapping, solute segregation, point-defect recombination, dislocation interactions and nucleation, and growth of voids in tungsten and iron based materials \cite{Becquart2006,Becquart2007,Becquart2012,Marinica2012,Kong2014,Zhou2014,Ventelon2015,Schuler2015,Senninger2016,Messina2020}. Recently, kinetic approaches have been used to model the behaviour of W-Re alloys under irradiation. DFT calculations suggest that the increased solute transport may result from strong Re and Os binding to point defects. While the  most stable self-interstitial configuration in W is an $\langle111\rangle$ dumbbell, which undergoes fast 1D migration \cite{nguyen2006,fitzgerald2008}, the   strong binding of Re and Os atoms to self-interstitial defects leads to the formation of W-solute mixed interstitial dumbbells,  which can undergo effective 3D migration via a series of translation and rotation processes \cite{becquart2010,Muzyk2011,suzudo2014,suzudo2015,gharaee2016,suzudo2016,Setyawan2017,Li2017,Li2019,castin2020}. Kinetic Monte Carlo (KMC) models where this transport process was implemented have predicted the formation of Re rich precipitates, facilitated primarily by W-Re mixed interstitial migration \cite{huang-2017,Lloyd2019b}. Neither of these studies correctly captured vacancy clustering because of the adopted pair interaction Ising model. The di-vacancy interaction in W is slightly repulsive, whereas the larger vacancy cluster binding energies are more attractive \cite{Muzyk2011,Wrobel2017,Mason2017,Mason2019,Messina2019}. So far, molecular dynamic simulations of radiation damages were carried only for the  W-Re binary alloys where inter-atomic potentials were constructed based on fitting to DFT data \cite{Bonny2017,Setyawan2018,Chen2018}  

Below, an alternative approach is developed and applied to model RIS of solutes in irradiated alloys within the framework of thermodynamic approach to modelling irradiation microstructures \cite{Russell1984,Bocquet1979}. This enables defining and predicting the free energies of various competing phases in the presence of irradiation-induced defects, with supersaturated vacancy and interstitial concentrations. The final microstructure would
be composed of new phases in steady-state
configurations corresponding to the lowest free energy of an driven alloy under radiation conditions \cite{Martin1996}. The approach makes it possible to understand what drives radiation-induced precipitation, using a thermodynamic viewpoint based on the free energy minimization, which is applied to an alloy containing defects produced by irradiation. The treatment overcomes major uncertainties encountered in the context of a complex kinetic approach, associated with ascertaining that a kinetic model, which requires information not only about energies of various configurations but also about the transition rates and defect mobilities, goes beyond an {\it ad hoc} explanation of experimental observations. Using a combined DFT and constrained cluster expansion (CE) Hamiltonian, the proposed approach has been used to simulate the formation of anomalous radiation induced precipitates in binary W-Re solid solutions. Both void formation and decoration with Re was observed in bcc-W at a finite temperature using the thermodynamic integration method in conjunction with exchange Monte Carlo (MC) simulations \cite{Wrobel2017}. As a proof of concept, vacancies were considered as an additional component of a binary W-Re system, mapping the alloy onto a ternary alloy system containing tungsten and rhenium atoms, as well as vacancies. 

Formation energies of interstitial defects in tungsten are about three times the vacancy formation energy \cite{Muzyk2011,nguyen2006,Derlet2007,Arakawa2020}. Hence it is natural from the thermodynamic perspective to first consider vacancies when evaluating the free energy of an alloy subjected to irradiation at a high temperature \cite{Kato2011,Semenov2012}. This approach is now generalized to investigate the origin of voids decorated by transmutation elements (Re,Os,Ta) in tungsten under neutron irradiation with a relatively low transmutation rate close to the DEMO fusion reactor condition.   

The paper is organized as follows: After the Introduction,  in Section II, a systematic presentation of the computational methodology is outlined. The subsection II.A. describes a new matrix formulation of  CE formalism which has been developed to model a general multi-component system \cite{AFC2019}. This subsection starts with the dentition of the CE Hamiltonian with cluster functions defined in the matrix formula (II.A.1) which allows us to determine efficiently the short-range order between various species at different composition (II.A.2).  The subsection ends with the definition of configuration entropy from thermodynamic integration technique (II.A.3). The new formulation has been applied to studying W and Fe based high-entropy alloys for fusion materials applications \cite{El-Atwani2019,Damian2020,Fedorov2020}. In subsection II.2, DFT data for the enthalpy of mixing and binding energies of vacancy clusters interacting with transmutation elements (Re,Os,Ta) are analyzed. The Effective Cluster Interactions (ECIs) obtained by mapping DFT into the CE Hamiltonian are presented in subsection II.3. The results from MC simulations are summarised in Section III. It starts with a comparison of simulations for different defective systems with Re, Os and Ta in subsection III.A. The effects of Os are investigated in subsection III.B in the context of treatment of a W-Re-Os-Vac alloy.  Subsection III.C analyses the dependence of microstructure on the short-range order (SRO)  parameters as a function of composition and temperature.  
Section IV focuses on a comparison between the modelling results derived from this work with experimental data on tungsten samples irradiated with neutrons at the High Flux Reactor. 
It starts with the formation of faceted voids of neutron irradiated W at high temperature (subsection IV.A). The distribution of Re and Os inside transmutation induced precipitation is analysed from TEM (subsection IV.B) and APT (subsection IV.C). Main conclusions are given in Section V.

\section{Methodology}

\subsection{Matrix formulation of cluster expansion formalism}
In this subsection, we start by outlining the matrix formulation of the cluster expansion method and its generalisation to a $K$-component system  \cite{AFC2017,AFC2019,Fedorov2020,Damian2020}.  It enables computing chemical short-range order parameters as functions of temperature for arbitrary compositions of multi-component alloys. 

\subsubsection{Enthalpy of mixing}

The enthalpy of mixing of an alloy with chosen configuration represented by a vector of configurational variables, $\vec{\sigma}$\textcolor{red}{,} can be calculated from DFT using the value of total energy per atom of the simulated alloy structure including defects,  $E^{lat}_{tot}(\vec{\sigma})$, and the corresponding pure elements,  $E^{lat}_{tot}(p)$, underlying the same lattice as the alloy structure as follows:
\begin{equation}
\Delta H_{mix}(\vec{\sigma})= E^{lat}_{tot}(\vec{\sigma}) - \sum\limits_{p=1}^{K}x_pE^{lat}_{tot}(p),
\label{eq:hmix}
\end{equation}
where K is the number of alloy components\textcolor{red}{,} $x_p$ are the average concentrations of each alloy component and $\vec{\sigma}$ denote the ensemble of occupation variables in the lattice. 

In the cluster expansion formalism, the enthalpy of mixing can be parametrized as a polynomial in the occupational variables \cite{VandeWalle2002}:
\begin{equation}
{\Delta H}_{mix}^{CE}( \vec{\sigma } )=
\sum\limits _{\omega,n,s} J^{(s)}_{\omega,n} m^{(s)}_{\omega,n}  \langle {\Gamma_{\omega',n'}^{(s')}(\vec{\sigma})}\rangle_{\omega,n,s}, 
\label{eq:deltahmixing}
\end{equation}
where the summation is performed over all the clusters, distinct under symmetry operations in the studied lattice, represented by the following parameters: $\omega$ and $n$ are the cluster size (the number of lattice points in the cluster) and its label (maximal distance between two atoms in the cluster in terms of coordination shells), respectively.
$m^{(s)}_{\omega,n}$ denotes the site multiplicity of the decorated clusters (in per-lattice-site units); and $J^{(s)}_{\omega,n}$ represents the Effective Cluster Interaction (ECI) energy corresponding to the same $(s)$ decorated cluster. A more detailed description on how to get the ECIs from the Density Function Theory calculations will be presented in the next subsection. Within the CE method, the energy Hamiltonian,  given by equation \ref{eq:deltahmixing} can be expended in term of the ECIs as follows:

\begin{equation}
\begin{array}{ll}
{\Delta H}_{mix}^{CE}( \vec{\sigma } )= &
J_{1,1}^{(0)}\left\langle \Gamma _{1,1}^{(0)} \right\rangle +\sum\limits_s J_{1,1}^{(s)} \left\langle \Gamma _{1,1}^{(s)} \right\rangle+ \\ \\ &
+\sum\limits_{n,s} m_{\text{2,n}}^{(s)} J_{\text{2,n}}^{(s)}  \left\langle \Gamma _{\text{2,n}}^{(s)} \right\rangle + 
\sum\limits_{n,s} m_{\text{3,n}}^{(s)} J_{\text{3,n}}^{(s)}  \left\langle \Gamma _{\text{3,n}}^{(s)} \right\rangle + \\ \\ &
+\sum\limits_{n,s}^{\text{higher-order}} \ldots,
\end{array}
\label{eq:deltahmixing2}
\end{equation}
where the two terms from first line of Eq.(\ref{eq:deltahmixing2}) denote the zero and point interactions, respectively; the next two terms of the second line from this equation: the two and three body interactions, respectively and in the last line: the higher order denote many-body terms from four-body cluster interactions.

In Eq.(\ref{eq:deltahmixing}),  $\langle {\Gamma_{\omega',n'}^{(s')}(\vec{\sigma})}\rangle_{\omega,n,s}$ denotes the cluster function, averaged over all the clusters of size, $\omega'$, and label, $n'$, decorated by the sequence of point functions, \textcolor{red}{,} $(s')$, 
that are equivalent by the symmetry to the cluster, $\omega, n$, and decorated by the same sequence of point functions, $(s)$. Later in the text, $\langle {\Gamma_{\omega',n'}^{(s')}(\vec{\sigma})}\rangle_{\omega,n,s}$ is referred to as 
$\langle {\Gamma_{\omega,n}^{(s)}(\vec{\sigma})}\rangle$,  for ease of notation. For a $K$ component system, the cluster function is not conventionally presented as a simple product of occupation variable\textcolor{red}{,} $\vec{\sigma}$, as for a binary system \cite{Muzyk2011}. Instead is more transparently defined as the product of orthonormal point functions of occupation variables \cite{deFontaine1994,Wrobel2015}: $\gamma_{j,K}(\sigma_i)$, on a specific cluster described by $\omega$ and $n$:
\begin{equation}
\Gamma_{\omega,n}^{(s)}(\vec{\omega})=\gamma_{j_1K}(\sigma_1)\gamma_{j_2K}(\sigma_2)\cdots \gamma_{j_{\omega}K}(\sigma_{\omega})
\label{eq:gamma}
\end{equation}



In this work, the point function is defined by the same formulation as in Ref.~\cite{VandeWalle2009}:
\begin{equation}
\gamma_{j, K}(\sigma_{i})=
\begin{cases}
1 & \textrm{ if }j=0\textrm{ }, \\
-\cos\left(2\pi\lceil\frac{j}{2}\rceil\frac{\sigma_{i}}{K}\right) & \textrm{ if }j>0\textrm{ and odd},  \\
-\sin\left(2\pi\lceil\frac{j}{2}\rceil\frac{\sigma_{i}}{K}\right) & \textrm{ if }j>0\textrm{ and even},
\end{cases}
\label{eq:gamma_def}
\end{equation}
where $i = 0,1,2,...(K-1)$, $j$ is the index of point functions $j=0,1,2,...(K-1)$ and $\lceil\frac{j}{2}\rceil$ stands for the ceiling function - rounding up to the closest integer. It is found recently that, from  mathematical point of view, the point functions $\gamma_{j,K}[\sigma_i]$ can be described in terms of a matrix formulation via the relationship \cite{AFC2019}:

\begin{equation}
(\bar{\bar{\tau}}_K)_{j i}\equiv\gamma_{j,K}[\sigma_{i}]
\label{eq:gamma_matrix}
\end{equation}

for matrix elements and the matrix $(\bar{\bar{\tau}}_{K})$ can be written as the following:
\begin{equation}
(\bar{\bar{{\tau}}}_{K})=\left( \begin{array}{cccc}
\gamma_{j=0,K}(\sigma_{i}=0)  & \cdots  & \gamma_{j=0,K}(\sigma_{i}=K-1) \\
\vdots & \ddots  & \vdots \\
\gamma_{j=K-1,K}(\sigma_{i}=0)  & \cdots & \gamma_{j=K-1,K}(\sigma_{i}=K-1)
\end{array} \right).
\label{tau_generalized_matrix}
\end{equation}

Matrix elements of the inverse of the $(\bar{\bar{\tau}}_{K})$ matrix can be obtained by the expression \cite{AFC2019}:

\begin{equation}
(\bar{\bar{\tau}}^{-1}_{K})_{ij}=\begin{cases}
\frac{1}{K} & \textrm{ if }j=0\textrm{ }, \\
-\frac{2}{K} \cos\left(2\pi\lceil\frac{j}{2}\rceil\frac{\sigma_i}{K}\right) & \textrm{ if }j>0\textrm{ and }j-1<K\textrm{ and j odd},  \\
-\frac{2}{K} \sin\left(2\pi\lceil\frac
{j}{2}\rceil\frac{\sigma_i}{K}\right) & \textrm{ if }j>0\textrm{ and j even}, \\
-\frac{1}{K} \cos\left(2\pi\lceil\frac{j}{2}\rceil\frac{\sigma_i}{K}\right) & \textrm{ if }j-1=K\textrm{ and j odd}.
\end{cases}
\label{eq:tauinverse}
\end{equation}
Eq. (\ref{eq:tauinverse}) represents the inverse of the  $\bar{\bar{\tau}}_{K}$ matrix to ensure that the basis set defined by Eq. (\ref{eq:gamma_def}) is rigorously orthonormal. The size of the $\bar{\bar{\tau}}_{K}$ matrix is K$\times$K, where K is the number of components. In particular for the case of four component, K=4, the matrices $\bar{\bar{\tau}}_{4}$ and its inverse matrix, $\bar{\bar{\tau}}_{4}^{-1}$, have the following expressions \cite{Fedorov2020}:

\begin{equation}
\begin{array}{ll}
\bar{\bar{\tau}}_{4}=
\left ( \begin{array}{rrrrr}
1  & 1  & 1  & 1 \\
-1 & 0  & 1  & 0 \\
0  & -1 & 0  & 1 \\
-1 & 1  & -1 & 1
\end{array}
\right )
& ~~~
\bar{\bar{\tau}}^{-1}_{4}=\dfrac{1}{4}
\left (
\begin{array}{rrrrr}
1 & -2 & 0   & -1 \\
1 & 0  & -2  & 1 \\
1 & 2  & 0   & -1 \\
1 & 0  & 2   & 1
\end{array}
\right )
\end{array}
\end{equation}

The general expression for the cluster correlation function averaged over all cluster configuration, can be rewritten using the matrix formulation as follows \cite{AFC2019,Fedorov2020}:
\begin{equation}
\langle \Gamma_{\omega,n}^{(s)}\rangle=\sum\limits _{A,B,\cdots}(\prod\limits_{\omega}\bar{\bar{\tau}}_{K})_{(s),A,B,\cdots} y_{\omega,n}^{A,B,\cdots}= \sum\limits_{A,B, \cdots}{\overbrace{(\bar{\bar{\tau}}^{}_{K}\otimes\cdots\otimes\bar{\bar{\tau}}^{}_{K})}^{\omega}}_{A,B,\cdots} y_{\omega,n}^{(A,B\cdots)}
\label{eq:Gamma_general_matrix}
\end{equation}
where $(\prod\limits_{\omega}\bar{\bar{\tau}}_{K})_{(s),A,B,\cdots}$ denotes the matrix direct product (Kronecker product); the summation is done over the atomic species composing the alloy; $y_{n}^{A,B,\cdots}$ denotes the $many$-body probability of finding atomic species $A,B,\cdots$ in the corresponding $\omega$ cluster with coordination shell, denoted by $n$. 
$\otimes$, denotes the matrix direct product.


From Eq.~(\ref{eq:Gamma_general_matrix}), the two-body and the three-body cluster function in Eq.(\ref{eq:deltahmixing2}) can be written by the following expressions:

\begin{equation}
\langle \Gamma _{2,n}^{(s)} \rangle= \sum\limits _{A,B}  (\bar{\bar{\tau}}_{K}\otimes\bar{\bar{\tau}}_{K})_{(s),A,B} y_{2,n}^{AB},
\label{eq:G_g_2}
\end{equation}
and 
\begin{equation}
\langle \Gamma _{3,n}^{(s)} \rangle=
\sum\limits _{A,B,C}  (\bar{\bar{\tau}}_{K}\otimes\bar{\bar{\tau}}_{K}\otimes\bar{\bar{\tau}}_{K})_{(s),A,B,C} y_{3,n}^{ABC},
\label{eq:G_g_3}
\end{equation}

respectively. $y_{2,n}^{AB}$ and $y_{3,n}^{ABC}$ denote the 2- and 3-body probability functions, respectively. Similarly, the higher-order cluster functions can also be given within the matrix formulation.


Using Eq.(\ref{eq:tauinverse}) of the matrix, $\tau_{K}^{-1}$, the cluster probability function can be expressed within a general form from the inverse matrix formulation of Eq.(\ref{eq:Gamma_general_matrix}) as follows:

\begin{equation}
y_{\omega,n}^{(AB\cdots)}={\overbrace{(\bar{\bar{\tau}}^{-1}_{K}\otimes\cdots\otimes\bar{\bar{\tau}}^{-1}_{K})}^{\omega}}_{AB\cdots,ij\cdots} {\langle \Gamma_{\omega,n}^{(ij\cdots)}\rangle}
\label{eq:clusterprob}
\end{equation}

\subsubsection{Chemical Short-Range Order}


From the Eq.(\ref{eq:clusterprob}), the point probability function is written as:
\begin{equation}
y^{A}_{1,1}=\sum\limits _{s} (\bar{\bar{\tau}}^{-1}_{K})_{A,(s)} \langle \Gamma_{1,1}^{(s)}\rangle,
\label{eq:point-probability}
\end{equation}
and the pair probability function is determined by the following formula :
\begin{equation}
y_{2,n}^{AB}= \sum\limits _{s} (\bar{\bar{\tau}}^{-1}_{K}\otimes\bar{\bar{\tau}}^{-1}_{K})_{A,B,(s)} \langle \Gamma_{2,n}^{(s)}\rangle.
\label{eq:pair-probability}
\end{equation}

Eq.(\ref{eq:pair-probability}) allows the two-body cluster probability to be linked with the Warren-Cowley short-range order (SRO) parameter, $\alpha_{2,n}^{(AB)}$,  via the definition \cite{Warren1990}:  

\begin{equation}
y_{2,n}^{AB}=x_{A} x_{B} ( \, 1-\alpha_{2,n}^{AB} ) \, 
\label{eq:WC}
\end{equation}

where $x_{A}$ and $x_{B}$ denote the bulk concentration of the chemical species A and B, respectively. In the case where  $\alpha_{\text{2,m}}^{\text{AB}}=0$, the pair probability is given by the product of their concentrations $\text{x}_{\text{B}}\text{x}_{\text{B}}$ corresponding to random configuration of A and B species in an alloy system.  In the case of $\alpha_{\text{2,m}}^{\text{AB}}>0$, clustering or segregation between A-A and B-B pairs is favoured and for	$\alpha_{\text{2,m}}^{\text{AB}}<0$, the chemical ordering of A-B pairs occurs. By combining Eqs.(\ref{eq:point-probability}),(\ref{eq:pair-probability} and (\ref{eq:WC}), the SRO parameters for $K$-component system can be calculated by the general expression as follows: 


\begin{equation}
\alpha_{2,n}^{AB} = 1- \dfrac{\sum\limits _{s} (\bar{\bar{\tau}}^{-1}_{K}\otimes\bar{\bar{\tau}}^{-1}_{K})_{A,B,(s)} \langle \Gamma_{2,n}^{(s)}\rangle}{\big( \sum\limits _{s} (\bar{\bar{\tau}}^{-1}_{K})_{A,(s)} \langle \Gamma_{1,1}^{(s)}\rangle \big) \big( \sum\limits _{s} (\bar{\bar{\tau}}^{-1}_{K})_{B,(s)} \langle \Gamma_{1,1}^{(s)}\rangle \big)}.
\label{eq:sromatrix}
\end{equation}

The expression to calculate the average SRO parameter for first and second nearest neighbours in a bcc lattice  is given by \cite{Mirebeau2010,AFC2017,Damian2020}:

\begin{equation}
\alpha_{avg}^{AB}=\frac{8\alpha_{2,1}^{AB}+6\alpha_{2,2}^{AB}}{14}
\label{eq:SRO_avg}
\end{equation}

\noindent where $\alpha_{2,1}^{AB}$ and $\alpha_{2,2}^{AB}$  are the first and second nearest neighbours SRO parameters, respectively.




\subsubsection{Configuration entropy as a function of temperature}
The cluster expansion Hamiltonian, defined by Eq. (\ref{eq:deltahmixing}), which takes into account all $many$-body cluster interactions, can be used to explicitly determine the configuration entropy of a $K$-component system via the thermodynamic integration method \cite{Newman1999,Lavrentiev2009}.  Here the entropy is computed from fluctuations of the enthalpy of mixing at a given temperature using the following formula

\begin{equation}
S_{conf}[T]=\int_{0}^{T} \dfrac{C_{conf}(T')}{T'} dT'=\int_{0}^{T} \dfrac{\langle [{\Delta H_{mix}^{CE}(T')}]^{2}\rangle-{\langle [\Delta H_{mix}^{CE}(T')]\rangle}^{2}}{{T'}^{3}} dT',
\label{eq:thermointegration}
\end{equation}
where ${\langle [\Delta H_{mix}^{CE}(T')]^{2}\rangle}$ and $\langle [\Delta H_{mix}^{CE}(T')]\rangle^{2}$ are the square of the mean and mean square enthalpies of mixing, respectively. The average over configurations at finite temperature in Eq.(\ref{eq:thermointegration}) can be performed by combining the CE formalism with Monte Carlo (MC) technique\textcolor{red}{,} from which all the simulation steps at the accumulation stage for a given temperature are taken into account. The accuracy of evaluation of configuration entropy from Eq.(\ref{eq:thermointegration}) depends on the size of temperature integration step and the number of MC steps performed at the accumulation stage \cite{Wrobel2015,Wrobel2017}.
In this work, semi-canonical exchange MC simulations were also performed using the ATAT package. For each composition of the $K$ component system, simulations were performed starting from a disordered, high-temperature state at $T$ = 3000~K. The system was then cooled down with a temperature step of ${\Delta}T$ = 10~K, with around 3000 equilibration and accumulation Monte Carlo passes at a given temperature within the thermodynamic integration method. 
With the definition of configuration entropy from Eq.(\ref{eq:thermointegration}, the hybrid CE-Monte Carlo method enables the evaluation of the free energy of mixing 
\begin{equation}
F_{mix}(T)=H_{mix}(T)-T S_{conf}(T).
\label{eq:free_energy}
\end{equation}

Monte Carlo simulations using the CE Hamiltonian from Eq.(\ref{eq:deltahmixing2}) produce the cluster correlation functions determined by  Eq.(\ref{eq:Gamma_general_matrix}) and the cluster probabilities from Eq.(\ref{eq:clusterprob}), for all the configurations found for each temperature.  Therefore the matrix formulation can be used to efficiently compute the free energy of a multi-component system as a function of temperature \cite{Fedorov2020}.

It is important to emphasise that by using Eq.(\ref{eq:clusterprob}) for a specific choice of the maximum cluster $\omega,n$, it is possible to express the configuration entropy within a mean field approximation or so-called Cluster Variation Method (CVM) \cite{Cenedese1991}, in an analytical form as functions of $y_{\omega,n}^{(AB\cdots)}$.  The expression for CVM configuration entropy can be written in a general form as follows\cite{Finel1992}: 

\begin{equation}
S_{\omega,n}([\vec{\sigma}], T)=k_{B}\sum_{ i=1}^{s[\omega,n]} \eta_{\omega_{i},n_{i}} \sum_{[(AB\cdots)_{\omega_{i},n_{i}}]}  y_{\omega_{i},n_{i}}^{(AB\cdots)_{\omega_{i},n_{i}}}([\vec{\sigma}], T) ln ( y_{\omega_{i},n_{i}}^{(AB\cdots)_{\omega_{i},n_{i}}}([\vec{\sigma}], T))
\label{eq:entropyCVM}
\end{equation}

that represents the factorization of entropy contributions from all sub-cluster $(\omega_i,n_i)$ to the CVM cluster $(\omega,n)$. The weights $\eta_{\omega_{i},n_{i}}$ in the first sum of Eq.(\ref{eq:entropyCVM}) are worked out following the iterative Barker formula \cite{Barker1953} in the formalism of the cluster variation method. A detailed discussion about how to calculate these weights can be found in \cite{AFC2019}. It is noted that the CVM approach is conventionally valid for describing configuration entropy at the high temperature limit. The simplest CVM approximation is for a completely disordered configuration for which the cluster probabilities can be written as a product of single-point probabilities or concentrations of species involved in the decorations of the cluster:

\begin{equation}
y_{\omega_i,n_i}^{(AB\cdots)_{\omega_i,n_i}}[\vec{\sigma}] \underset{T\to \infty}{\to} \prod_{j=1}^{\omega_{i}} y_{1,1}^{{ ((AB\cdots)_{j})_{\omega_{i},n_{i}} }}[\vec{\sigma}]=\prod_{j=1}^{\omega_{i}} x_{{((AB\cdots)_{j})_{\omega_{i},n_{i}}}}[\vec{\sigma}]
\label{eq:HTmultibodyprob}
\end{equation}

where the probability of single point cluster (see Eq.(\ref{eq:point-probability})) can be represented in terms of the concentration $x_{(AB\cdots_{j})_{\omega_{i},n_{i}}}[\vec{\sigma}]$. Using Eq.(\ref{eq:HTmultibodyprob}, the configuration entropy for a random configuration can be rewritten from (Eq.(\ref{eq:entropyCVM})) as follows: 

\begin{equation}
S_{conf}^{random}=-k_{B}\sum\limits _{p=1}^{K} x_{p} \ln (x_{p})
\label{eq:random}
\end{equation}

At $T\to \infty$, the $S_{conf}^{random}$ is therefore independent of the temperature. The temperature dependence of the cluster contribution to the entropy starts with the pair correlation functions, which can be determined from MC simulations. For example, using Eq.(\ref{eq:point-probability}) the configuration entropy contribution for the first nearest neighbor (1NN) pairs in a bcc lattice is expressed as:

\begin{equation}
\begin{split}
S_{conf}^{2,1} (\vec{\sigma},T) =& 7 \sum_{s} y_{1,1}^{(s)} (\vec{\sigma},T) \ln (y_{1,1}^{(s)} (\vec{\sigma}, T)) \\ &-4 \sum_{s} y_{2,1}^{(s)} (\vec{\sigma}, T) \ln (y_{2,1}^{(s)} (\vec{\sigma}, T))
\end{split}
\label{eq:1nn_pair}
\end{equation}

whereas those for the second nearest-neighbour (2NN) pairs is: 

\begin{equation}
\begin{split}
S_{conf}^{2,2} (\vec{\sigma},T) =& 5 \sum_{s} y_{1,2}^{(s)} (\vec{\sigma},T) \ln (y_{1,2}^{(s)} (\vec{\sigma}, T)) \\ & -3 \sum_{s} y_{2,2}^{(s)} (\vec{\sigma}, T) \ln (y_{2,2}^{(s)} (\vec{\sigma}, T))
\end{split}
\label{eq:2nn_pair}
\end{equation}

Importantly, the SRO Warren-Cowley parameters, defined from the pair probabilities from  Eqs.(\ref{eq:WC}) and (\ref{eq:sromatrix}), for the 1NN and 2NN can be also evaluated from MC simulations at a given temperature T and for an arbitrary set of concentration ensemble, ${\vec{\sigma}}$.  


\subsection{DFT calculations}

The DFT calculations for selected configurations of vacancy-(Re,Os,Ta) clusters in W were performed with the Vienna Ab-initio Simulation Package (VASP) \cite{Kresse1996a,Kresse1996}. The interaction between ions and electrons is described by using the projector augmented waves (PAW) method \cite{Blochl1994}. Exchange and correlation were treated in the generalized gradient approximation GGA-PBE \cite{Perdew1996}, with PAW potentials for W, Re, Os and Ta containing semi-core $p$ electron contributions. To treat clusters containing from 2 to 47 sites with solute atoms and vacancies, orthogonal bcc super-cells containing 128 and 250 sites were used in calculations. Total energies were calculated using the Monkhorst-Pack mesh\cite{MonkhorstPack1976} of $k$-points in the Brillouin zone, with the $k$-mesh spacing of 0.15 \AA$^{-1}$.
This corresponds to 4$\times$4$\times$4 or 3$\times$3$\times$3 $k$-point meshes for a bcc super-cell of 4$\times$4$\times$4 or 5$\times$5$\times$5 bcc structural units, respectively. The plane wave cut-off energy was 400 eV. The total energy convergence criterion was set to 10$^{-6}$ eV/cell, and force components were relaxed to 10$^{-3}$ eV/\AA. Super-cell calculations were performed considering vacancy clusters interacting with solute atoms (Re, Os and Ta) in bcc W lattice under constant pressure conditions, with structures optimized by relaxing both atomic positions as well as the shape and volume of the super-cell.

\begin{figure*}
\subfloat{
		\includegraphics[width=.5\linewidth]{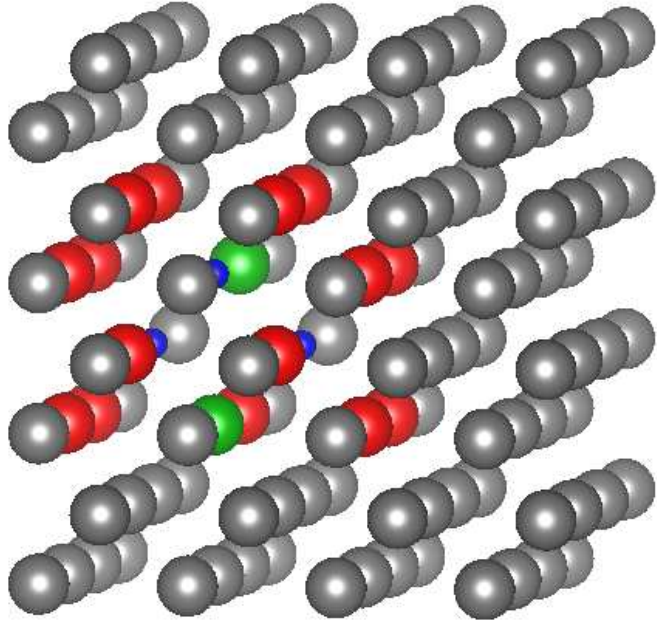}}
\caption{\label{W15Re2Os3Vac} Representative structure for a three vacancy cluster (blue) decorated by 15 Re (red), 2 Os atoms (green) in a 4$\times$4$\times$4 super-cell with W (grey) atoms }
	\label{fig:W15Re2Os3Vac}
\end{figure*}

Besides the data obtained from the large supercell structures, the DFT calculations were also performed for  2$\times$2$\times$2 supercell structures generated from decoration of one vacancy and other W, Re, Os and Ta atoms from 58  bcc-like ordered structures originally used for binary system \cite{NMD2017,Muzyk2011}. In addition, the 2$\times$2$\times$2 special quasi-random structures (SQS), created from Alloy Theoretic Automated Toolkit (ATAT) code \cite{vandeWalle2002b}, for configurations with one vacancy and different W and solute atoms were also employed to provide DFT data of the enthalpy of mixing in the present work, via the definition from Eq.(\ref{eq:hmix}). It is noted that there is a difference between the definition of enthalpy of mixing and enthalpy of formation in the presence of Re or Os, as can be seen from the Supplementary Materials figure $S1$.  In total, there are 148, 174, 641 values of input DFT energies for the present investigation of the ternary W-Os-Vac, W-Ta-Vac and the quaternary W-Re-Os-Vac systems, respectively. Specifically for the case of W-Re-Os-Vac system, the 224 DFT values generated from our previous study for the sub-ternary system W-Re-Vac \cite{Wrobel2017} are also included. Fig.\ref{fig:W15Re2Os3Vac} shows a representative structure in a 4$\times$4$\times$4 bcc super-cell with a tri-vacancy cluster initially taken from the configuration shown in Fig.5a of \cite{Wrobel2017} but decorated by both Re and Os atoms. All the new DFT data of enthalpy of mixing for W-Ta-Vac and W-Re-Os-Vac including W-Os-Vac as the subsystem are shown in the table S2 and S3 of the Supplementary Materials, respectively. 

DFT calculations also enable determining the binding energy of defect clusters. The binding energy of each defect configuration is computed using the following definition: 

\begin{equation}
E_{b}(A_{1},A_{2}\cdots,A_{N}) =\sum\limits_{i}^{N}E(A_{i}) -[E(\sum\limits_{i}^{N}A_{i}) - (N-1)E_{ref}] 
\label{eq:Eb}
\end{equation}

where $A_{i}$ marks each component of the N-component cluster. E($\sum\limits_{i}^{N}A_{i}$), E($A_{i}$), and $E_{ref}$ are respectively the energy of the supercell containing the cluster, that containing the isolated component $A_{i}$, and the perfect W atom super-cell. With this convention, positive binding energies stand for attractive interactions: for example, the binding energy of the configuration, shown in Fig.\ref{fig:W15Re2Os3Vac}, equals 3.04 eV demonstrating the strong interaction between a vacancy cluster with solute Re and Os atoms in W. 

\begin{figure*}
\subfloat{
		\includegraphics[width=.6\linewidth]{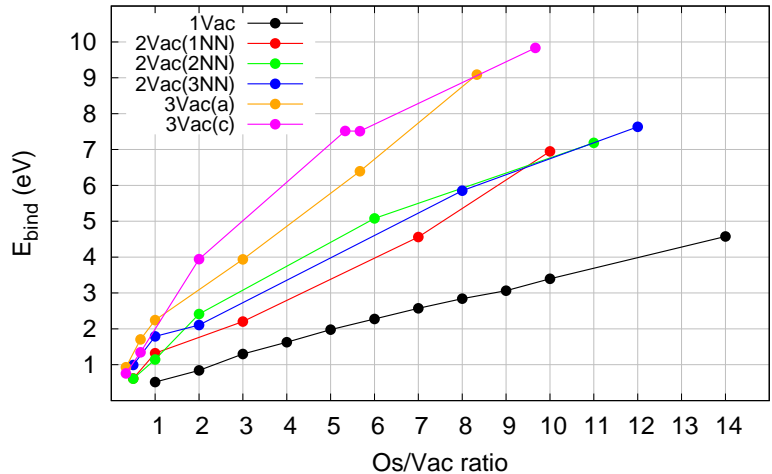}}
\caption{\label{Os_Vac} Binding energies for various  Os-vacancy configurations in W-Os-Vac system as a function of ratio between Os atoms and vacancies.  "3Vac(a)" and "3Vac(c)" refer to the structures 5a and 5b in \cite{Wrobel2017}}
	\label{fig:Os_Vac}
\end{figure*}

Fig.\ref{fig:Os_Vac} shows the dependence of binding energy as a function of Os to vacancy ratio in W-Os-Vac for single, di-vacancy and tri-vacancy configurations decorated by different Os atoms. Comparing with similar calculations shown in Fig.4a and Fig.4b of \cite{Wrobel2017},  for W-Re-vacancy system, where the maximum binding energy as a function of ratio between Re and vacancy is around 1.5 eV, it is found that the binding energy between Os and vacancy is much more strongly attractive. In particular for the tri-vacancy configurations shown in Figs. 5a and 5c of \cite{Wrobel2017}, the binding energies for Os interacting with tri-vacancies can be more than 6 times higher than those for the corresponding W-Re-Vac system. 

\begin{figure*}
\subfloat{
		\includegraphics[width=.8\linewidth]{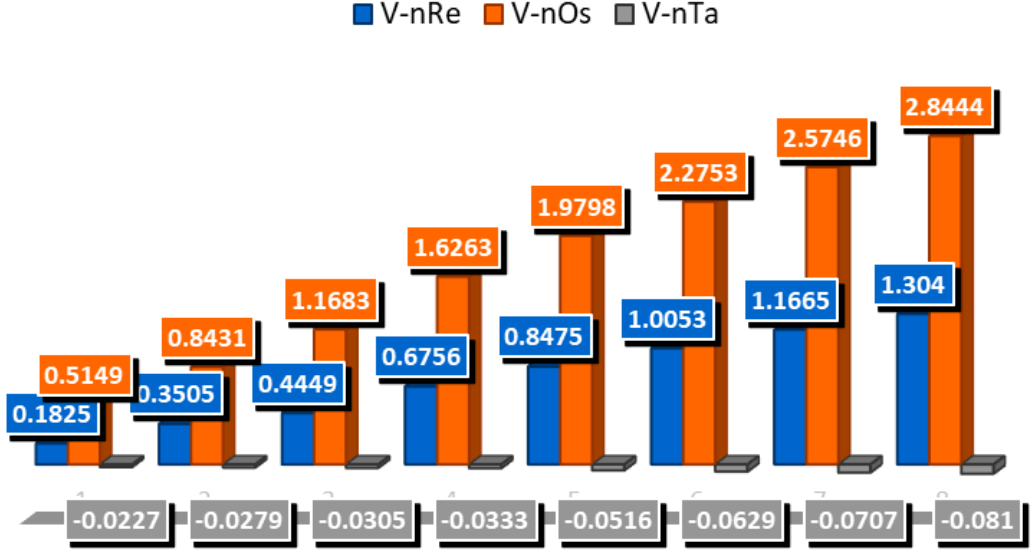}}
\caption{\label{VReOsTa} Binding energies of a single vacancy interacting with the first nearest neighbour atoms from 1 to 8 for separated solute Re, Os and Ta elements in bcc-W}
	\label{fig:VReOsTa}
\end{figure*}

Fig.\ref{fig:VReOsTa} provides a comparison of DFT trend of binding energies of a single vacancy in bcc-W interacting with three solute atoms Re, Os and Ta occupying n (n=1--8) first nearest neighbour positions. It is striking to see the different behaviours of the negative binding energies in Ta with respect to Re and Os. We will see in the next section, that this prediction plays an important role in explaining the decoration of Re and Os to vacancy clusters, whereas solute Ta atoms didn't have any effects in anomalous segregation under irradiation \cite{Xu2015,Xu2017}.       

To cross check the accuracy of our CE study, an independent DFT study of binding energies for selected defective configurations of vacancy-(Re)-(Os) clusters in W has also been performed within the VASP \cite{Messina2019}. Each configuration with 1 to 4 vacancies and a variable amount of Re and Os atoms (up to 4 components in total) was embedded in the center of a 5$\times$5$\times$5 super-cell with a bcc crystal structure and periodic boundary conditions on all sides. The simulations were performed with the standard projector-augmented wave (PAW) full-core pseudopotentials (i.e. without semi-core electrons), and the Perdew-Burke-Erzernhof (PBE) parametrization of the generalized-gradient approximation (GGA) was applied to the exchange-correlation function, with a plane-wave cutoff of 350 eV. A Monkhorst-Pack scheme with a 3x3x3 k-point grid was used to sample the Brillouin zone. For each relaxation, the cell volume and shape were kept fixed to the equilibrium lattice parameter of $a_{0}$ = 3.1725\AA obtained with the above parametrization. For each cluster composition, the configurations were picked semi-randomly to favor the most compact ones, and the maximum distance between one cluster component and the others could not exceed $\surd 3$$a_{0}$.

\subsection{Effective cluster interactions}
Mapping of the DFT data of enthalpy of mixing to the CE Hamiltonian in Eq. (\ref{eq:deltahmixing}) was performed using the ATAT package \cite{VandeWalle2002,VandeWalle2002a,VandeWalle2009,Hart2008}. The effective cluster interactions were obtained using the structure inversion method \cite{Connolly1983}. The many-body interactions $J_{|\omega|,n}^{(s)}$ are given by an inner product of cluster function $\Gamma_{\omega,n}^{(s)}$ and the corresponding energy \cite{Asta1991,Wolverton1994}, namely

\begin{equation}
J_{\omega,n}^{(s)}=\langle \Gamma_{\omega,n}^{(s)}(\vec{\sigma}), E(\vec{\sigma}) \rangle.
\label{eq:ECI_def}
\end{equation}

To arrive at a suitable choice of Hamiltonian (Eq. (\ref{eq:deltahmixing2})), the fitness of a cluster expansion can be, in general, quantified by means of leave-many-out cross-validation (LMO-CV) \cite{Hart2005}. The set of structures $\vec{\sigma}$ for which we carry out first-principles (DFT) energy calculations, $\Delta E_{DFT}(\vec{\sigma}_{input})$ is subdivided into a fitting set, $\{\vec{\sigma}\}_{fit}$\textcolor{red}{,} and a prediction set\textcolor{red}{,} $\{\vec{\sigma}\}_{pred}$.
We determine ECIs ($J_{|\omega|,n}^{(s)}$) by fitting to $\Delta E_{DFT}(\vec{\sigma})$ in $\{\vec{\sigma}\}_{fit}$. The predicted values of $\Delta E_{CE}(\vec{\sigma})$ for each structure, $\{\vec{\sigma}\}_{fit}$, are then compared with their known first-principles counterparts, $\Delta E_{DFT}(\vec{\sigma})$. This process can be iterated $M$ times for various subdivisions of $\{\vec{\sigma}\}_{input}$ into fitting and prediction sets,  $\{{\vec{\sigma}_{fit}}\}^{(m)}$ and $\{{\vec{\sigma}_{pred}}\}^{(m)}$ $(m=1...N)$. The values of the LMO-CV score is defined as the average prediction error of this iterative procedure,
\begin{equation}
(CV)^{2} = \frac{1}{M}\frac{1}{N_{pred}}\sum_{m=1}^{M}\sum_{\{\vec{\sigma}\}^{(m)}_{pred}} {\left[\Delta (E)^{(m)}_{CE}(\vec{\sigma})
-\Delta E_{DFT}(\vec{\sigma})\right]}{^2} ,
\label{eq:CV}
\end{equation}
In this work, the CE Hamiltonian defined by Eq. \ref{eq:deltahmixing2} is applied to a defective vacancy-rich W(Re,Os,Ta) solid solution where, as opposed to a conventional ternary non-defective alloy \cite{Wrobel2015}, there are no further ground-state stable structures that can be found in comparison with the ones predicted from equilibrium phase stability conditions. Therefore, the LMO-CV score can be reduced to the leave-one-out cross validation (LOO-CV) score with $M=1$  in Eq. (\ref{eq:CV}).
The LOO-CV score characterizing the quality of agreement between DFT and CE has been assessed using the input structures representing interactions in various W(Re,Os,Ta)-Vac systems. The overall values of the LOO-CV error are around 10 meV, confirming the very good accuracy of the set of ECIs used in the simulations.

\begin{figure*}
\subfloat{
		\includegraphics[width=.9\linewidth]{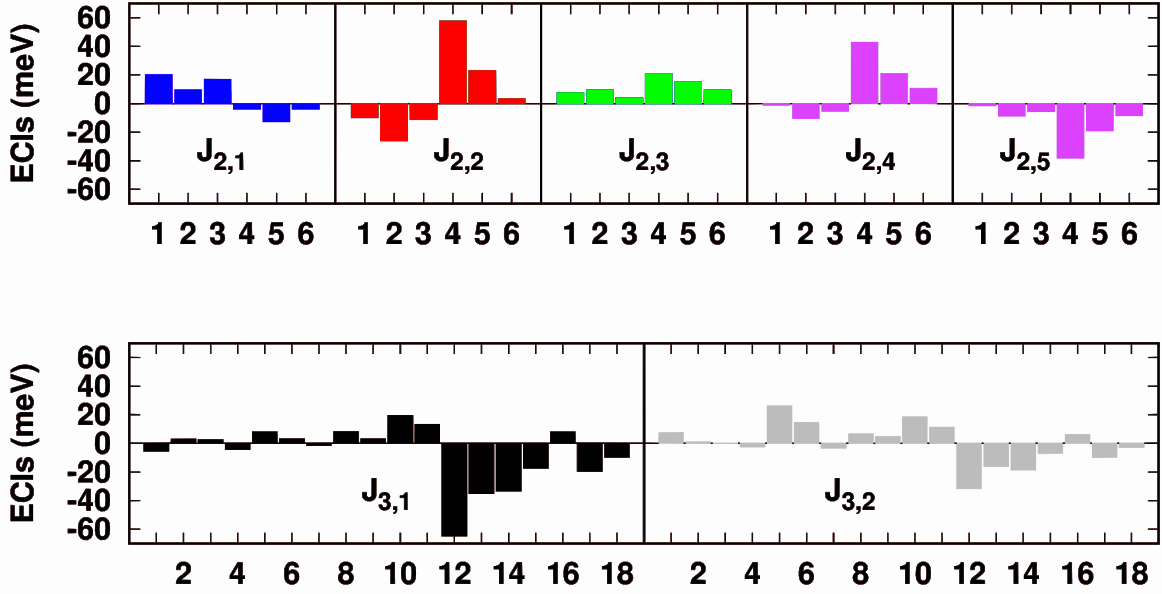}}
\caption{\label{quaternary-eci} Effective cluster interaction energies for quaternary W-Re-Os-Vac system.}
	\label{fig:quaternary-eci}
\end{figure*}

The values of ECIs ($J_{|\omega|,n}^{(s)}$) for different W(Re,Os,Ta)-Vac investigated systems are derived by mapping the calculated DFT energies onto corresponding CE Hamiltonian. According to the definition of cluster functions (Eq. (\ref{eq:gamma})) and point functions (Eq. (\ref{eq:gamma_def})), each cluster can be decorated in various ways for each nearest-neighbour shell of interactions. In this work, different sets of two-body and many-body interactions have been evaluated and analysed from the LOO-CV scores determined by Eq.(\ref{eq:CV}), to determine a optimized ECIs set for the quaternary W-Re-Os-Vac and two other ternary systems (W-Os-Vac and W-Ta-Vac). For the case of W-Re-Os-Vac system, using 641 DFT energy data described in the previous subsection, the optimized set for 30 two-body interactions within 5 nearest-neighbour shells and 36 three-body within the two nearest-neighbour shells in bcc lattice has been found with a very good LOO-CV score of 7.88 meV.  Values of all the optimized ECIs for ternary W-Re-Os-Vac alloys are shown in Fig. \ref{fig:quaternary-eci} and also in the last column of Table \ref{tab:ECI_quaternary}. A more detailed set of information about the ECIs in the two ternary systems: W-Ta-Vac and W-Os-Vac, can be found in the table S3 of the Supplementary Materials. 5 shells for the two-body interactions and 3 shells for three-body interactions were also optimized in these systems in the similar way as it has been discussed for the W-Re-Os-Vac system. The general expression of the CE Hamiltonian used in the present paper can be written in the form of Eq.(\ref{eq:deltahmixing2}) where the high-order many-body interactions starting from four-body ones were not included into the consideration.  
\newpage
\begin{table*}
\begin{scriptsize}
\caption{\footnotesize Size $|\omega|$, label $n$, decoration $(s)$, multiplicity $m_{|\omega|,n}^{(s)}$ and coordinates of points in clusters on bcc lattice. $J_{|\omega|,n}^{(s)}$ (in meV) are effective cluster interactions (ECIs) calculated in the framework of CE for bcc ternary W-Re-Os-vacancy system. Index $(s)$ is the same as the sequence of points in the corresponding cluster.
        \label{tab:ECI_quaternary}}
\begin{ruledtabular}
\begin{tabular}{cccccc}
    $|\omega|$ & $n$ & ($s$) & Coordinates & $m_{|\omega|,n}^{(s)}$ & $J_{|\omega|,n}^{(s)}$  \\ 
\hline
    0     &       & (0)   &       & 1     & 417.812 \\
    1     &       & (1)   & \multicolumn{1}{l}{$(0,0,0)$} & 1     & -436.091 \\
    1     &       & (2)   &       & 1     & 1516.983 \\
    1     &       & (3)   &       & 1     & 785.518 \\
    2     & 1     & (1,1) & \multicolumn{1}{l}{$(0,0,0; \frac{1}{2},\frac{1}{2},\frac{1}{2})$} & 4     & 20.138 \\
          &       & (1,2) &       & 8     & 9.511 \\
          &       & (1,3) &       & 8     & 16.855 \\
          &       & (2,2) &       & 4     & -3.851 \\
          &       & (2,3) &       & 8     & -12.648 \\
          &       & (3,3) &       & 4     & -3.812 \\
    2     & 2     & (1,1) & \multicolumn{1}{l}{$(0,0,0; 1,0,0)$} & 3     & -9.602 \\
          &       & (1,2) &       & 8     & -25.909 \\
          &       & (1,3) &       & 8     & -11.196 \\
          &       & (2,2) &       & 4     & 57.626 \\
          &       & (2,3) &       & 8     & 23.074 \\
          &       & (3,3) &       & 4     & 3.306 \\
    2     & 3     & (1,1) & \multicolumn{1}{l}{$(0,0,0; 1,0,1)$} & 6     & 7.353 \\
          &       & (1,2) &       & 8     & 9.532 \\
          &       & (1,3) &       & 8     & 3.979 \\
          &       & (2,2) &       & 4     & 20.860 \\
          &       & (2,3) &       & 8     & 15.293 \\
          &       & (3,3) &       & 4     & 9.420 \\
    2     & 4     & (1,1) & \multicolumn{1}{l}{$(0,0,0; 1\frac{1}{2},\frac{1}{2},\frac{1}{2})$} & 12    & -1.235 \\
          &       & (1,2) &       & 8     & -10.263 \\
          &       & (1,3) &       & 8     & -5.422 \\
          &       & (2,2) &       & 4     & 42.922 \\
          &       & (2,3) &       & 8     & 20.977 \\
          &       & (3,3) &       & 4     & 10.532 \\
    2     & 5     & (1,1) & \multicolumn{1}{l}{$(0,0,0; 1,1,1)$} & 4     & -1.533 \\
          &       & (1,2) &       & 8     & -8.546 \\
          &       & (1,3) &       & 8     & -5.637 \\
          &       & (2,2) &       & 4     & -37.997 \\
          &       & (2,3) &       & 8     & -18.990 \\
          &       & (3,3) &       & 4     & -8.440 \\
    3     & 1     & (1,1,1) & \multicolumn{1}{l}{$(1,0,0; \frac{1}{2},\frac{1}{2},\frac{1}{2};$} & 12    & -5.273 \\
          &       & (2,1,1) & \multicolumn{1}{l}{$0,0,0)$} & 24    & 2.784 \\
          &       & (3,1,1) &       & 24    & 2.490 \\
          &       & (1,2,1) &       & 12    & -4.196 \\
          &       & (2,2,1) &       & 24    & 7.930 \\
          &       & (3,2,1) &       & 24    & 3.238 \\
          &       & (1,3,1) &       & 12    & -1.427 \\
          &       & (2,3,1) &       & 24    & 8.100 \\
          &       & (3,3,1) &       & 24    & 2.923 \\
          &       & (2,1,2) &       & 12    & 19.352 \\
          &       & (3,1,2) &       & 24    & 13.219 \\
          &       & (2,2,2) &       & 12    & -64.614 \\
          &       & (3,2,2) &       & 24    & -34.958 \\
          &       & (2,3,2) &       & 12    & -33.269 \\
          &       & (3,3,2) &       & 24    & -17.291 \\
          &       & (3,1,3) &       & 12    & 7.910 \\
          &       & (3,2,3) &       & 12    & -19.429 \\
          &       & (3,3,3) &       & 12    & -9.533 \\
          \end{tabular}
\end{ruledtabular}
\end{scriptsize}
\end{table*}

\setcounter{table}{0}
\begin{table*}
\begin{scriptsize}
\caption{\footnotesize (Continue) Size $|\omega|$, label $n$, decoration $(s)$, multiplicity $m_{|\omega|,n}^{(s)}$ and coordinates of points in clusters on bcc lattice. $J_{|\omega|,n}^{(s)}$ (in meV) are effective cluster interactions (ECIs) calculated in the framework of CE for bcc ternary W-Re-Os-vacancy system. Index $(s)$ is the same as the sequence of points in the corresponding cluster.}
\begin{ruledtabular}
 \begin{tabular}{cccccc}
$|\omega|$ & $n$ & ($s$) & Coordinates & $m_{|\omega|,n}^{(s)}$ & $J_{|\omega|,n}^{(s)}$  \\
\hline
    3     & 2     & (1,1,1) & \multicolumn{1}{l}{$(\frac{1}{2},-\frac{1}{2},-\frac{1}{2}; 0,0,0;$} & 12    & 7.270 \\
          &       & (2,1,1) & \multicolumn{1}{l}{$-\frac{1}{2},-\frac{1}{2},\frac{1}{2})$} & 24    & 1.042 \\
          &       & (3,1,1) &       & 24    & -0.293 \\
          &       & (1,2,1) &       & 12    & -2.316 \\
          &       & (2,2,1) &       & 24    & 26.156 \\
          &       & (3,2,1) &       & 24    & 14.325 \\
          &       & (1,3,1) &       & 12    & -3.307 \\
          &       & (2,3,1) &       & 24    & 6.749 \\
          &       & (3,3,1) &       & 24    & 4.755 \\
          &       & (2,1,2) &       & 12    & 18.534 \\
          &       & (3,1,2) &       & 24    & 11.105 \\
          &       & (2,2,2) &       & 12    & -31.691 \\
          &       & (3,2,2) &       & 24    & -16.092 \\
          &       & (2,3,2) &       & 12    & -18.567 \\
          &       & (3,3,2) &       & 24    & -6.818 \\
          &       & (3,1,3) &       & 12    & 6.169 \\
          &       & (3,2,3) &       & 12    & -9.816 \\
          &       & (3,3,3) &       & 12    & -2.806 \\
    \end{tabular}
\end{ruledtabular}
\end{scriptsize}
\end{table*}

\begin{figure*}
\subfloat{
		\includegraphics[width=.8\linewidth]{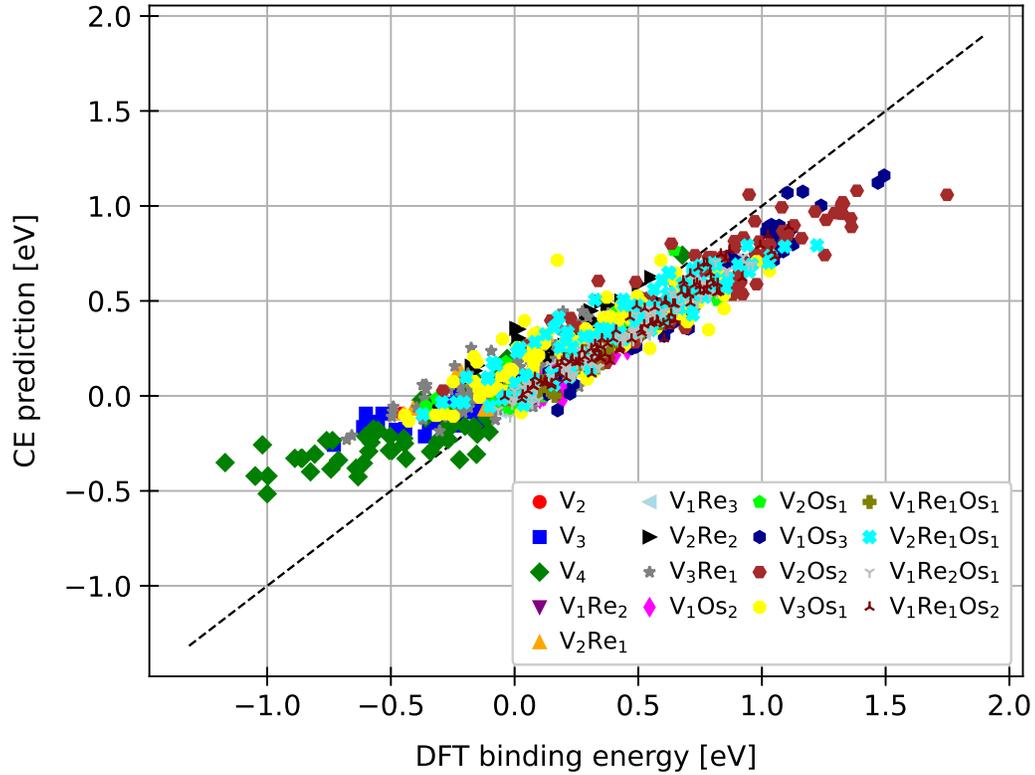}}
\caption{A comparison of binding energies between the present CE model and DFT data performed independently for various configurations of  defective Vac-Re-Os clusters in 5$\times$5$\times$5 bcc super-cell with W. V denotes vacancy component}
	\label{fig:CV_DFT}
\end{figure*}

The predictive power of the obtained ECIs for W-Re-Os-Vac can be checked by comparing binding energy values calculated by the present CE model with those from an independent set of DFT calculations for semi-random configurations (see the previous subsection) of Vac-Re-Os defective clusters. These calculations were performed for 1379 configurations of  Re-Os-Vac clusters generated with the fifth nearest neighbour distance in a super-cell of 250 sites. The list of configurations is then split into about 80$\%$ of compact and 20$\%$ in the less compact ones for different cluster composition. The result from a general comparison, shown in Fig.\ref{fig:CV_DFT}, demonstrates that there is a good  agreement between our CE model and the DFT data for new defect configurations over all configuration with the average error of 0.168 eV, despite the fact that there are uncertainties of the two DFT sets of calculations in terms of employed PAWs, cut-off energies and the relaxation methods. It is also important to note that the binding energies of vacancy clusters in bcc-W are very sensitive to detailed DFT calculations, as it has been discussed elsewhere \cite{Muzyk2011,Mason2017,Mason2019}. Therefore the higher values obtained from the CE model for four-vacancy clusters are understandable.       

\section{Results from Monte-Carlo simulations}

Within the framework of the present constrained CE model for investigating steady-state of alloys containing high concentration of vacancies, the finite-temperature properties of W(Re,Os,Ta)-Vac system were analysed using quasi-canonical MC simulations and ECIs derived from DFT calculations. 

\subsection{Comparison between solute Re, Os and Ta} 

In this sub-section, MC simulations at various temperatures were performed to understand the anomalous segregation of different solute atoms (Re,Os and Ta), which are the main transmutation elements occurring in W under neutron irradiation conditions. It allows us to understand meta-stable alloy configurations with defects at irradiated conditions with a constant temperature.

\begin{figure*}
	\subfloat{
		\includegraphics[width=0.30\linewidth]{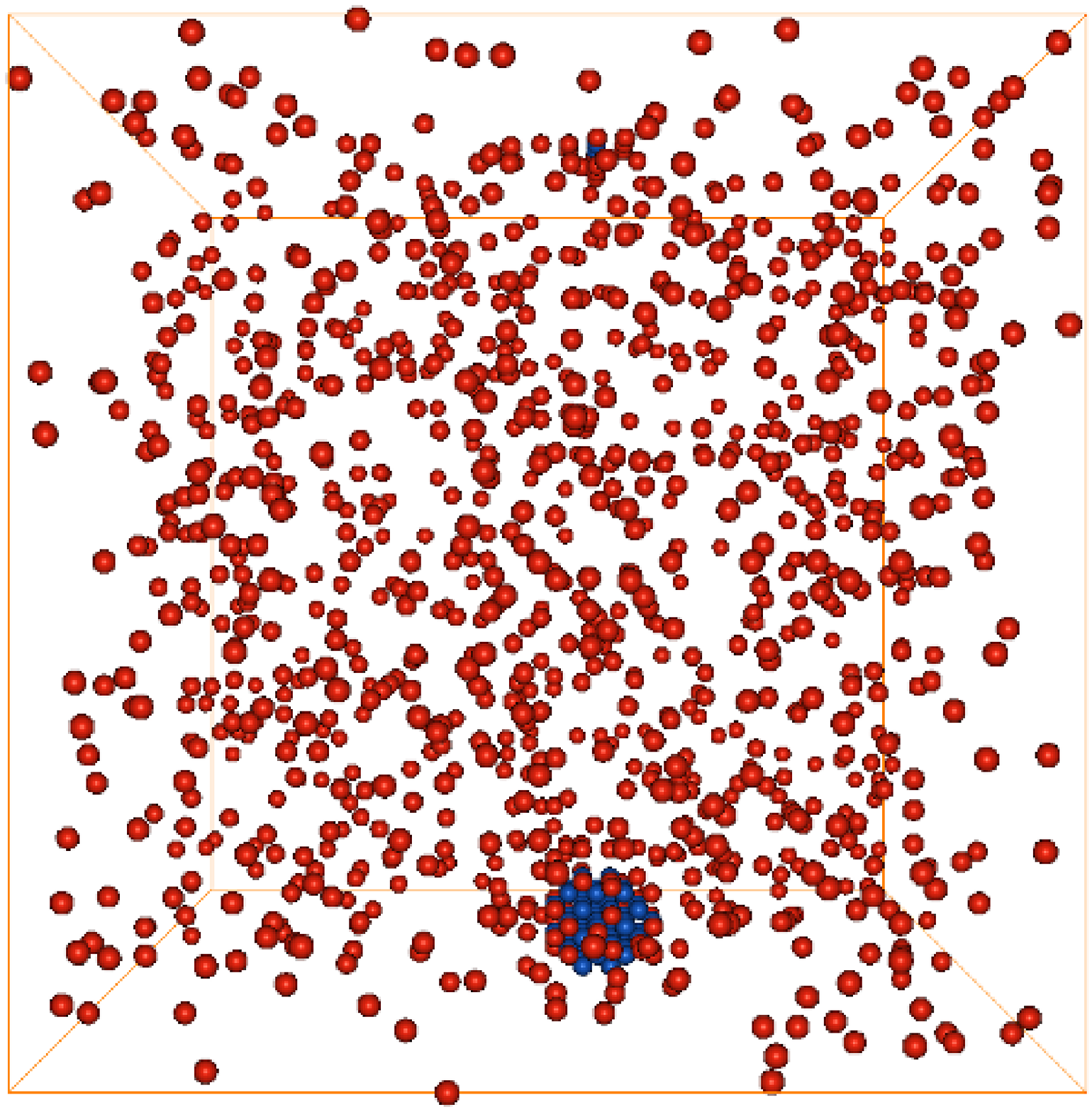}}
	\subfloat{
		\includegraphics[width=0.30\linewidth]{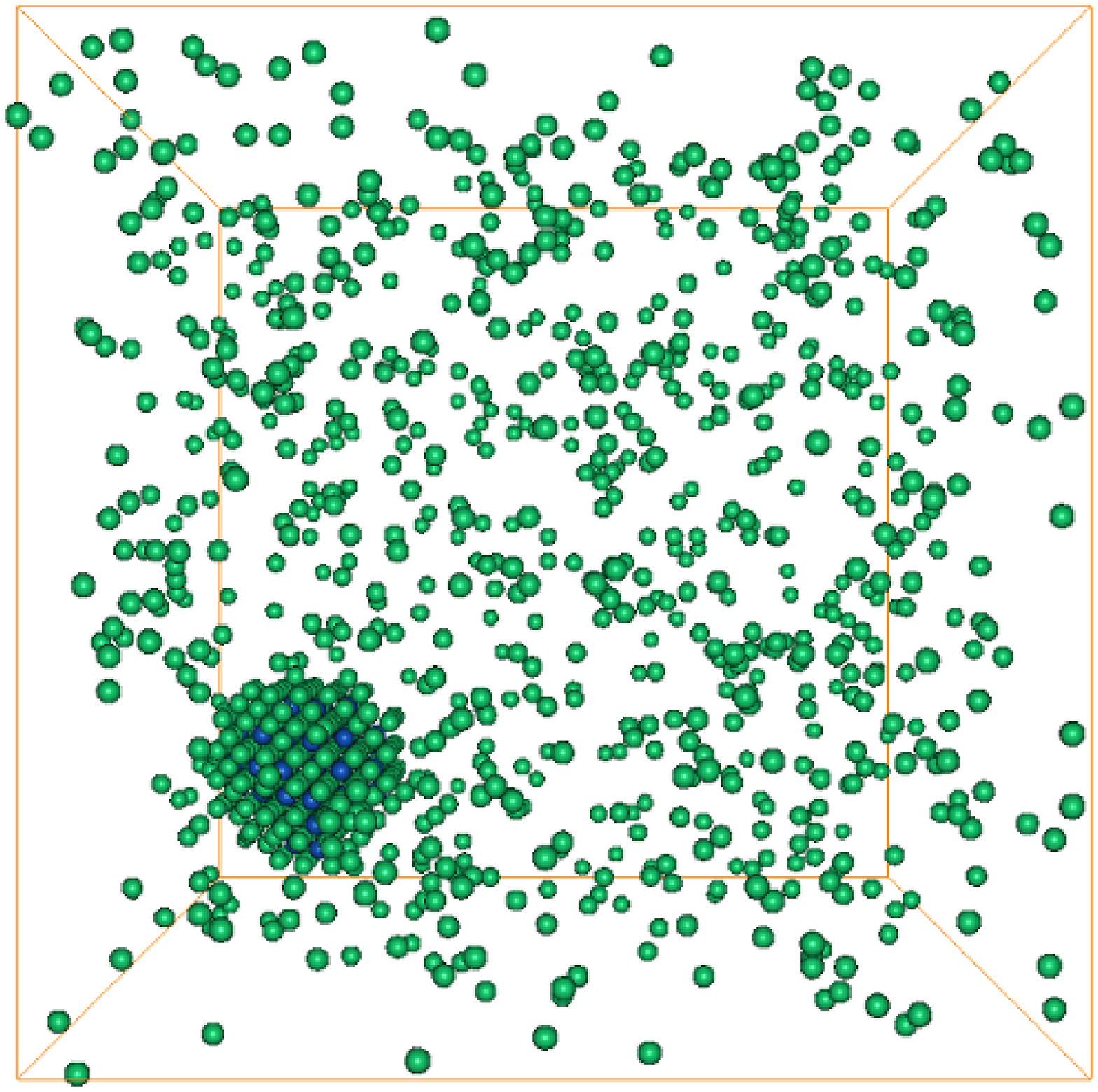}}
	\subfloat{
		\includegraphics[width=0.30\linewidth]{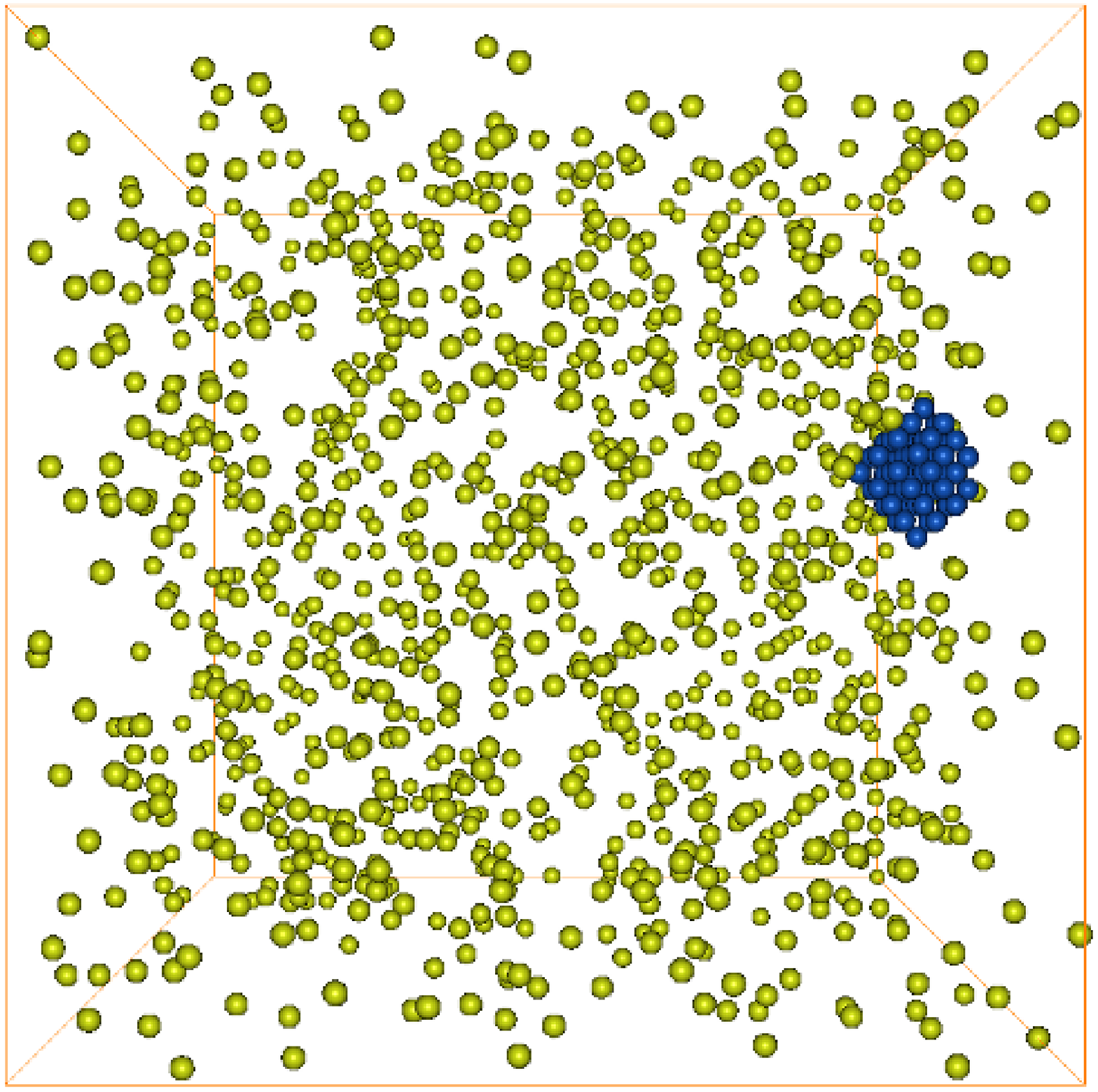}}\\
	\subfloat{
		\includegraphics[width=0.30\linewidth]{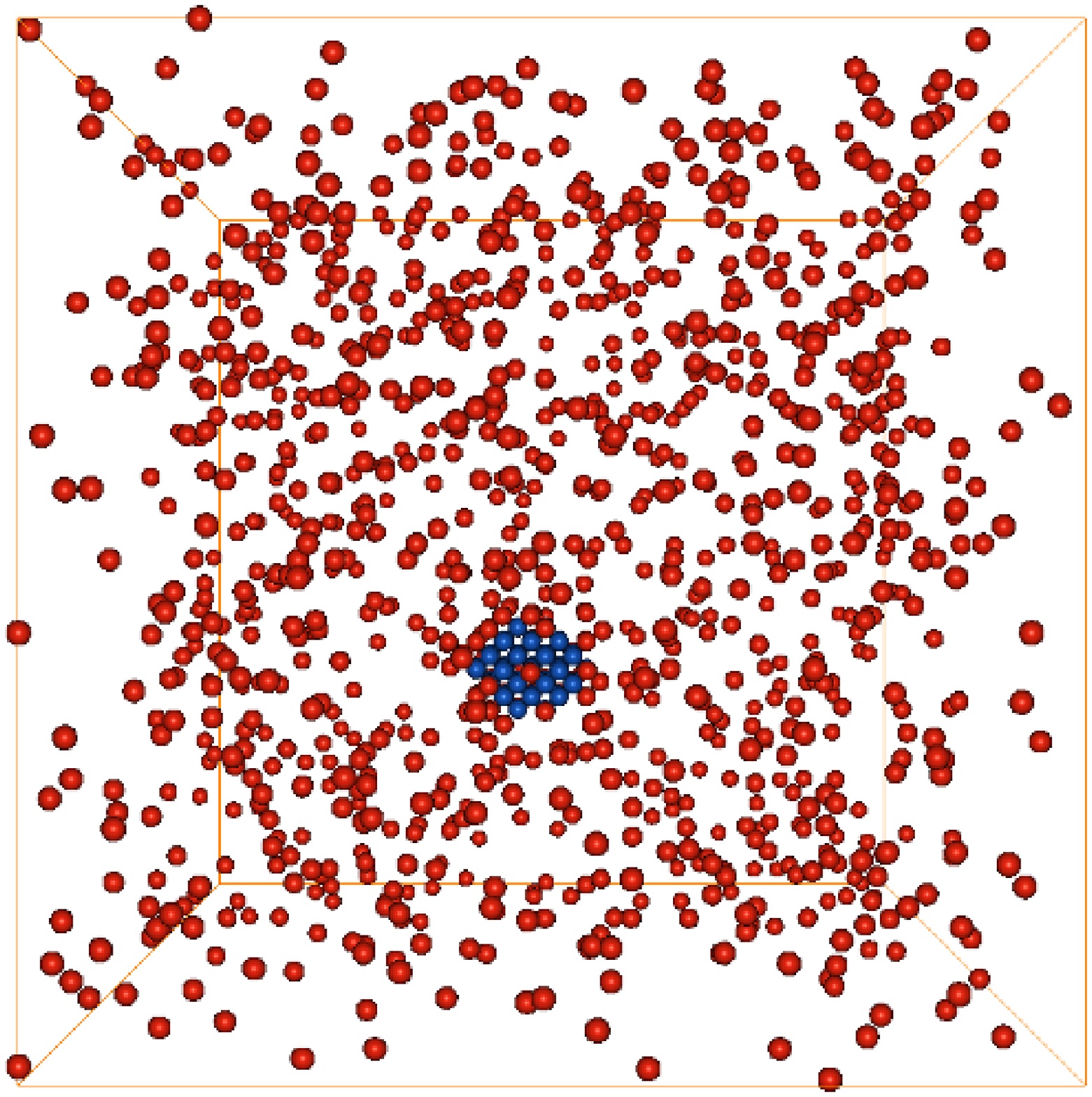}}
	\subfloat{
		\includegraphics[width=0.30\linewidth]{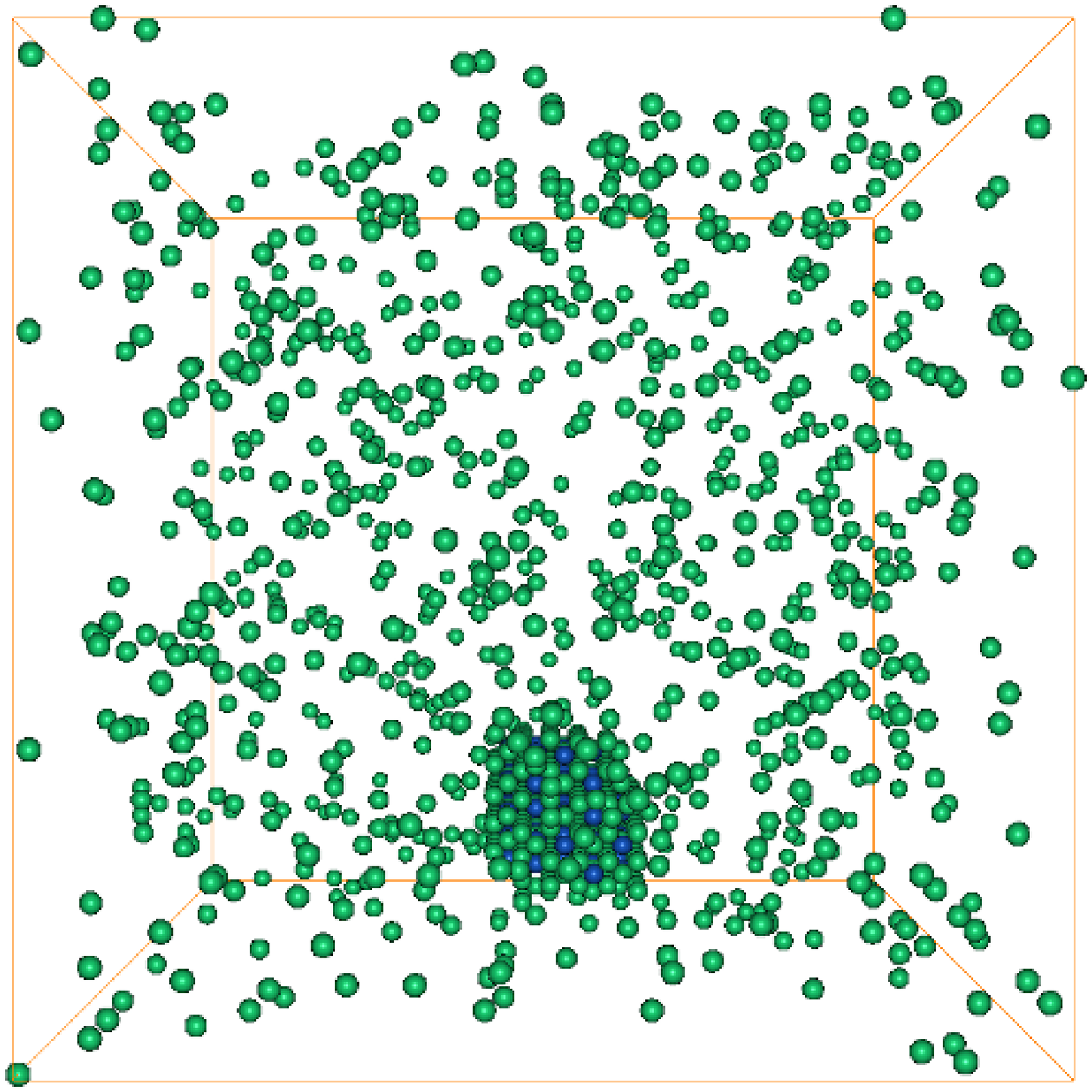}}
	\subfloat{
		\includegraphics[width=0.30\linewidth]{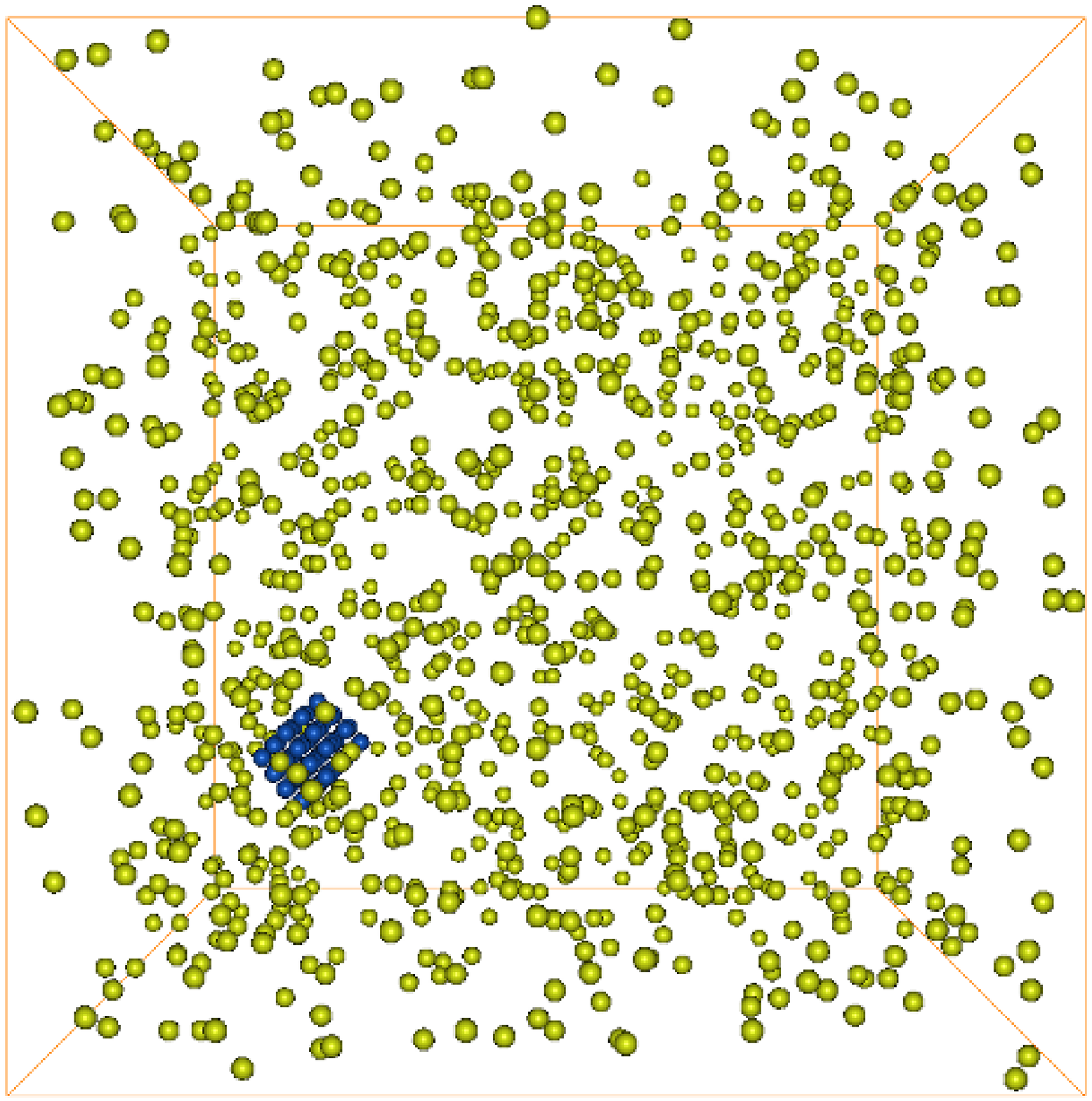}}
\caption{\label{ReOsTa} A comparison of MC simulations for W-Re-Vac (left), W-Os-Vac (middle) and W-Ta-Vac (right) systems at T=800K (top) and T=1200K (bottom).}
\label{fig:ReOsTa}
\end{figure*}

Fig.\ref{fig:ReOsTa} shows structures of W-Re-Vac, W-Os-Vac and W-Ta-Vac alloys with 2 \% at.Re, Os and Ta, obtained from MC simulations performed assuming $T=$800 K, 1200 K  and 0.1 \% of vacancy concentration. The screen shots in this figure were taken 20000 MC steps per atom inside a 30$\times$3$\times$30 simulation box. The result for solute Re case at $T$=800 K is identical  to the one investigated previously (see Table 2 from \cite{Wrobel2017}) from which clustering of vacancies in the form of voids, or vacancy clusters decorated by Re atoms is observed. The choice of the other temperature value, $T$=1200 K, is motivated by comparing our simulations with the available experimental results for neutron irradiated W \cite{Klimenkov2016,Lloyd2019}.  

In the W-2\%Os alloy at 800 K, we observe either the formation of voids surrounded by Os with much stronger formation of voids, or clusters of vacancies, with a pronounced concentration of solute Os in a comparison with those for Re. The situation is completely different for the case with W-2$\%$Ta, for which Ta atoms have a tendency to move away from voids formed in bcc-W. The different behaviour observed in MC simulations is consistent with DFT predictions of the binding energy trend for vacancy interacting with Re, Os and Ta as shown in Fig.(\ref{fig:VReOsTa}). A stronger attraction of Os to the voids in a comparison with Re is also supported by the analysis of much higher binding energy values as a function of ratio of solute to vacancy concentration displayed in Fig.(\ref{fig:Os_Vac}).  At $T$=1200 K, both Re and Os atoms are more weakly bound to the voids\textcolor{red}{,} whereas again there is  no segregation of Ta atoms to the vacancy clusters. This finding is in a very good agreement with the finding from Atom Probe Tomography experimental observation \cite{Xu2015,Xu2017} that there are stronger effects of radiation induced segregation of Os in a comparison with Re, whereas there is no effect of Ta precipitation under self-ion irradiation in W.

\begin{figure*}
\subfloat{
		\includegraphics[width=.8\linewidth]{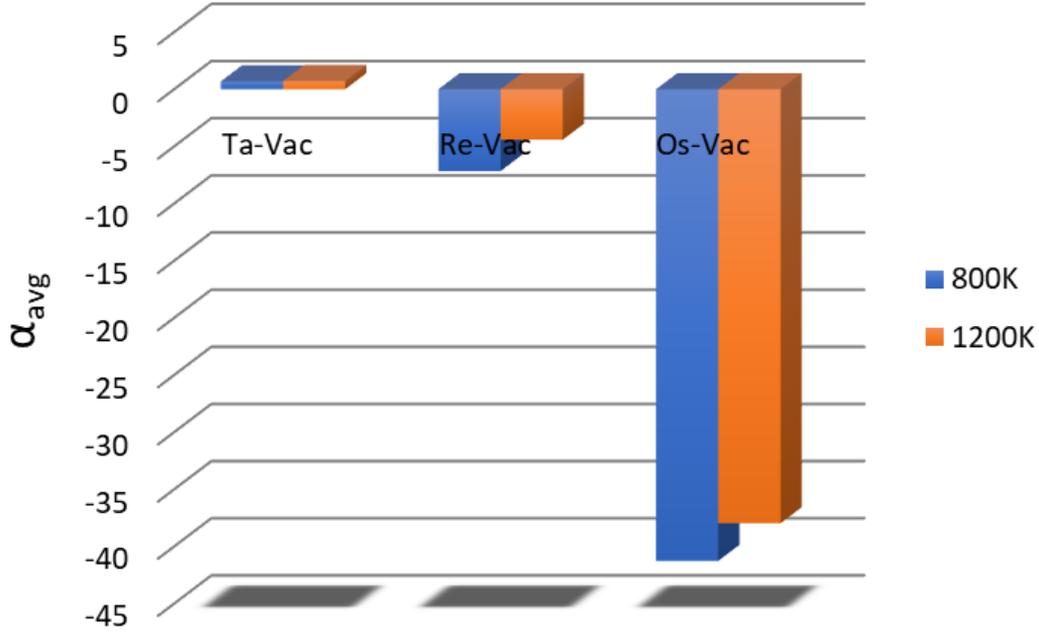}}
\caption{\label{SROReOsTa} A comparison of average short-range order, (see Eq. \ref{eq:SRO_avg}),  parameters between vacancy and solute atoms Re, Os and Ta at two different temperatures}
	\label{fig:SROReOsTa}
\end{figure*}

The distinct behaviour between the 3 solute cases can be explained by analysing the short-range order parameters evaluated from MC simulations in the regions close to the voids of vacancy cluster formation.  
Fig.(\ref{fig:SROReOsTa}) shows SRO parameters for three Re-Vac, Os-Vac and Ta-Vac pairs calculated at two different temperatures (T=800K and 1200K). The SRO parameters are computed from the  cluster correlation functions via Eqs.(\ref{eq:pair-probability}),(\ref{eq:WC}), (\ref{eq:sromatrix}) and (\ref{eq:SRO_avg}) over the whole simulation cell. It is noted from Fig.(\ref{fig:ReOsTa}) all the vacancies in structures simulated at 800 K and 1200K are located inside the vacancy clusters. Therefore, the X-Vac SRO parameters refer only to the chemical properties of solute elements nearby to voids or vacancy clusters. For Ta-Vac pairs it is found that the SRO parameters are positive confirming the segregation of Ta from vacancy in the investigated  MC simulations. On the contrary, the SRO parameters are negative for W-Re pairs and they are strongly negative for Os-Vac pairs demonstrating different degree of attractive interaction between these atoms with with vacancy clusters. The present analysis using SRO parameters derived from  the matrix formation of CE is very efficient for understanding properties of multi-component system at different temperatures, as well as functions of arbitrary composition including those of defect concentration. It is also consistent with the previous analysis using the chemical pair interactions derived from the CE Hamiltonian within the two-body approximation. For the ternary system like W-X-Vac with X=Re,Os,Ta, the energy formula is:  

\begin{eqnarray}
\Delta H_{CE}(\vec{\sigma}) &=& J_1^{(0)}+J_1^{(1)}\left(1-3x_W\right) + J_1^{(2)}\frac{\sqrt{3}}{2}\left(x_X-x_{Vac}\right) \nonumber \\
&-&4\sum_{n}^{pairs} \left(V_n^{WX}y_n^{WX} + V_n^{WVac}y_n^{WVac}+V_n^{XVac}y_n^{XVac}\right) + \sum_{n}^{multibody} \ldots ,
\label{eq:CE_vs_V}
\end{eqnarray}

It is found from our construction of CE model, with only two-body cluster interactions in a similar way as it has been done for the W-Re-Vac system \cite{Wrobel2017}, that the values of chemical pair interaction for $V^{TaVac}$ are negative whereas those $V^{OsVac}$ are positive for first and second nearest neighbours. From Eq.(\ref{eq:CE_vs_V}), the results lead to the repulsive interaction between Ta and vacancy and attractive ones of Os with vacancy that is in agreement with the above SRO analysis at finite temperature.

\subsection{Os effects in W-Re-Os-Vac system} 

Taking into account the result from the previous subsection that the contribution of Ta into radiation induced segregation is insignificant, the focus of this section is on the effects of both Re and Os elements in defective bcc-W system at finite temperature.  
Table \ref{tab:MC_results_const_Re} shows the results of Monte-Carlo simulations at two different temperatures (T=800K and 1200K) for W-Re1.5$\%$ alloy with 4 different values of Os concentration (0.25$\%$, 0.50$\%$, 0.75$\%$ and 1.00$\%$) and two values of vacancy concentrations (0.25$\%$ and 0.50$\%$). Although MC simulations can be performed also with different Re concentration, the choice of constant value of Re1.5$\%$ is again motivated by a comparison with the transmutation products from neutron irradiated W samples investigated in experimental works \cite{Klimenkov2016,Lloyd2019} which will be discussed more in detail in the next subsection. 

Overall, as can be seen from Table \ref{tab:MC_results_const_Re}, Os has a significant role in the formation of sponge-like Os-vacancy clusters, formed with higher concentrations of Os. Voids decorated by both Re and Os can be observed even at low Os concentration. Despite a higher value of Re concentration in a comparison with Os, the present study shows that formation of Re-Vac clusters are not observed within the investigated W-Re-Os-Vac system, in a striking variance with the W-Re-Vac without Os (see results from Fig(.\ref{fig:ReOsTa}) and in \cite{Wrobel2017}). With regard to vacancy concentration and temperature dependence, the sponge-like clusters are the only form of precipitation in the alloys with larger concentration of Os and smaller concentration of vacancies, namely for Os1.0$\%$Vac0.2$\%$ at 800K and 1200 K, Os0.75$\%$Vac0.2$\%$ at 800K and 1200 K and Os0.5$\%$Vac0.2$\%$ at 1200K. From the other side, they are not formed for alloys with smaller concentrations of Os and larger concentrations of vacancies: Os0.25$\%$Vac0.2$\%$ at 1200K, Os0.25$\%$Vac0.5$\%$ at 800K and 1200K, Os0.5$\%$Vac0.5$\%$ at 1200K and Os0.75$\%$Vac0.5$\%$ at 1200K. The sponge-like structures form more easily at lower temperatures. For Vac0.5$\%$, they are observed for considered alloys at 800K\textcolor{red}{,} whereas they are not formed at 1200K (with an exception of Os1.0$\%$Vac0.5$\%$ alloy). Voids form more easily for larger concentrations of vacancies. For Vac0.5$\%$, they are formed for all considered alloy compositions, whereas for Vac0.2$\%$ they are formed only for alloys with smaller concentrations of Os (Os0.25$\%$Vac0.2$\%$ and Os0.5$\%$Vac.5$\%$). 

It is important to note that the formation of voids and vacancy clusters in presence of Re in bcc-W has been systematically investigated from DFT calculations. According to previous studies \cite{Muzyk2011,Wrobel2017,Mason2017,Mason2019} void configurations, starting from three-vacancy clusters in the 1NN and 2NN configurations, to a 15-vacancy cluster, are stable in a pure W matrix. When the Re/vacancy ratio increases, configurations with vacancies in the 3NN coordination shell become more stable, similarly to how it was found in the tri-vacancy case with the same relative positions of vacancies. DFT calculations show that rhenium-vacancy binding energies can be as high as 1.5 eV,  if the Re/vacancy ratio is in the range from 2.4 to 6.6. The present DFT study of Os interactions with single and clusters of vacancies has demonstrated that the corresponding binding energies, as shown in Figs.\ref{fig:VReOsTa} and (\ref{fig:Os_Vac}), are much stronger than those for Re-vacancy interactions.

\begin{table*}
\caption{Results of Monte Carlo simulations as functions of Os and vacancy concentration and temperature in W-Re-Os-Vac system. Concentration of Re atoms is fixed and equal to 1.5\%. MC simulations were performed at various temperatures and screen shots were taken after 20000 MC steps per atom. Again, red color represents Re atoms, green: Os atoms and blue:vacancies 
        \label{tab:MC_results_const_Re}}
\begin{ruledtabular}
    \begin{tabular}{|c|c|c|c|c|}
              & WRe1.5\%Os0.25\%Vac0.2\% & WRe1.5\%Os0.5\%Vac0.2\% & WRe1.5\%Os0.75\%Vac0.2\%  &  WRe1.5\%Os1.0\%Vac0.2\%\\
     \hline
     800 K & \includegraphics[width=.2\linewidth]{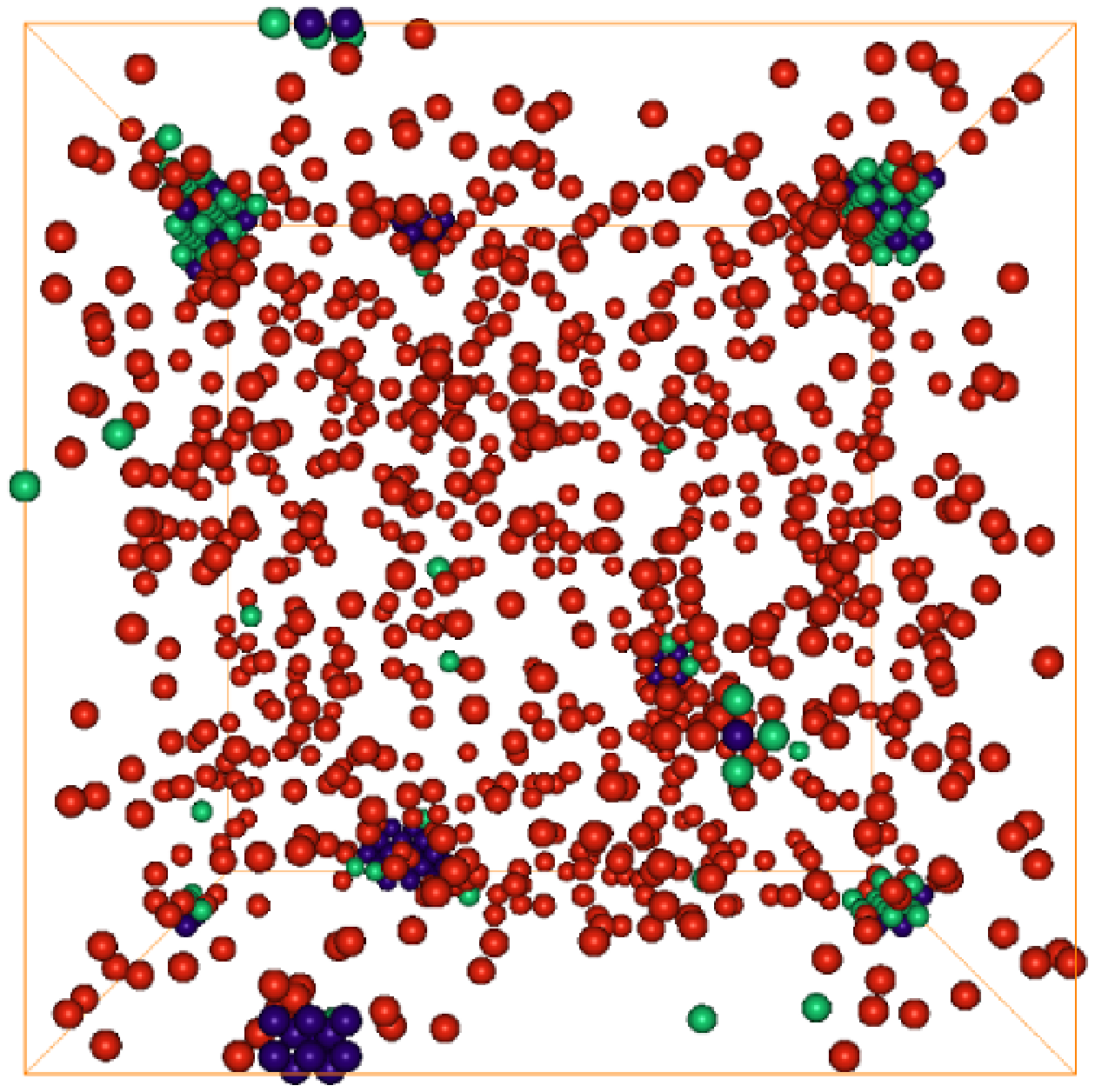} & \includegraphics[width=.2\linewidth]{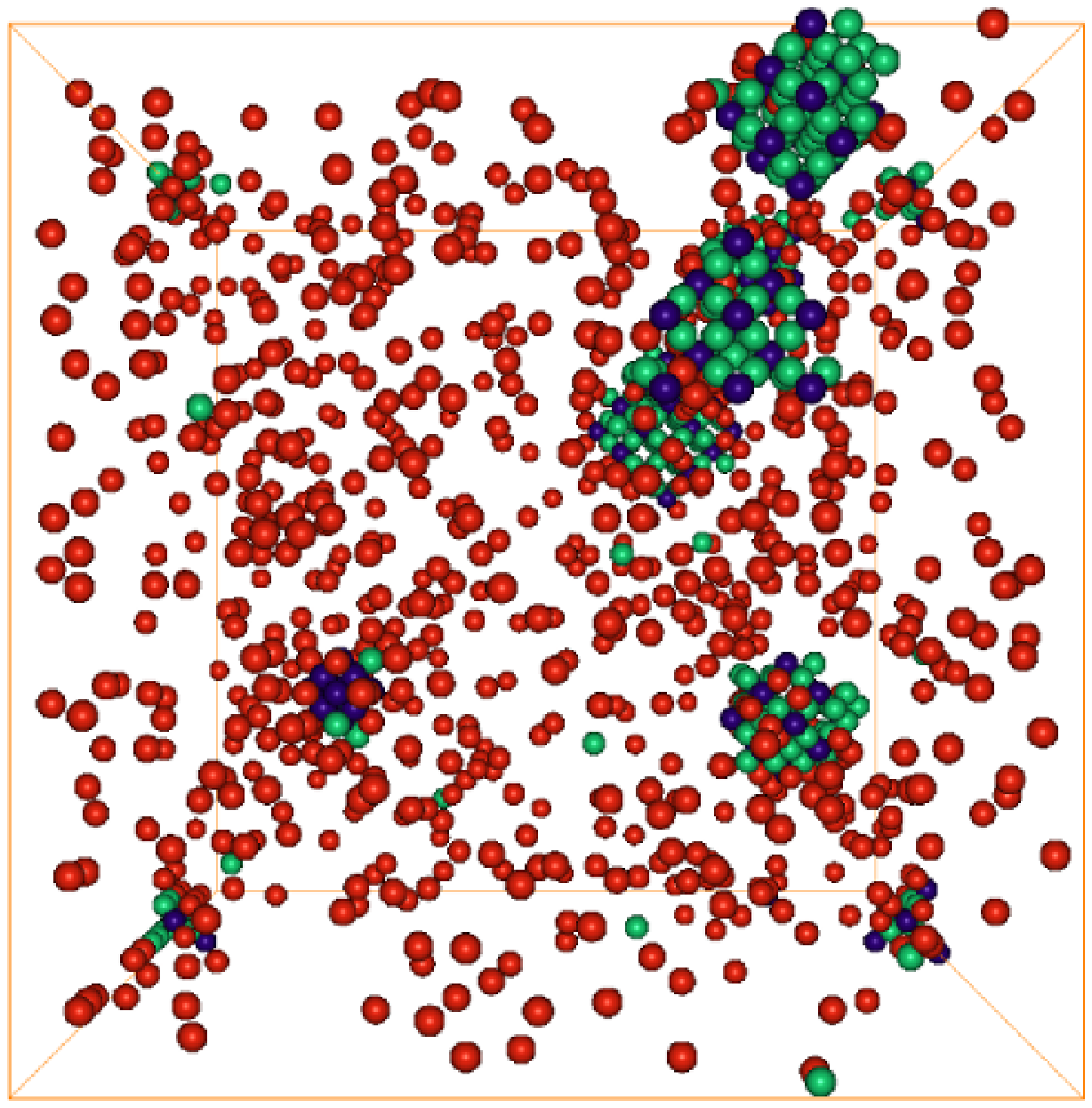} & \includegraphics[width=.2\linewidth]{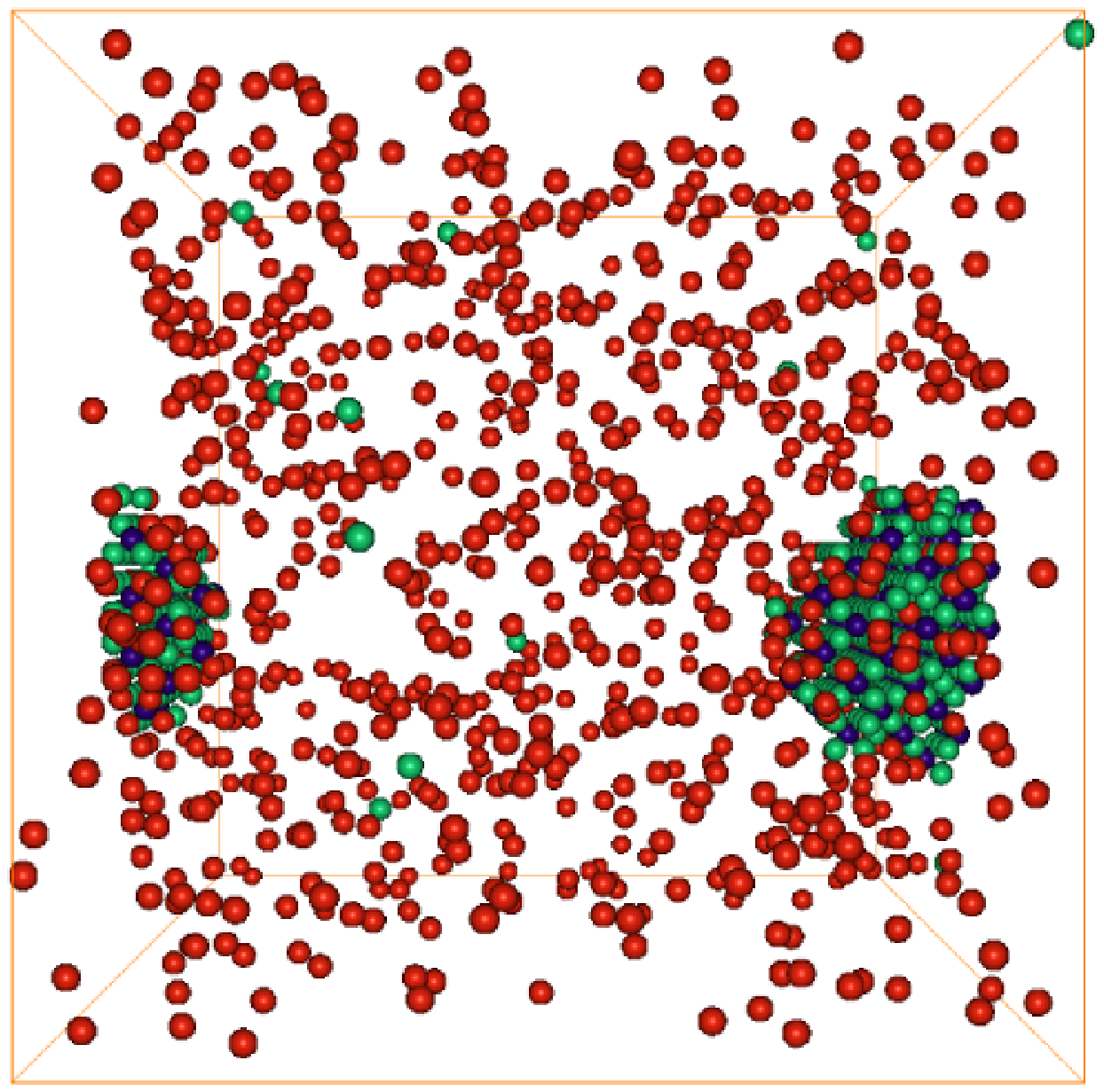} & \includegraphics[width=.2\linewidth]{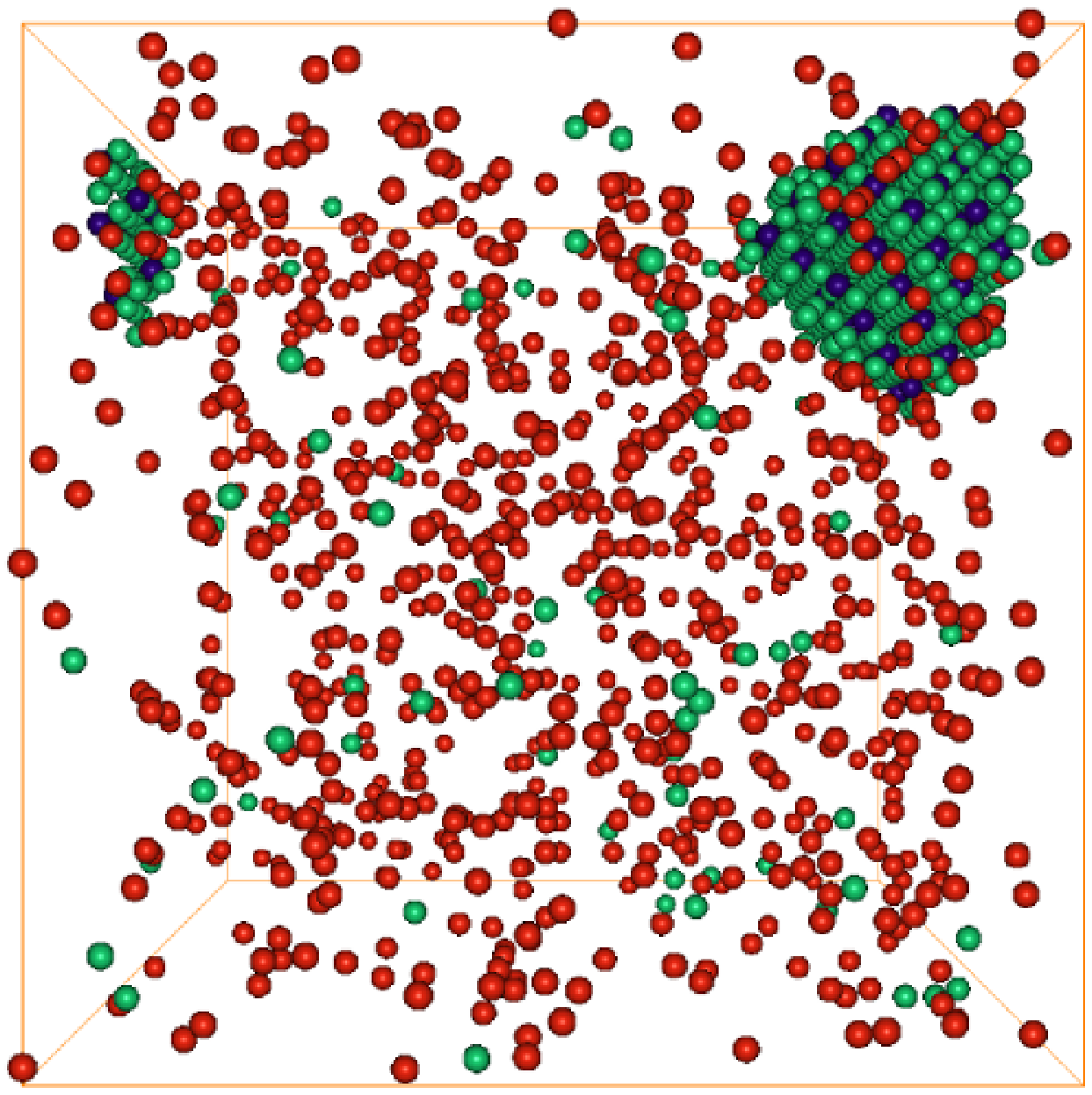} \\
     \hline
     1200 K & \includegraphics[width=.2\linewidth]{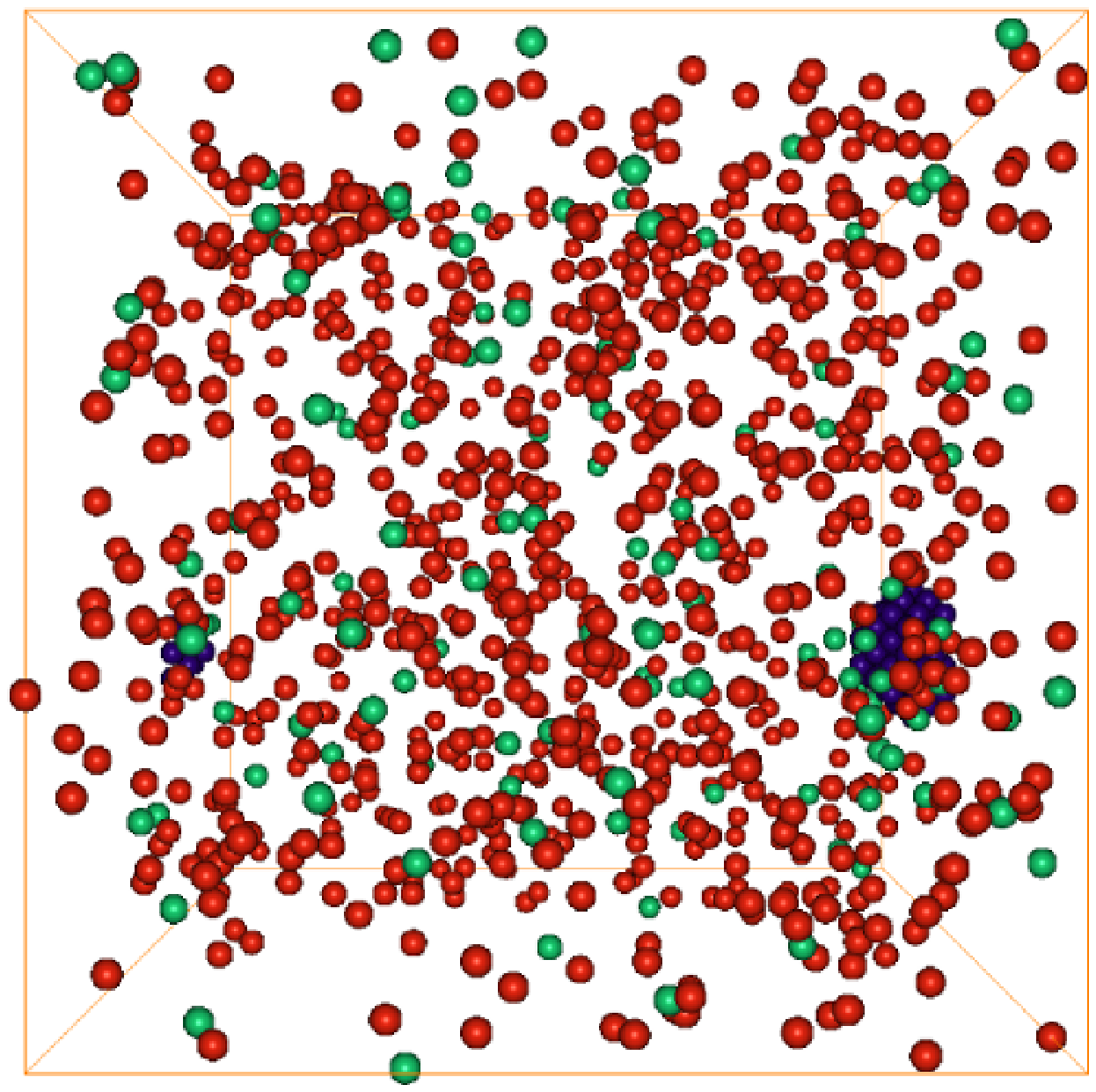} & \includegraphics[width=.2\linewidth]{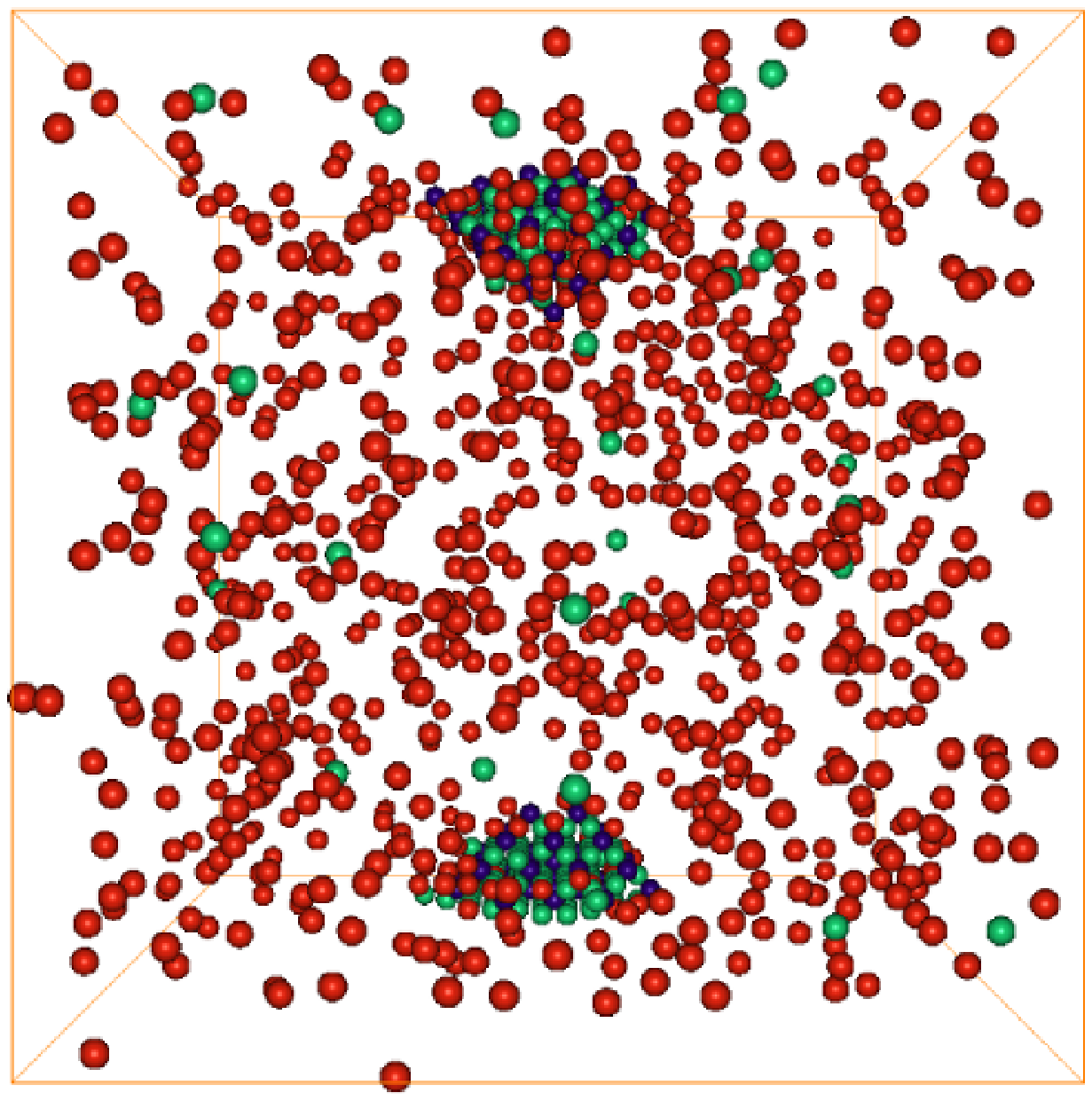} &  \includegraphics[width=.2\linewidth]{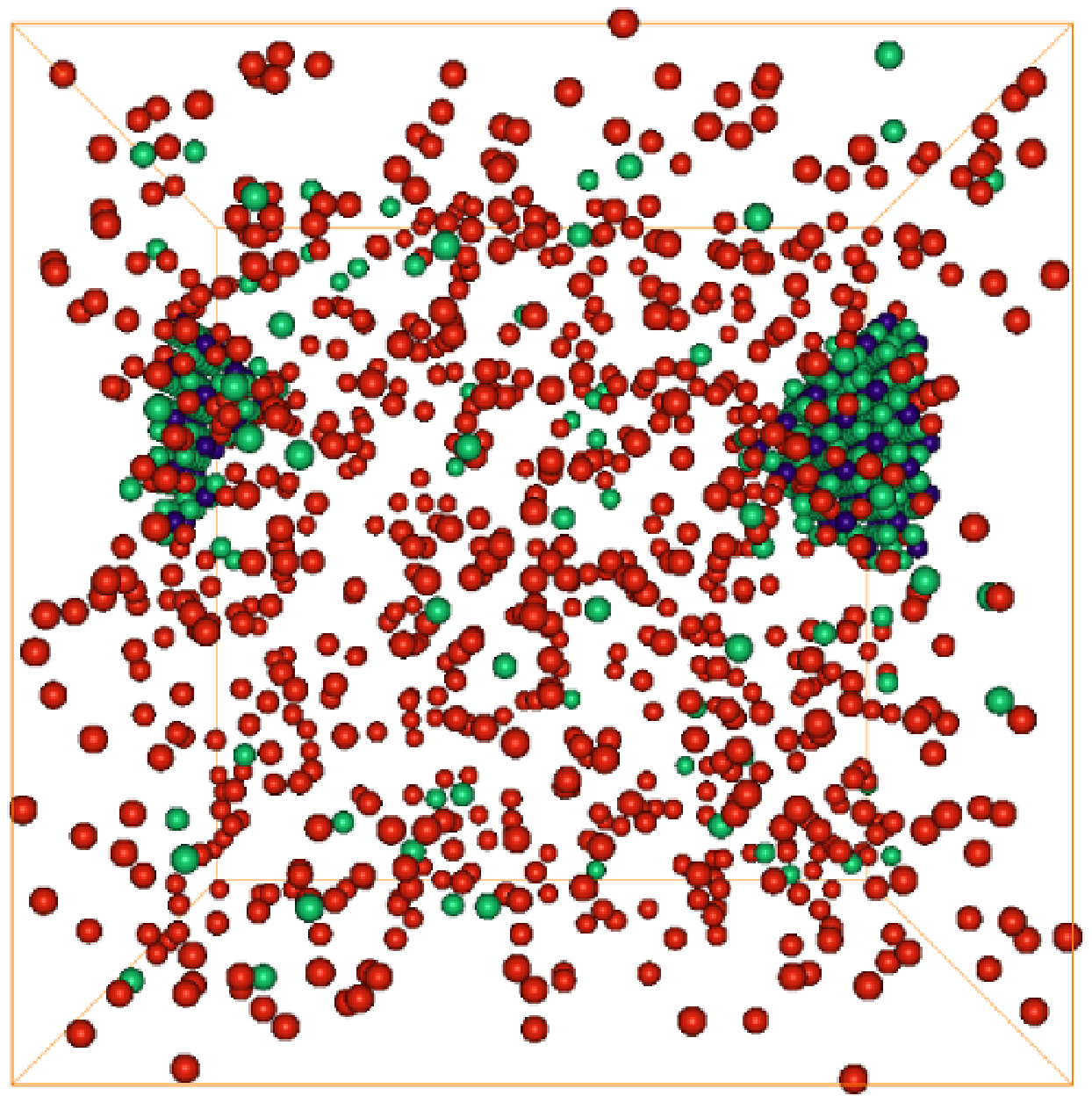}  & \includegraphics[width=.2\linewidth]{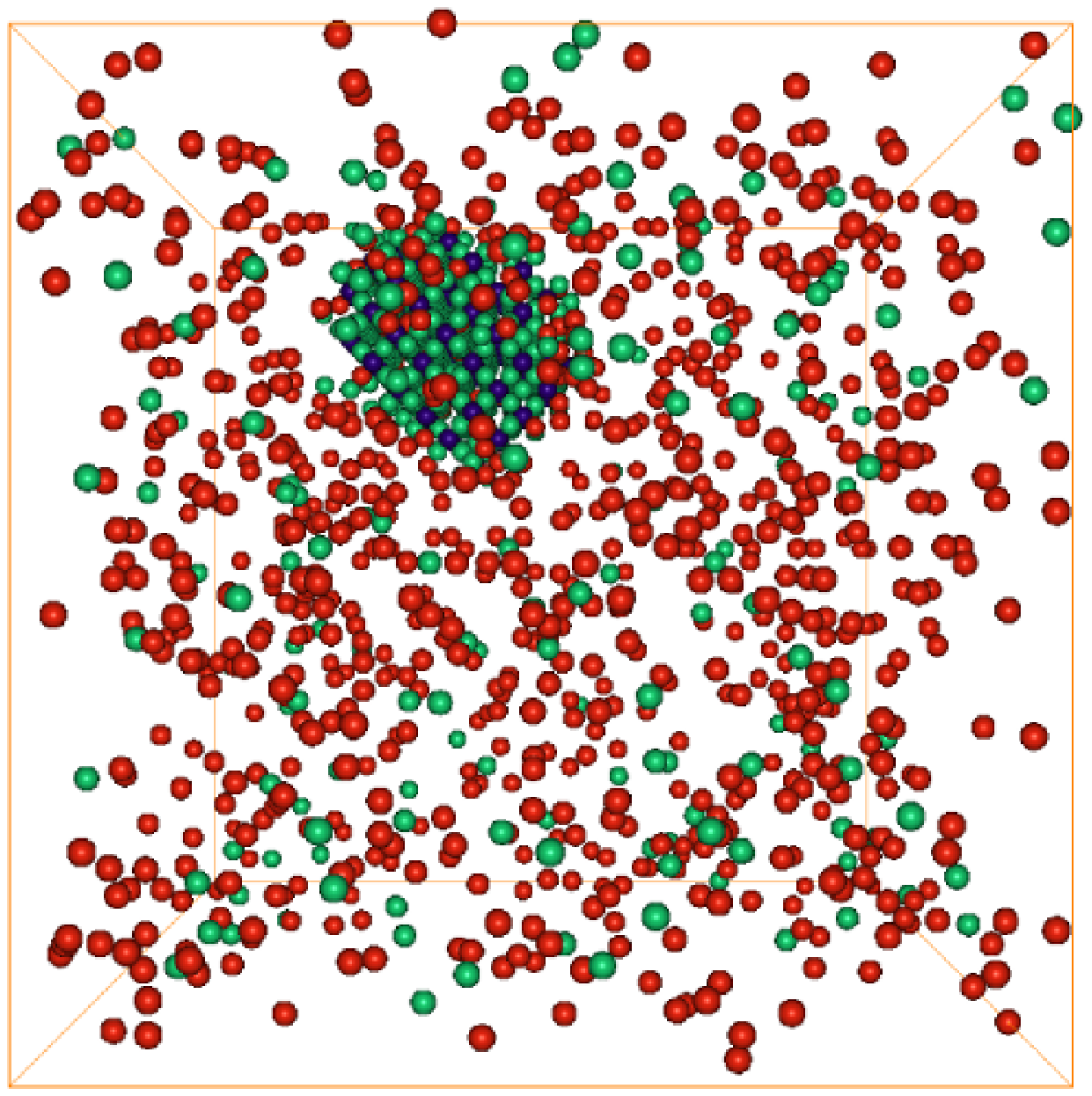} \\
     \hline
              & WRe1.5\%Os0.25\%Vac0.5\% & WRe1.5\%Os0.5\%Vac0.5\% & 
              WRe1.5\%Os0.75\%Vac0.5\%  &  
              WRe1.5\%Os1.0\%Vac0.5\%\\
     \hline
     800 K & \includegraphics[width=.2\linewidth]{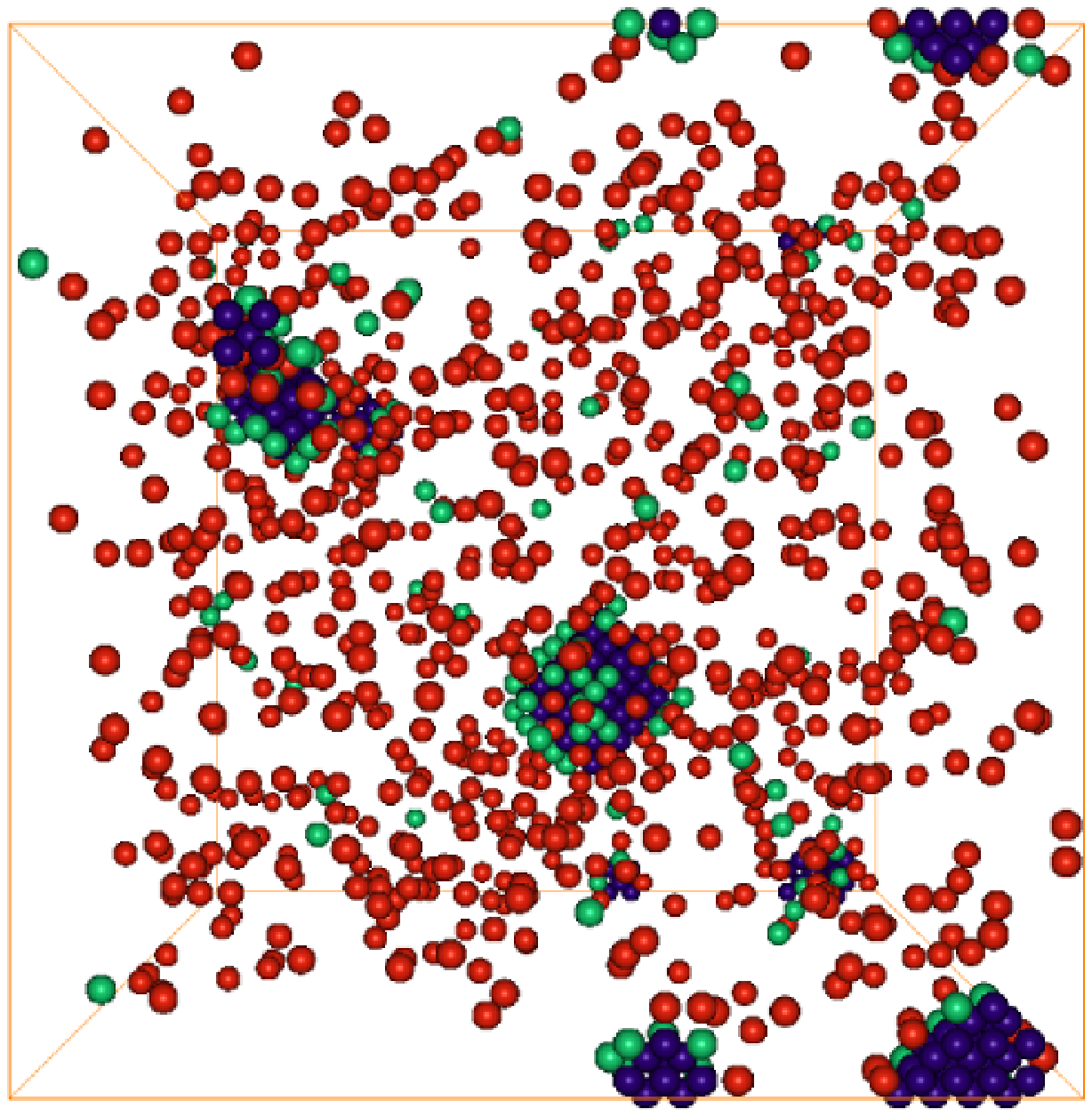} & \includegraphics[width=.2\linewidth]{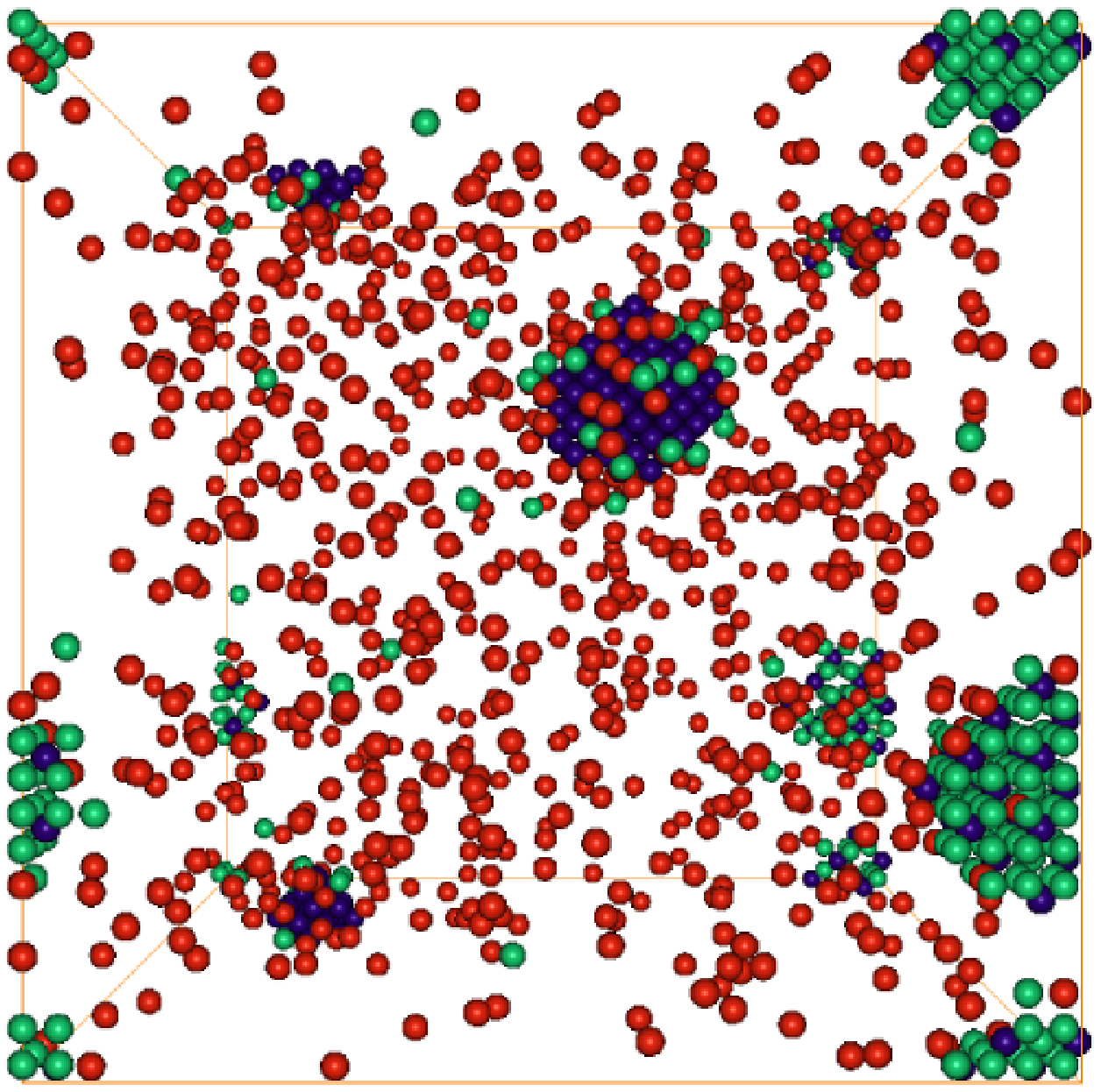} & \includegraphics[width=.2\linewidth]{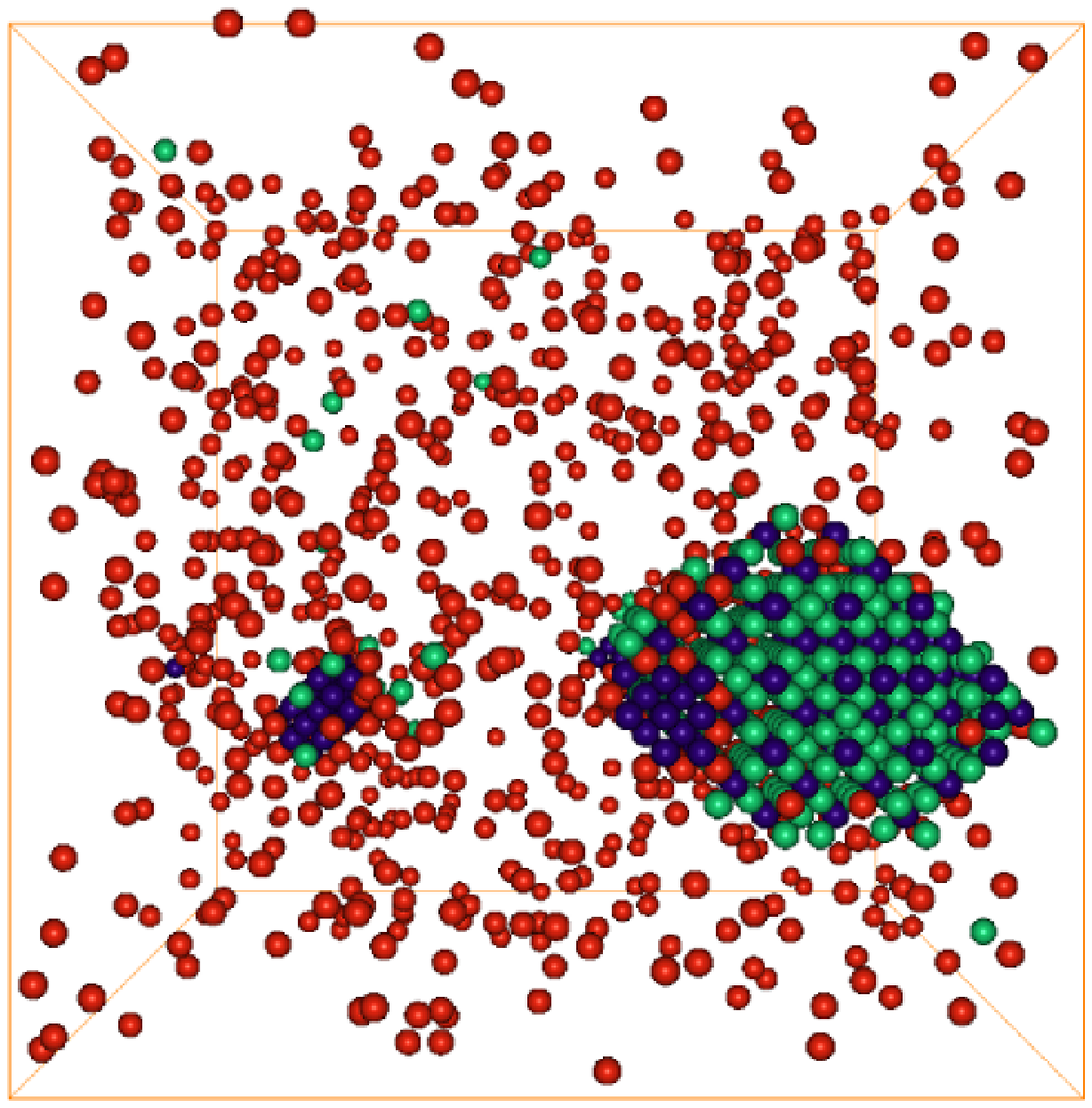} & \includegraphics[width=.2\linewidth]{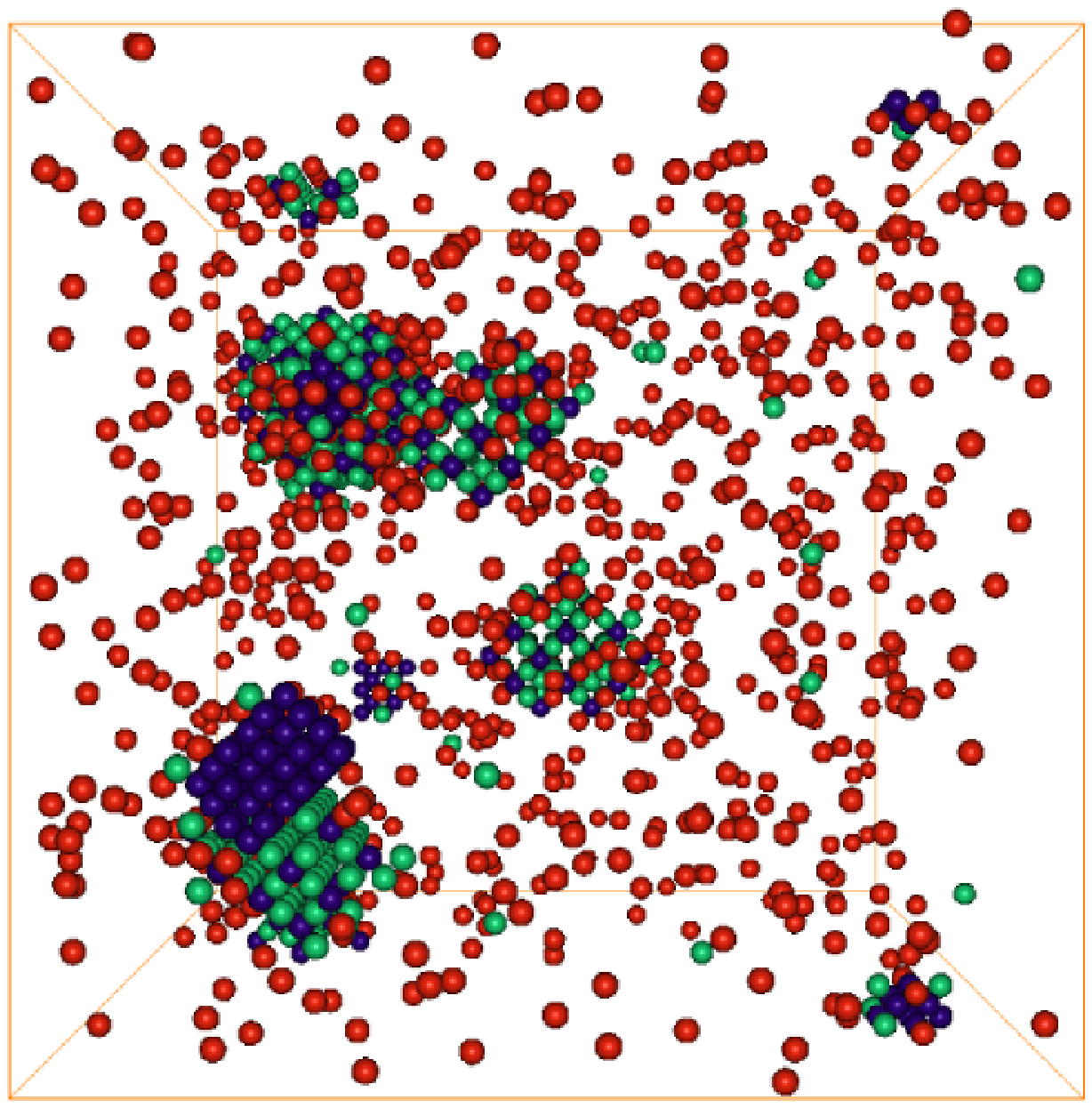} \\
     \hline
     1200 K & \includegraphics[width=.2\linewidth]{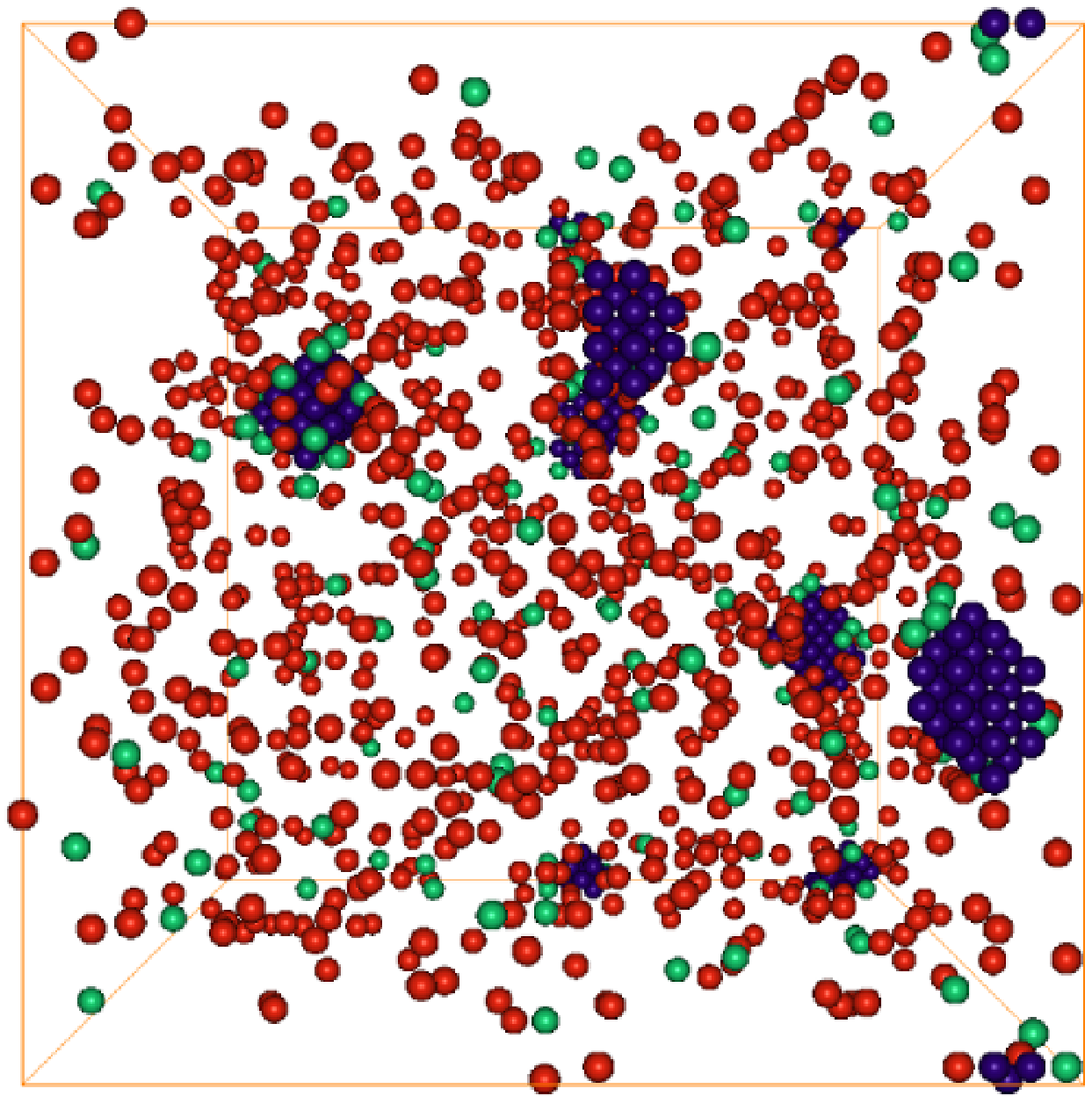} & \includegraphics[width=.2\linewidth]{WRe1.5Os0.25Vac0.5_1200K.eps} & \includegraphics[width=.2\linewidth]{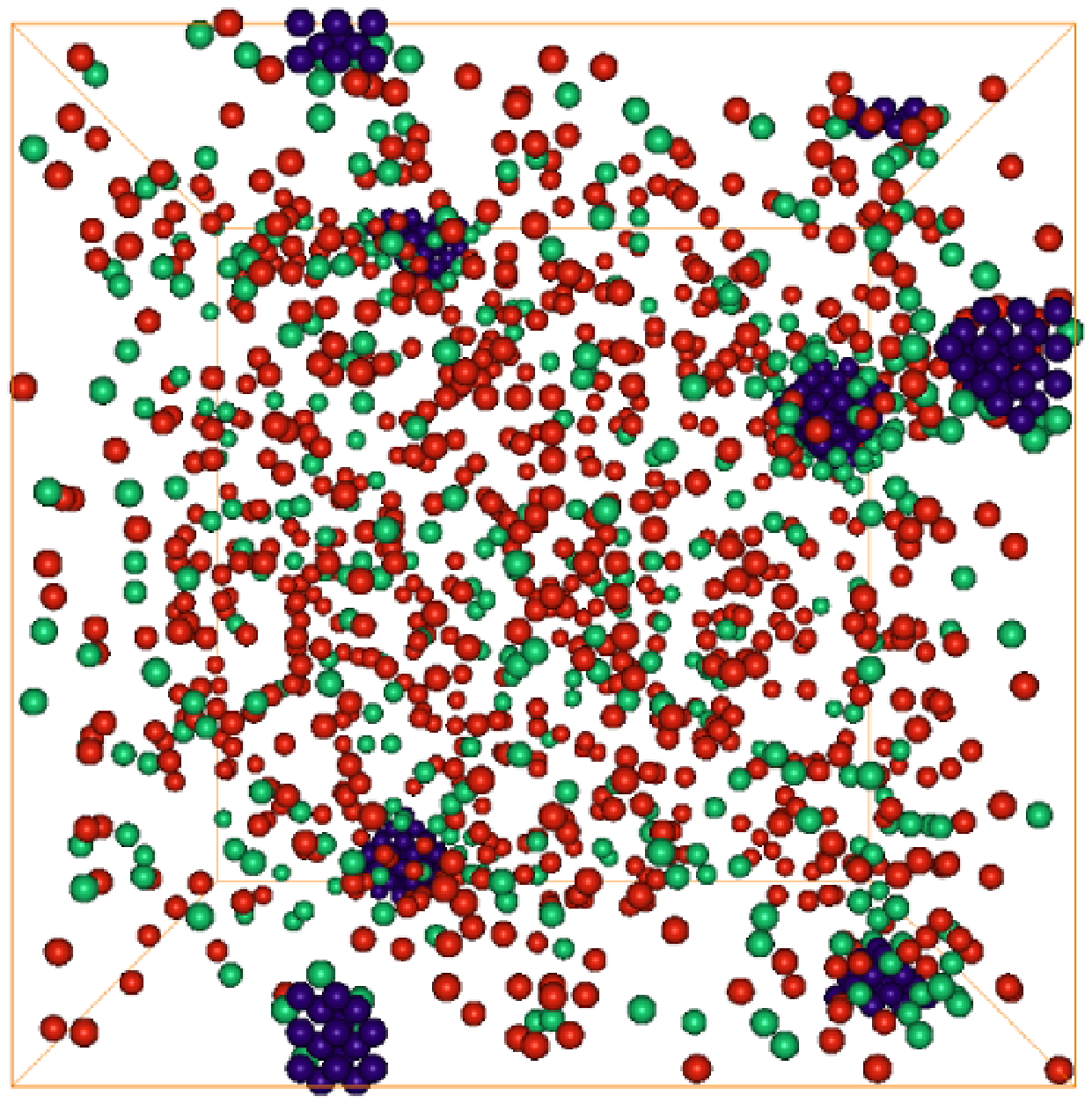}  & \includegraphics[width=.2\linewidth]{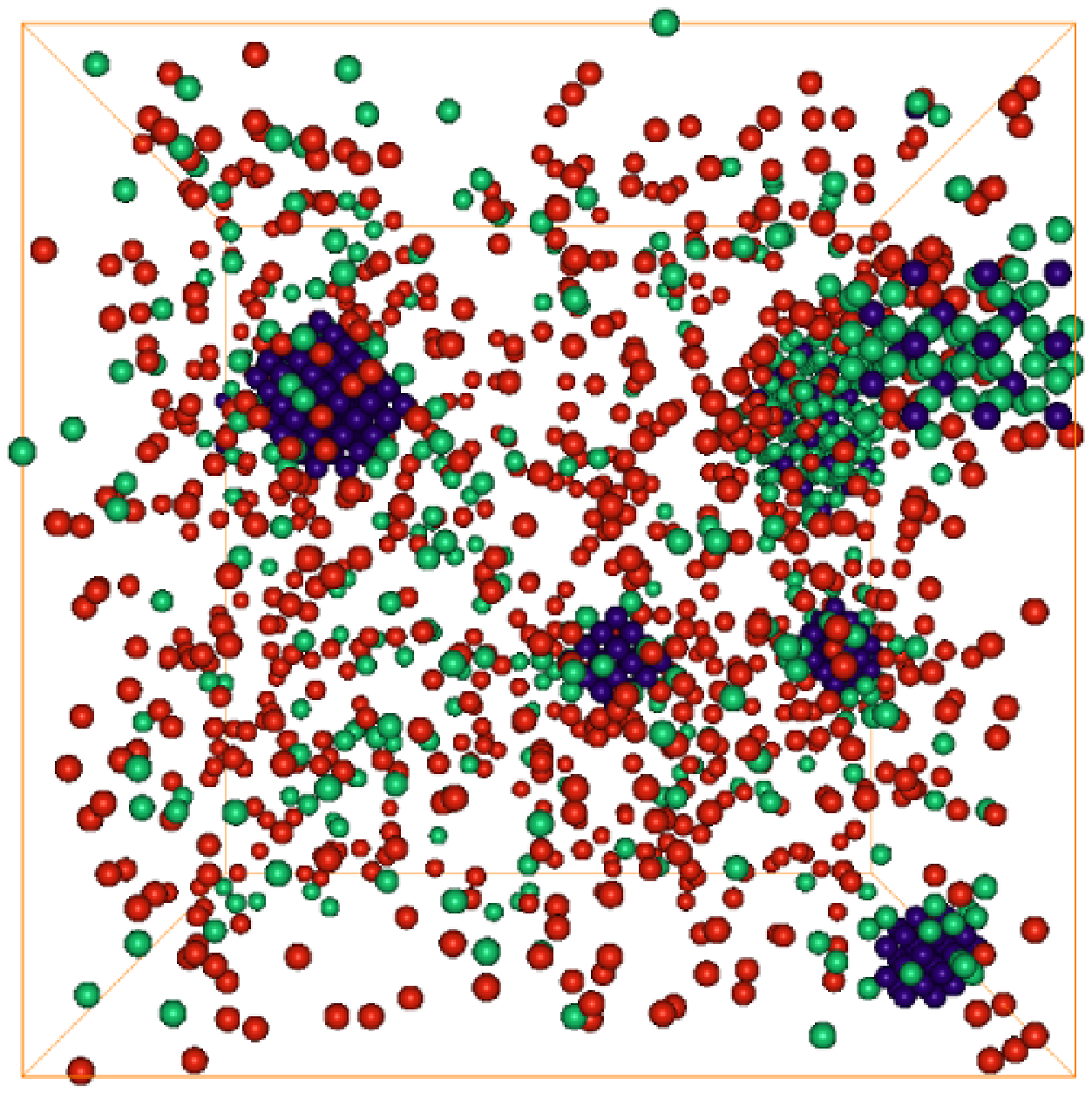} \\
    \end{tabular}%
\end{ruledtabular}
\end{table*}

\begin{figure*}
\raisebox{12\height}{(a)}\includegraphics[width=0.42\linewidth]{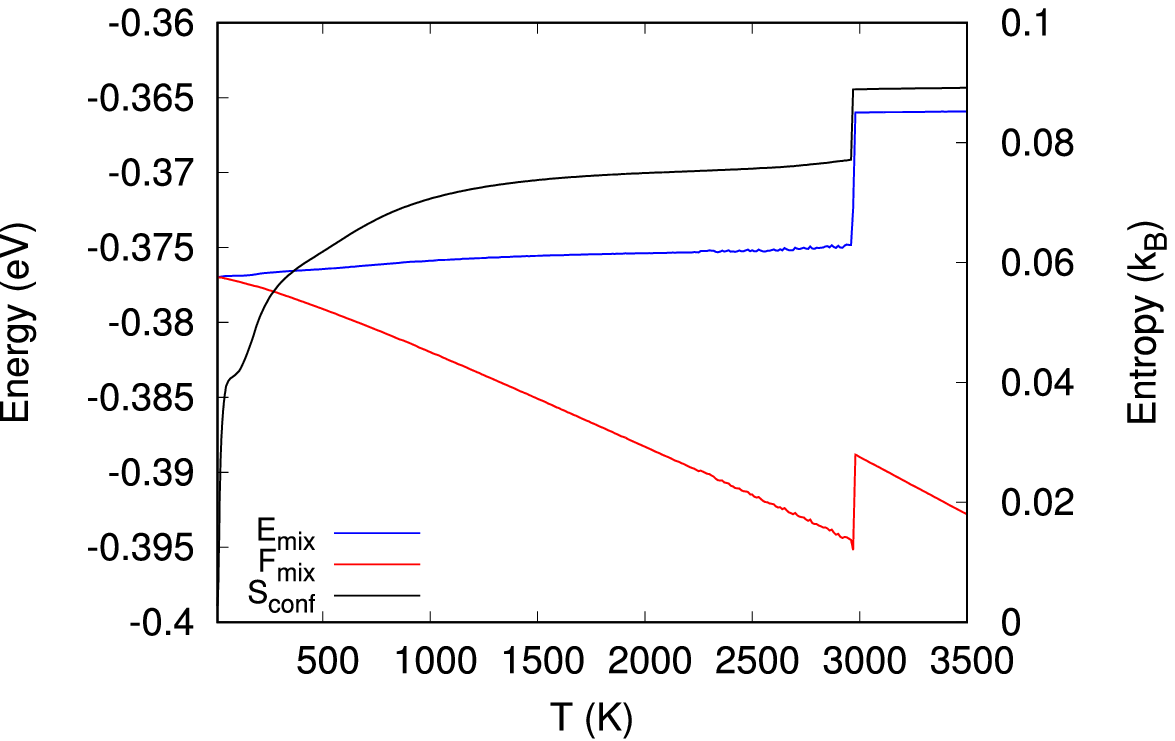}
\raisebox{12\height}{(b)}\includegraphics[width=0.42\linewidth]{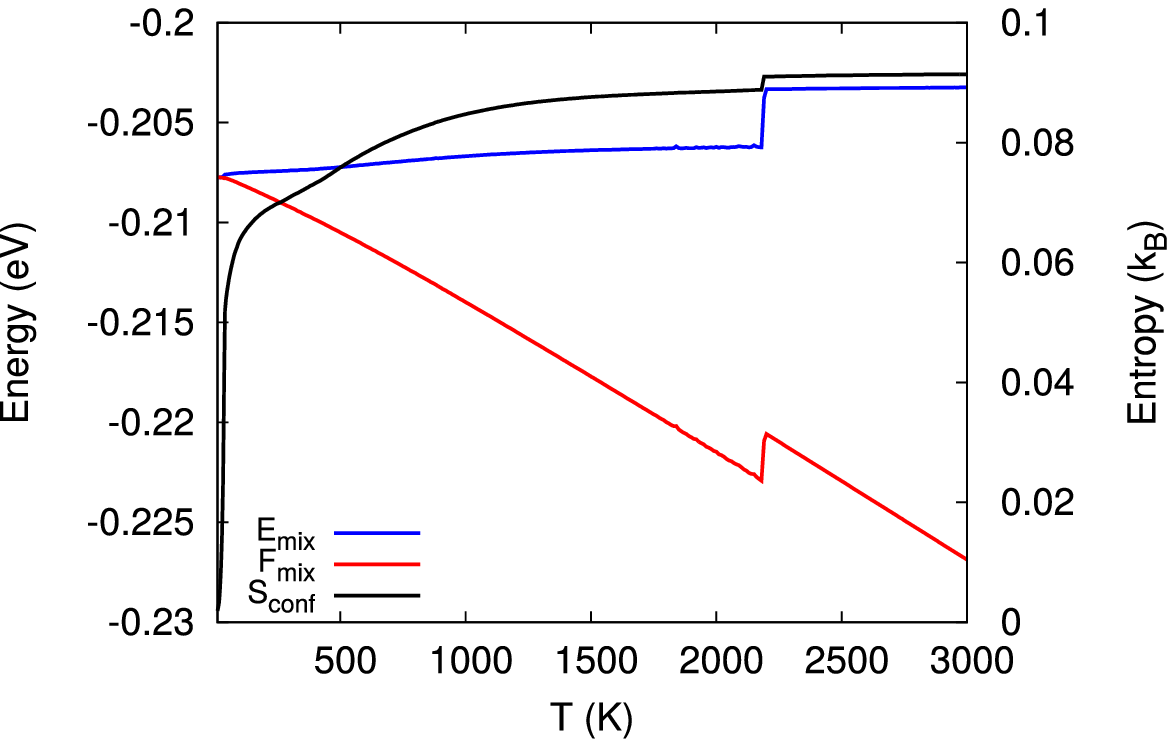}
\caption{\label{FE} Free energy as a function of temperature from thermodynamic integration: (a) WRe1.5$\%$Vac0.2$\%$ (b) WRe1.5$\%$Os0.25$\%$Vac0.2$\%$}
\label{fig:FE}
\end{figure*}

To understand the origin of steady-state configurations in a constrained system with defects such as the void formation in bcc-W in presence of both Re and Os solute atoms at finite temperature, it is better to perform quenched Monte Carlo simulations starting from results high temperatures starting from 3000K. The constrained alloys were cooled down to the low temperature of with 3000 MC steps per atom performed both at thermalisation and accumulation stages and with the temperature step of $\Delta T = 10$ K.
With such MC simulations, the configuration entropy determined from Eq.(\ref{eq:thermointegration}) and therefore the free energy of a system (Eq.(\ref{eq:free_energy})) can be properly investigated as a function of temperature and solute concentration. Fig.S2 from the Supplementary Materials shows the quenched MC simulation results at four different temperatures: 2400~K, 1800~K, 1200~K and 800~K for the WRe1.5$\%$Os0.25$\%$Vac0.2$\%$. Fig.(\ref{fig:FE}) provides different contribution to free energy calculations from enthalpy of mixing and configuration entropy as a function of temperature for the ternary: WRe1.5$\%$Vac0.2$\%$ and the quaternary with Os addition:  WRe1.5$\%$Os0.25$\%$Vac0.2$\%$. The dependence of configuration entropy as a function of temperature for both cases shows a similar trend, namely a clear jump displaying the transitions from ideal random solid solutions into the precipitation regime with the formation of voids. However, it is also visible that the effect of Os in decreasing down the transition temperature from 2900K in WRe1.5$\%$Vac0.2$\%$ to 2200K for WRe1.5$\%$Os0.25$\%$Vac0.2$\%$. Therefore, in a comparison with solute Re, Os atoms play a more significant role in anomalous precipitations formed by its strong interaction with vacancy defect clusters at high temperature in irradiated W even at small concentration range.  

\subsection{Short-range order analysis for W-Re-Os-Vac system}

\begin{figure*}
	\subfloat{
		\raisebox{20\height}{(a)}\includegraphics[width=0.40\linewidth]{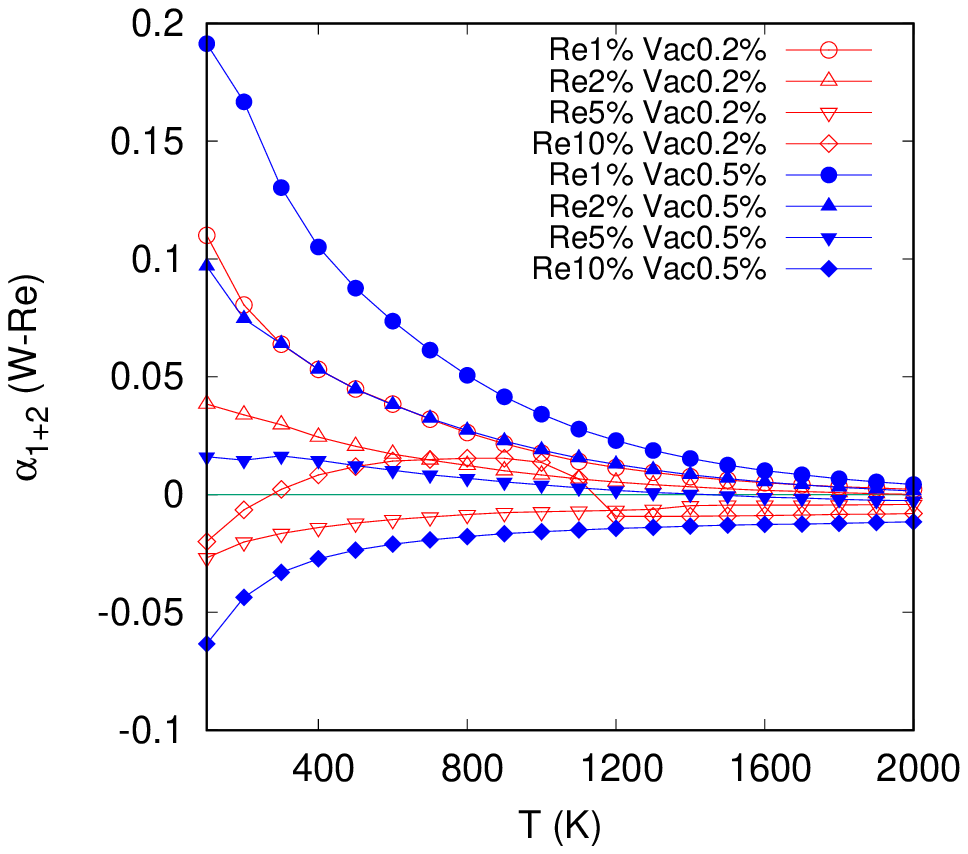}}
	\subfloat{
		\raisebox{20\height}{(b)}\includegraphics[width=0.40\linewidth]{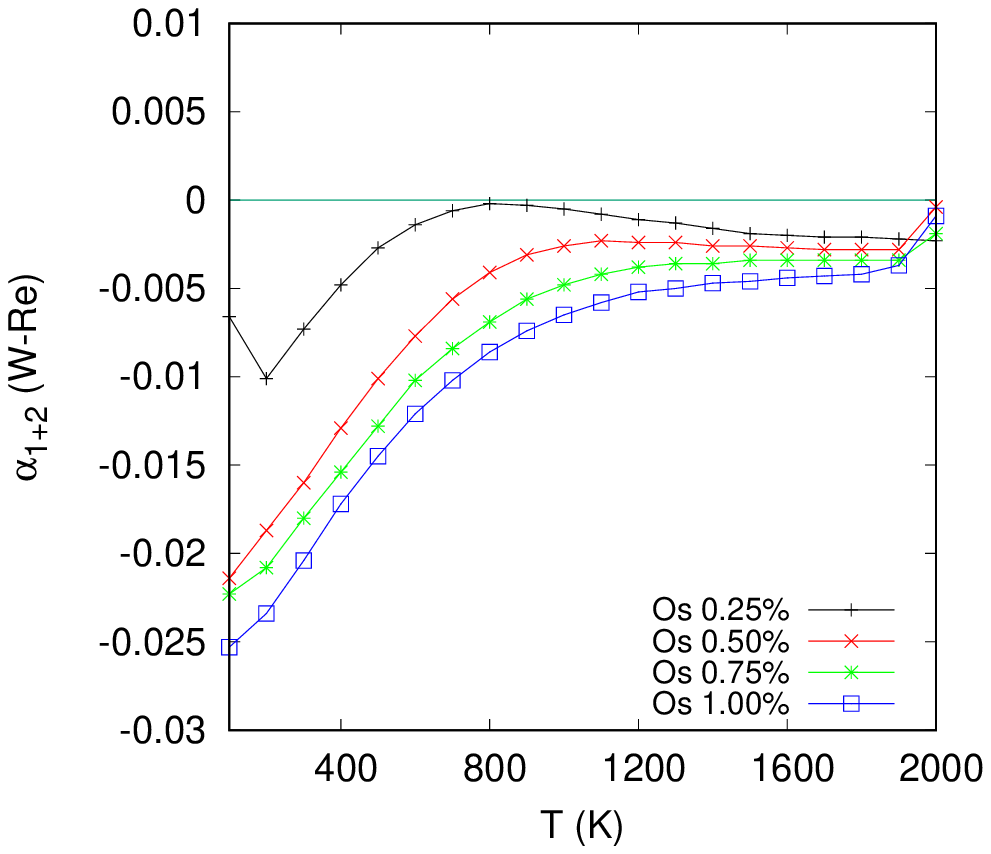}}\\
	\subfloat{
		\raisebox{20\height}{(c)}\includegraphics[width=0.40\linewidth]{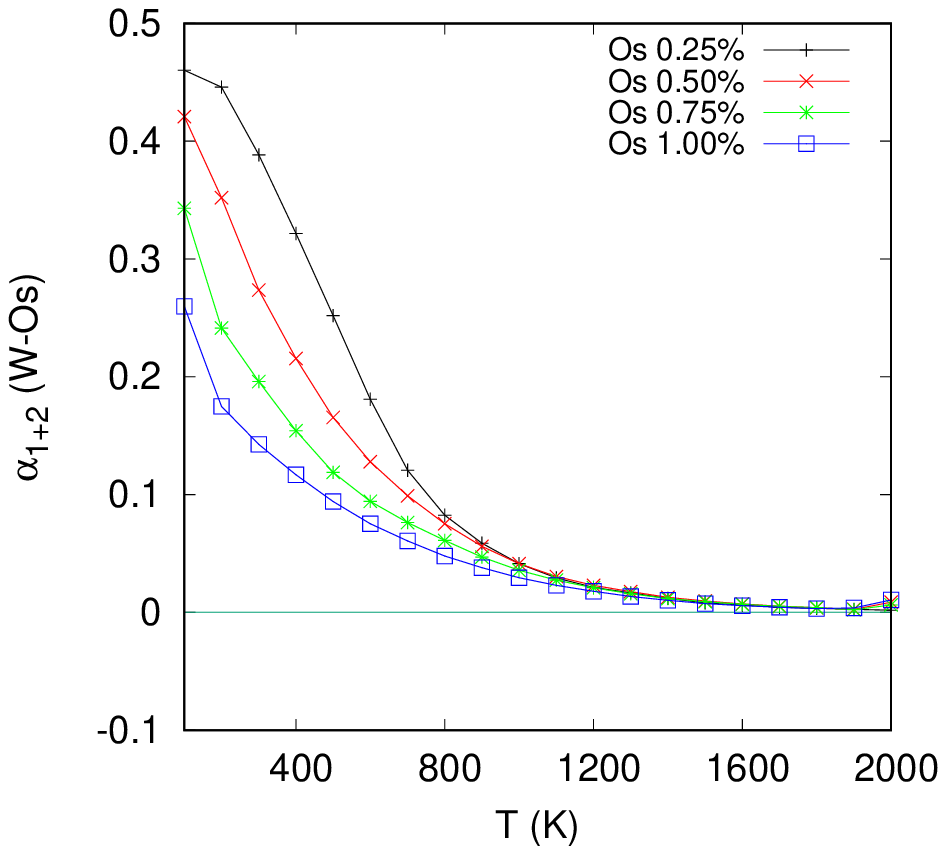}}
	\subfloat{
		\raisebox{20\height}{(d)}\includegraphics[width=0.40\linewidth]{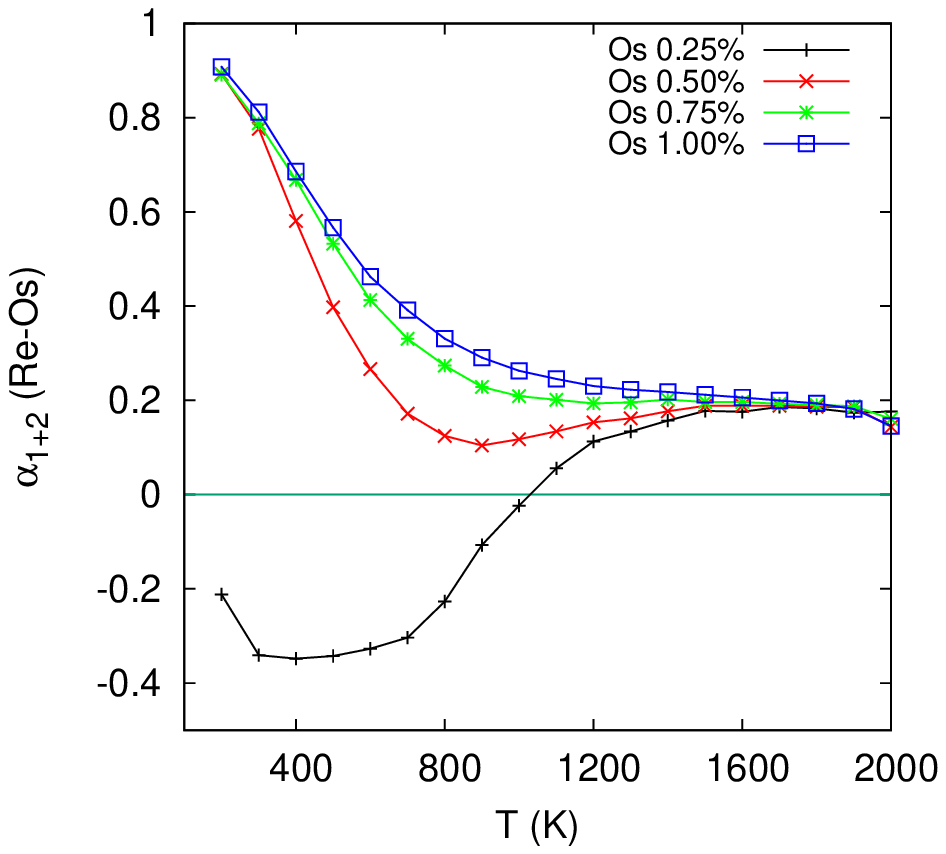}}
\caption{\label{SRO_Os} SRO dependence as a function of temperature with different solute (Re,Os) and vacancy concentration for ternary W-Re-Vac and quaternary W-Re-Os-Vac.}
\label{fig:SRO_Os}
\end{figure*}

To clarify the role of Os effect at finite temperature, a systematic study of short-range order dependence as a function of temperature and solute concentrations is carried out from quenched MC simulations. Again Eqs.(\ref{eq:sromatrix}) and (\ref{eq:SRO_avg}) are used to compute the average SRO over the whole cell for the complex quaternary W-Re-Os-Vac system. Unlike in the previous subsection, here the SRO analysis is more focused on the pair interactions between W and solute elements as well as between Re and Os. Fig.\ref{fig:SRO_Os} shows the dependence of average SRO parameters between W-Re (in both ternary W-Re-Vac and quaternary W-Re-Os-Vac), W-Os and Re-Os in the quaternary W-Re-Os-Vac system. 

Fig.\ref{fig:SRO_Os}a displays the dependence of W-Re SRO parameters as function of temperature with different Re and vacancy concentrations in the ternary -Re-Vac system. At small Re concentrations, especially for Re1$\%$ and Re2$\%$ cases with Vac0.2$\%$ and Vac0.5$\%$, it is found that W-Re SRO parameters are positive for the whole range of temperatures. This behaviour demonstrates the segregation of dilute Re from W in presence of vacancy defects. The segregation behaviour of Re from W in presence of vacancy is completely different to the case of W-Re binary without vacancy defects. In the latter due to the negative enthalpy of mixing (see Fig.S1a from Supplementary Materials) solute Re atoms are more attractive to W in the bcc lattice. For the cases with higher Re concentrations (Re5$\%$ and Re10$\%$), the W-Re SRO displays the transition from positive to negative values. This means that the mixing effect between Re and W is increased with Re concentration, despite the presence of small vacancy concentrations. Fig.S1a also shows the decreasing positive enthalpy of mixing trend in W-Re for different vacancy concentrations as a function of Re concentration. 

The next two Fig.\ref{fig:SRO_Os}b and \ref{fig:SRO_Os}c show the evolution of average SRO parameters for W-Re, W-Os pairs, respectively, as a function of temperature with four different Os concentrations (Os0.25$\%$, Os0.5$\%$, Os0.75$\%$ and Os1$\%$) with Vac0.2$\%$. Comparing the SRO behaviour for W-Re pairs between Figs.\ref{fig:SRO_Os}b and \ref{fig:SRO_Os}a, it can see that the presence of Os dramatically changes the behaviour of solute Re in the quaternary W-Re-Os-Vac system at small vacancy concentration. Re atoms are now more likely to be present in the W matrix due to negative SRO values in all range of temperature. Fig.\ref{fig:SRO_Os}c shows the opposite behavior of Os in a comparison with Re. They are unlikely to be present in W matrix due to the positive SRO values for W-Os pairs. The finding of negative SRO values for W-Re and positive ones for W-Os pairs is is consistent with the overall observation seen in Table \ref{tab:MC_results_const_Re} from MC results at fixed temperatures for the Vac0.2$\%$, where, unlike Os, Re atoms are more visible within the W matrix.    

Last but not least, Fig.\ref{fig:SRO_Os}d shows the most interesting result of the present analysis for the average SRO parameters between the two solute species: Re and Os. The Re-Os  pairs have positive SRO parameters for high Os concentration range, namely for Os1$\%$, Os0.75$\%$ and Os0.5$\%$ demonstrating that they do not like each other within the 1NN and 2NN of bcc matrix for W-Re-Os-Vac system. Even in the ternary system of W-Re-Os without defects, the presence of Os surrounded by Re atoms in the 1NN gives a negative binding energy, whereas for an Re atoms surround by Os atoms in the 1NN, the calculated binding energies are also mostly negative, except for some specific configurations for which they become positive (see Fig.S3 from Supplementary Materials). A similar situation has also been seen for the dependence of Re-Os SRO as a function of temperature for the case with low Os0.25$\%$ shown in Fig.\ref{fig:SRO_Os}d. It is found that for this case, the SRO for Re-Os can change the sign from positive to negative values at the temperature range below 1200 K. Since most of Os atoms were integrating strongly with voids and vacancy clusters, the negative values of Re-Os SRO lead to the conclusion that both of these solute atoms can be simultaneously segregated to the precipitation with vacancy clusters at low Os concentration range.

\section{Comparison with experimental data on neutron irradiated tungsten}

Modelling results given above enable a direct comparison with experimental observations carried out using both Transmission Electron Microscopy (TEM) and Atom Probe Tomography (ATP) on tungsten samples neutron irradiated at high temperature.    

\subsection{Modeling results for alloys containing Re1.5$\%$ and Os0.1$\%$}  

\begin{figure*}
	\subfloat{
		\raisebox{20\height}{(a)}\includegraphics[width=0.40\linewidth]{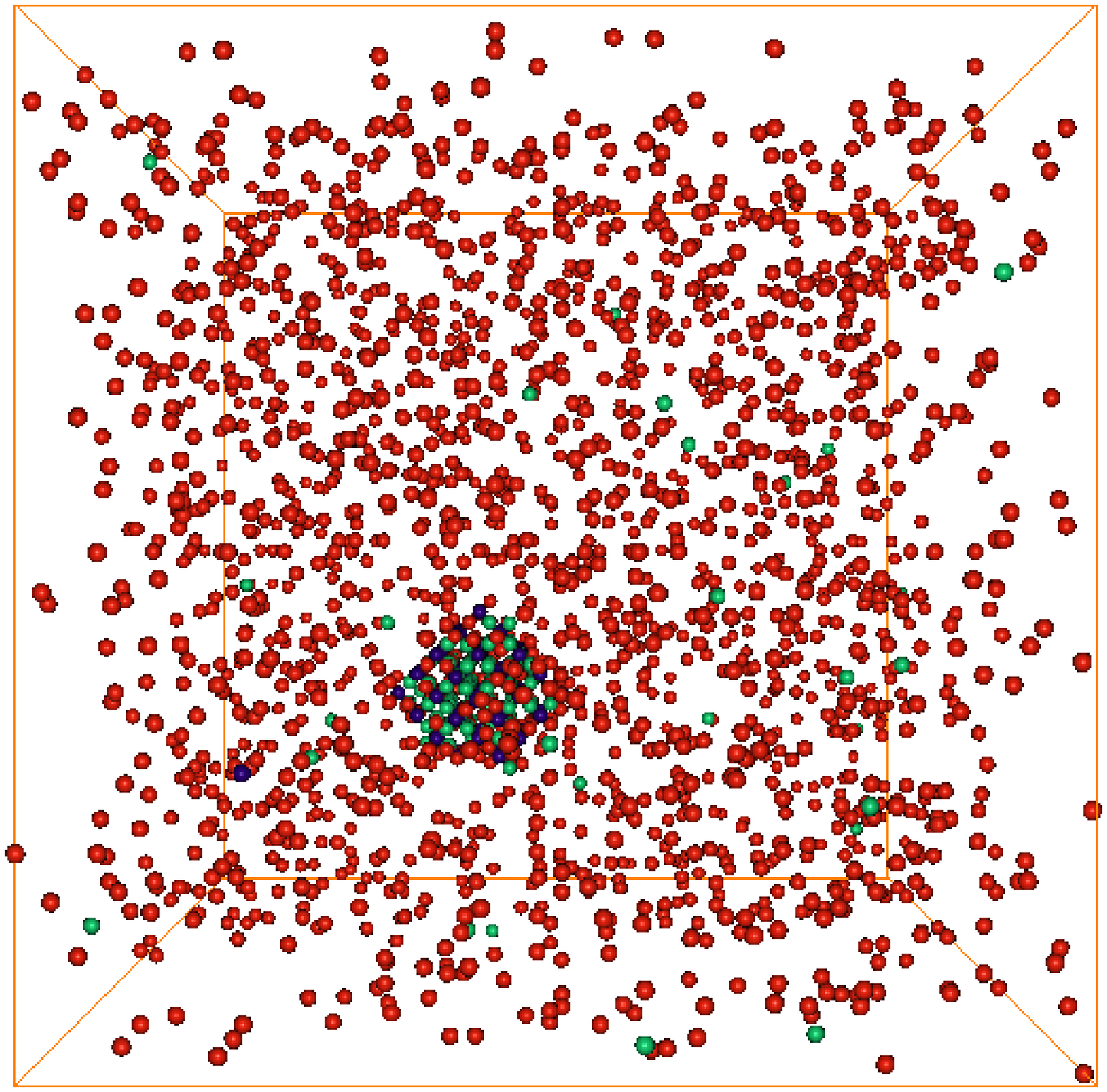}}
	\subfloat{
		\raisebox{20\height}{(b)}\includegraphics[width=0.40\linewidth]{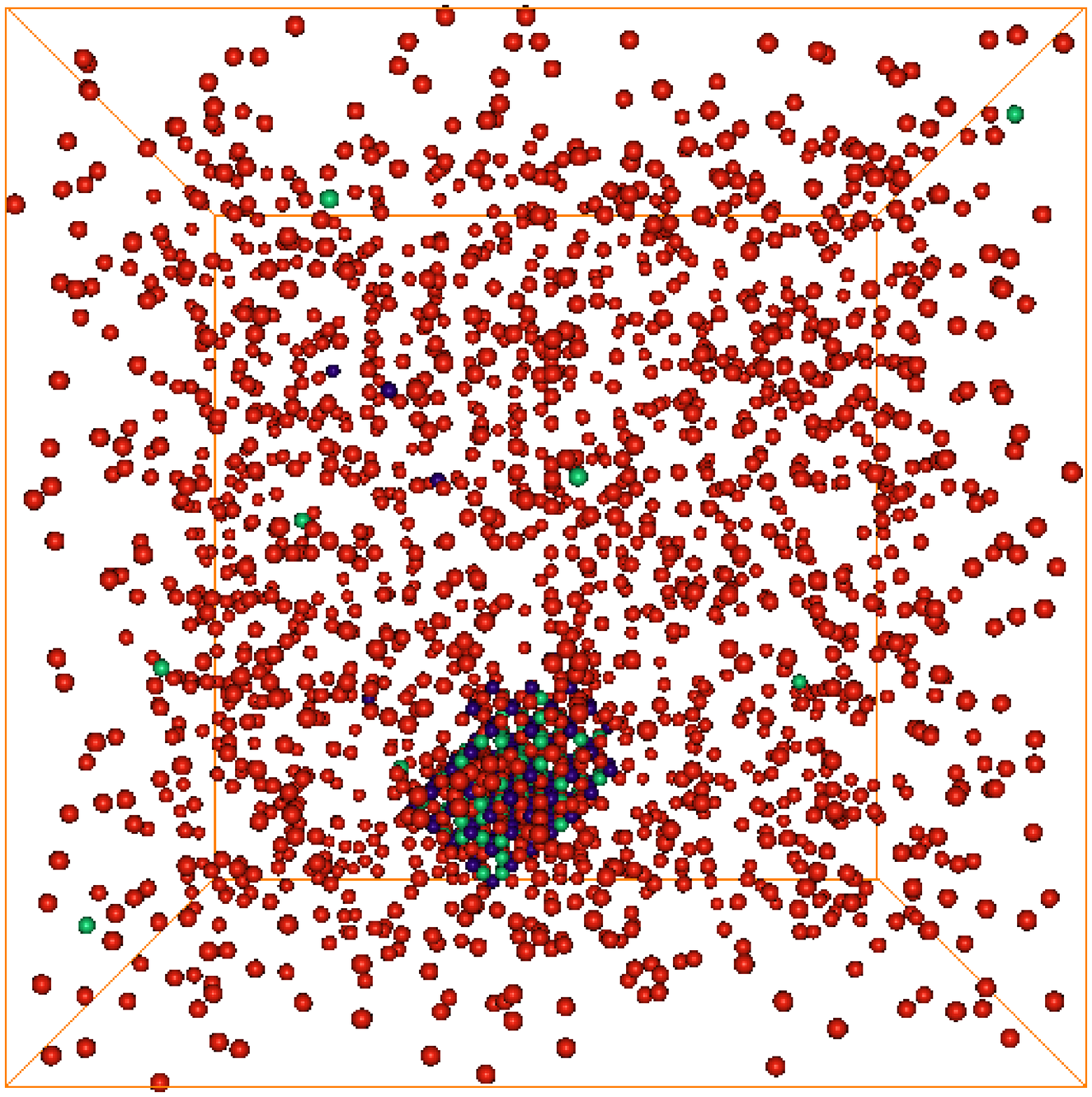}}\\
	\subfloat{
		\raisebox{20\height}{(c)}\includegraphics[width=0.40\linewidth]{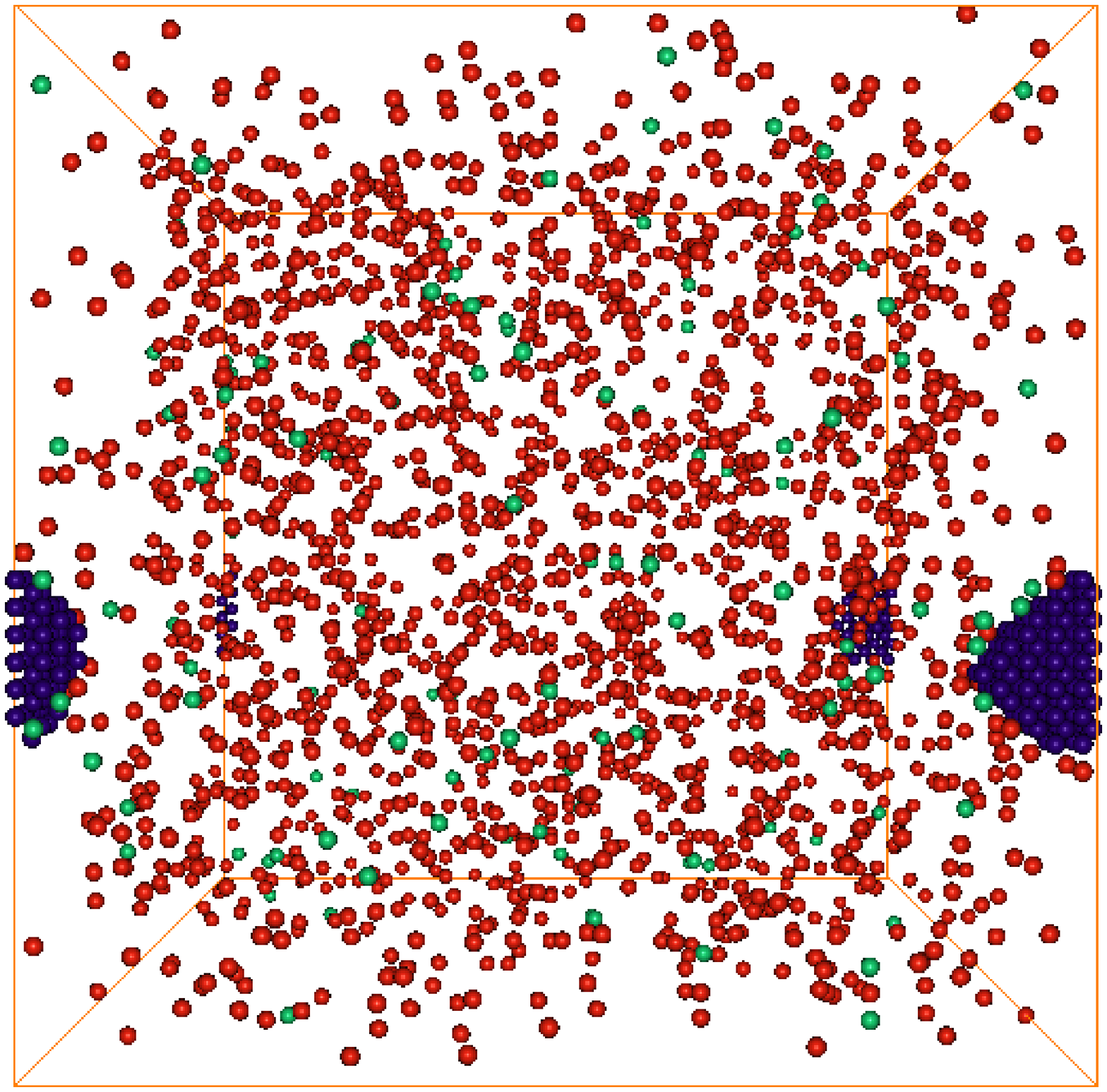}}
	\subfloat{
		\raisebox{20\height}{(d)}\includegraphics[width=0.40\linewidth]{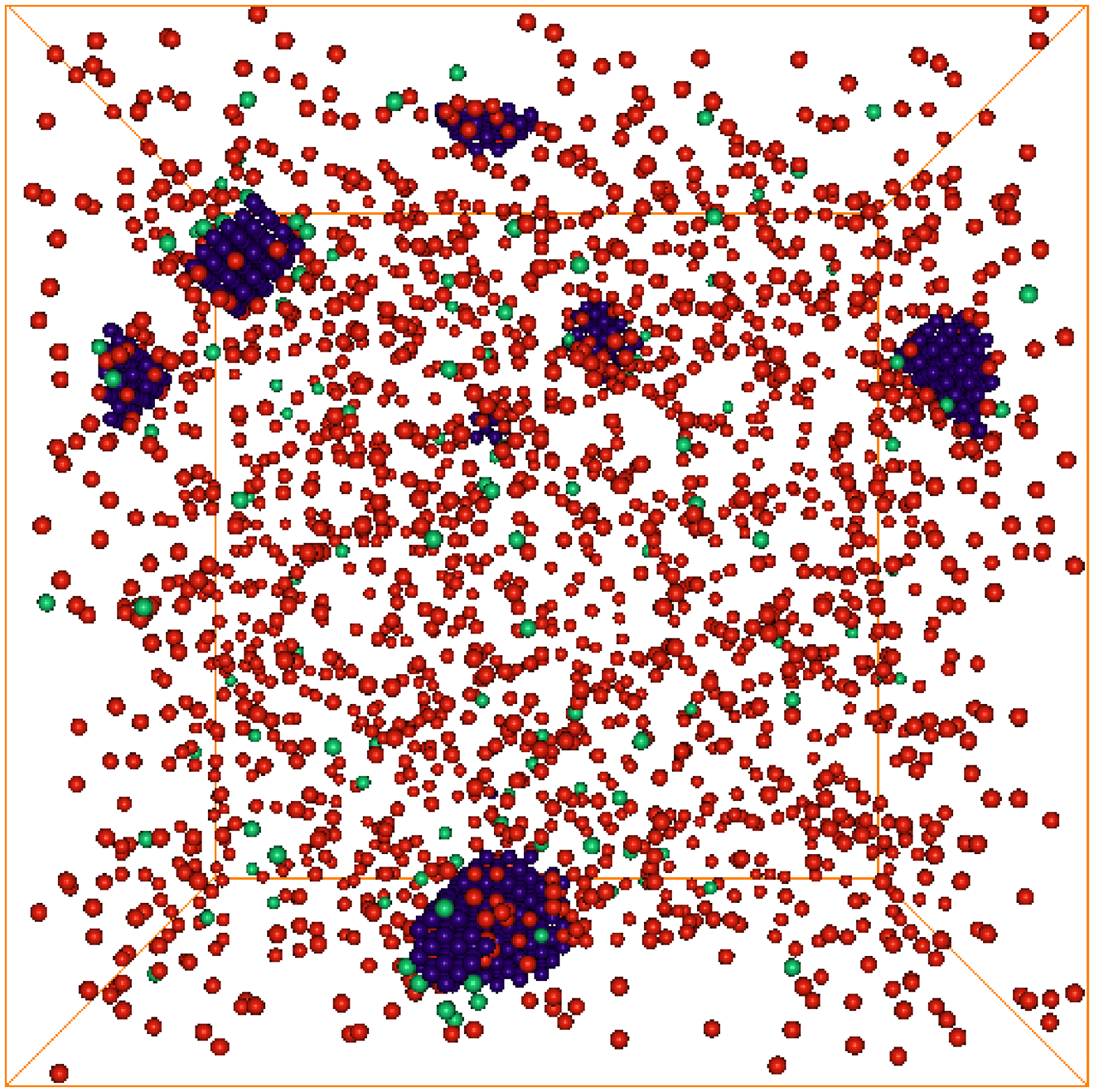}}
\caption{\label{WRe15Os1} MC simulations for WRe1.5Os0.1 with different vacancy concentrations: a) 0.05$\%$, b) 0.1$\%$, c) 0.2$\%$ and d) 0.5$\%$}    
\label{fig:WRe15Os1}
\end{figure*}

Fig.\ref{fig:SRO_Os}d shows that at Os concentration below 0.25$\%$, there is a strong possibility that both Re and Os are segregated to precipitates formed from voids or vacancy clusters. Fig.\ref{fig:WRe15Os1} shows results from quenched MC simulations performed using a 40$\times$40$\times$40 bcc supercell box for alloys containing Re1.5$\%$ and Os0.1$\%$ at four different vacancy concentration. These results were derived for simulations performed for $T=$1200 K, which is close to the irradiation temperature of 1173 K.  Figs. \ref{fig:WRe15Os1}c and \ref{fig:WRe15Os1}d show that at high vacancy concentration, void feature is more dominant in formation of precipitates whereas both Re and Os atoms were less attached to the void. At lower vacancy concentrations shown in Fig.\ref{fig:WRe15Os1}a and \ref{fig:WRe15Os1}b, the voids are decorated by a higher concentrations of both Re and Os atoms.  

\begin{figure*}
\raisebox{12\height}{(a)}\includegraphics[width=0.42\linewidth]{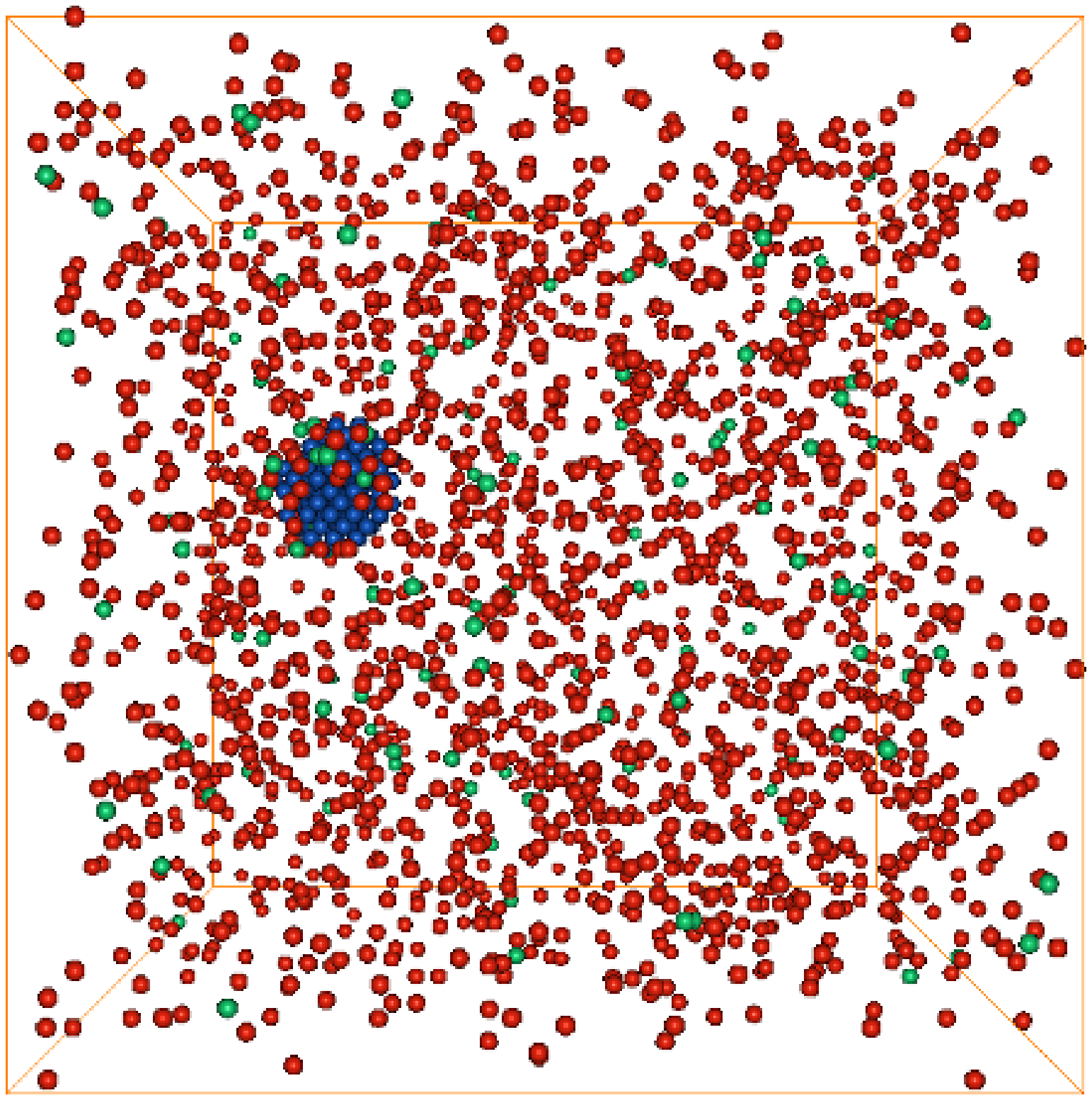}
\raisebox{12\height}{(b)}\includegraphics[width=0.42\linewidth]{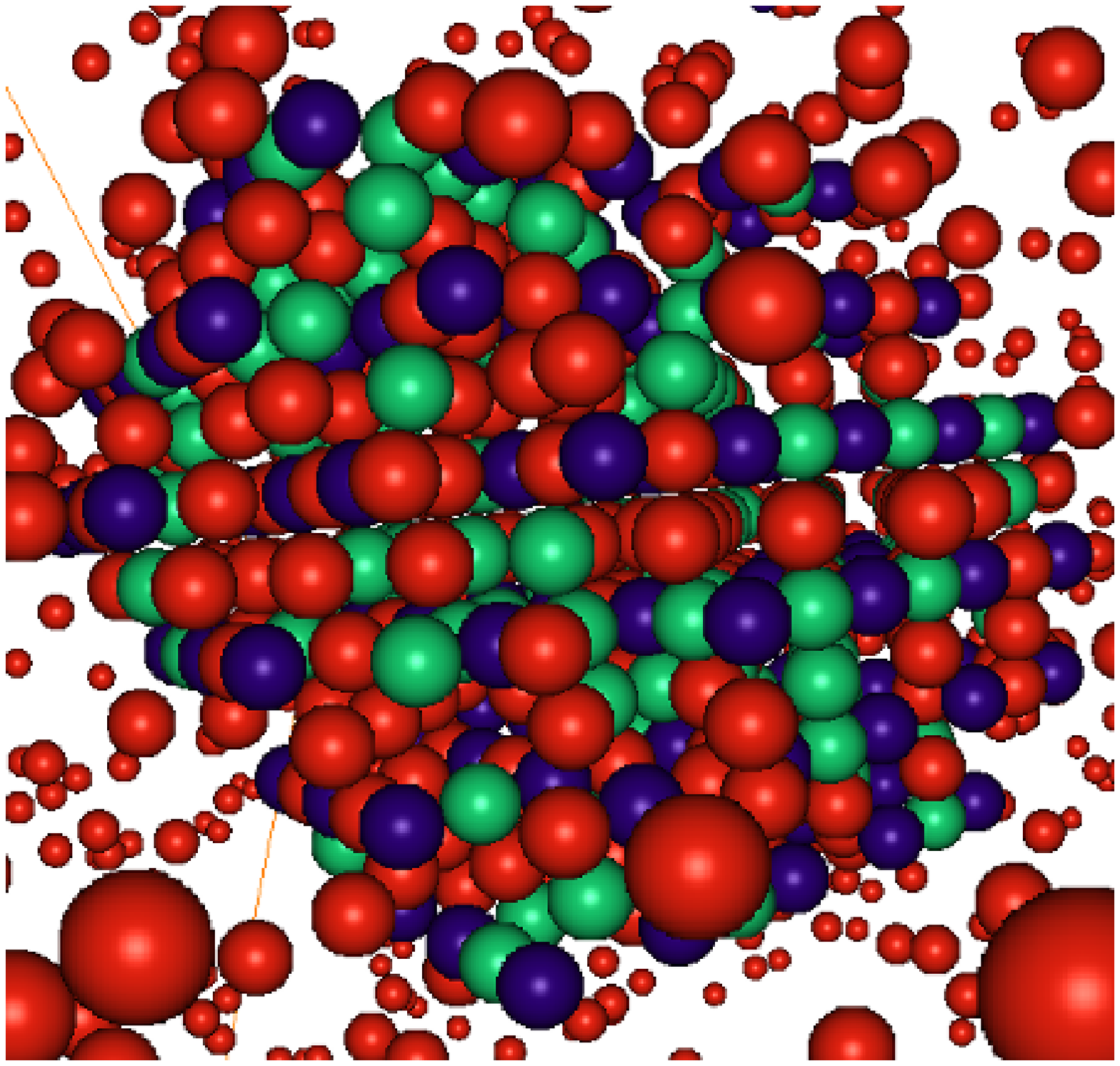}
\caption{\label{cluster} Clustering of Re and Os with vacancies from quenched simulations at 1200K (a) the zoom of cluster structure obtained from MC simulations at fixed temperature 1200 K}
\label{fig:cluster}
\end{figure*}

The SRO calculations of SRO parameter for Re-Os pairs give negative values of -4.784, -9.520, 0.087 and -0.238 for 0.05$\%$, 0.1$\%$, 0.2$\%$ and 0.5$\%$ of vacancy concentration, respectively. The strong negative values for the first two cases  confirmed a favorable correlation between Re and Os to be present within the precipitates. For interpreting experimental results of neutron irradiated with a relatively high dose, the case with Vac0.1$\%$ is more likely to be a representative structure from our simulations. Fig.\ref{fig:cluster}a shows again the configuration obtained for Re1.5$\%$Os0.1$\%$Vac0.1$\%$ alloy similar identical to those of Fig.\ref{fig:WRe15Os1}b but taken from different angle to display the void in the central core of the precipitate. The thermodynamic integration technique allows to predict the jump or transition of configurations entropy at around 1700K to confirm the void formation at lower temperature. The enthalpy of mixing evaluated at 1200K is found to be negative (-0.208 eV) for a stable steady-state of this configuration. These results for configuration entropy and enthalpy of mixing for Re1.5$\%$Os0.1$\%$Vac0.1$\%$ are consistent with the free-energy trend of its dependence as a function of temperature plotted in Fig.\ref{fig:FE} for other alloy composition of Re1.5$\%$Os0.25$\%$Vac0.2$\%$. Fig.\ref{fig:cluster}b shows a zoomed image of the cluster of precipitates. Using the definition of SRO parameter from Eq.(\ref{eq:WC}) the evaluated concentrations of each species show that there are around 47$\%$ of W, 25$\%$ of Re, 18$\%$ of Os and 11$\%$ of vacancies inside of this cluster. Importantly, it is found that the average concentration of Re within the precipitate is higher than those of Os due to the presence of SRO effects in the presence of vacancy defects and within the low range of nominal concentration for solute Os atoms ($\le $ 0.25 $\%$).

To gain more insight into the properties of voids simulated in this study, larger MC simulations with 250000 sites (50$\times$50$\times$50 super-cell) were performed (see Fig.S4 from the Supplementary Materials). Three voids were observed in these simulations and they were formed at 2000 K for the considered W1.5Re0.5Vac system. The shape and size of these voids are analysed and projected into three different xy, yz and zx planes as shown in Table \ref{fig:voids}. It is found that all the three voids are faceted on the (110) equivalent planes that are in an excellent agreement with the High Resolution Transmission Electron Microscopy (HRTEM) experimental observations \cite{Klimenkov2016}. The averaged void sizes obtained in present MC simulations are varied between 3nm and 5 nm that are also comparable with different experimental observations \cite{Klimenkov2016,Lloyd2019}. 

\begin{table*}
\caption{Void shapes obtained from the quenched MC simulations at 1200K and projected into three different planes. The unit of x, y and z axis is in Armstrong \AA}
\begin{ruledtabular}
    \begin{tabular}{|c|c|c|c|}
              & xy & yz & zx \\
     \hline
     Void 1 & \includegraphics[width=.3\linewidth]{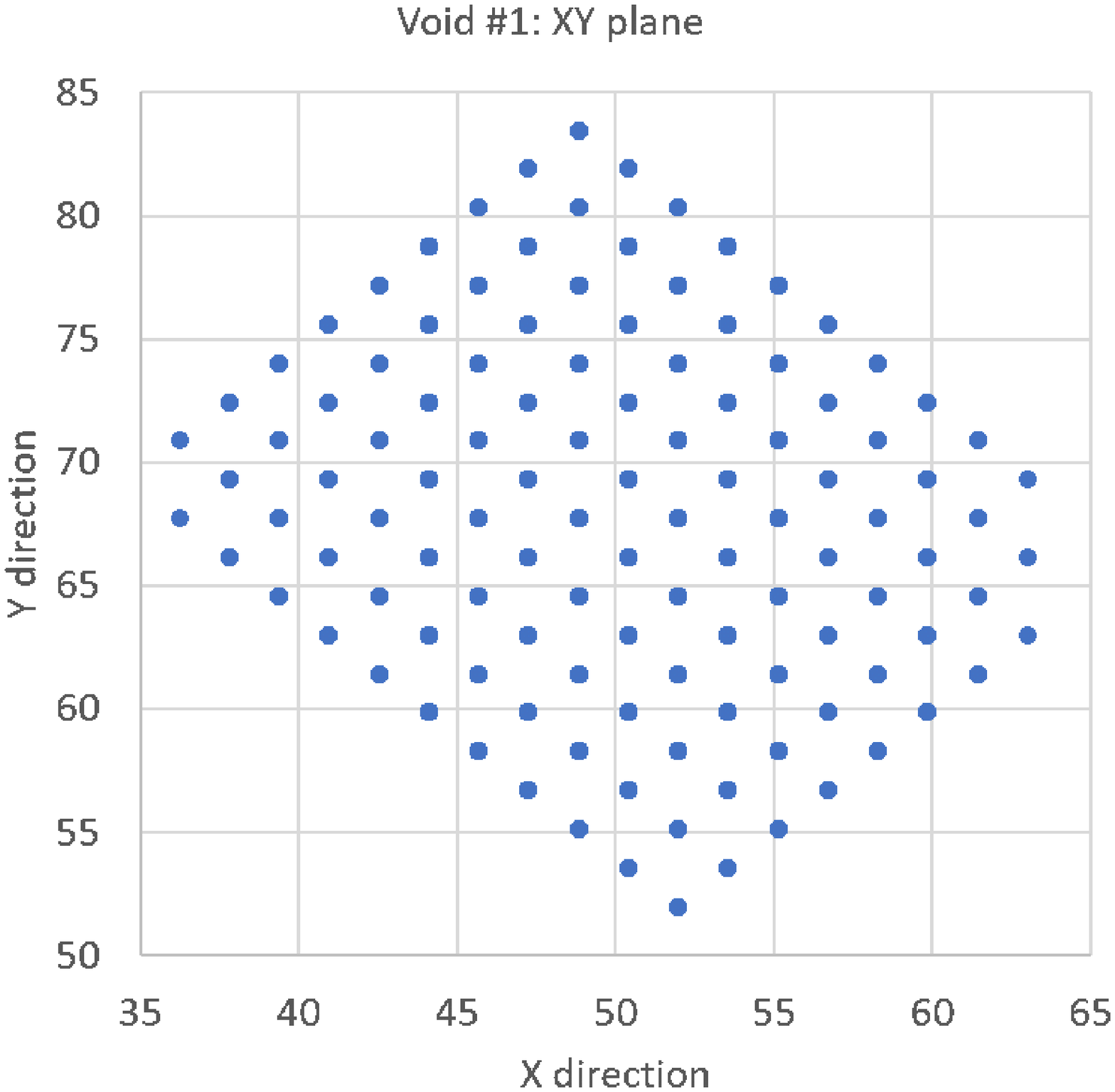} & \includegraphics[width=.3\linewidth]{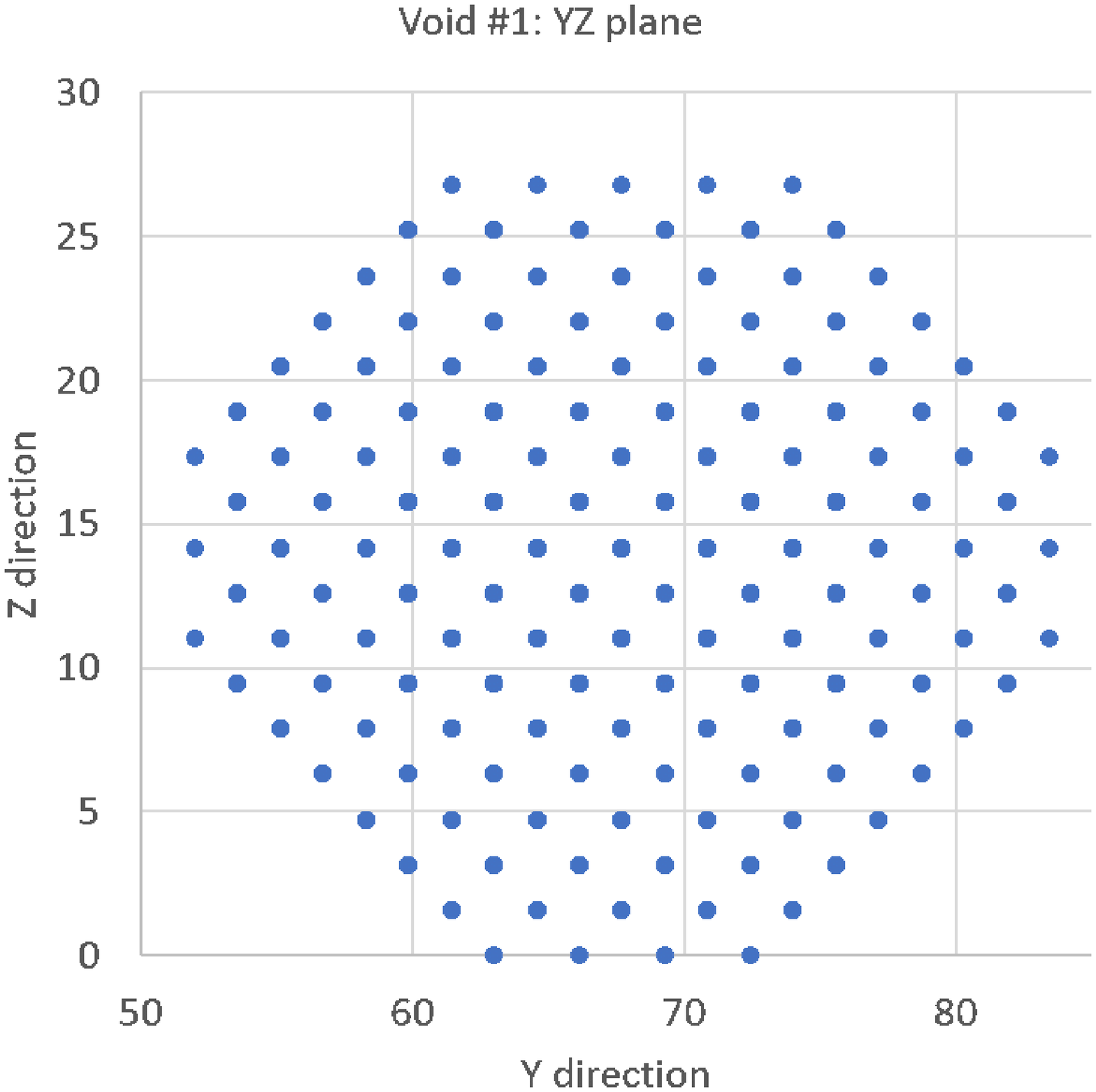} & \includegraphics[width=.3\linewidth]{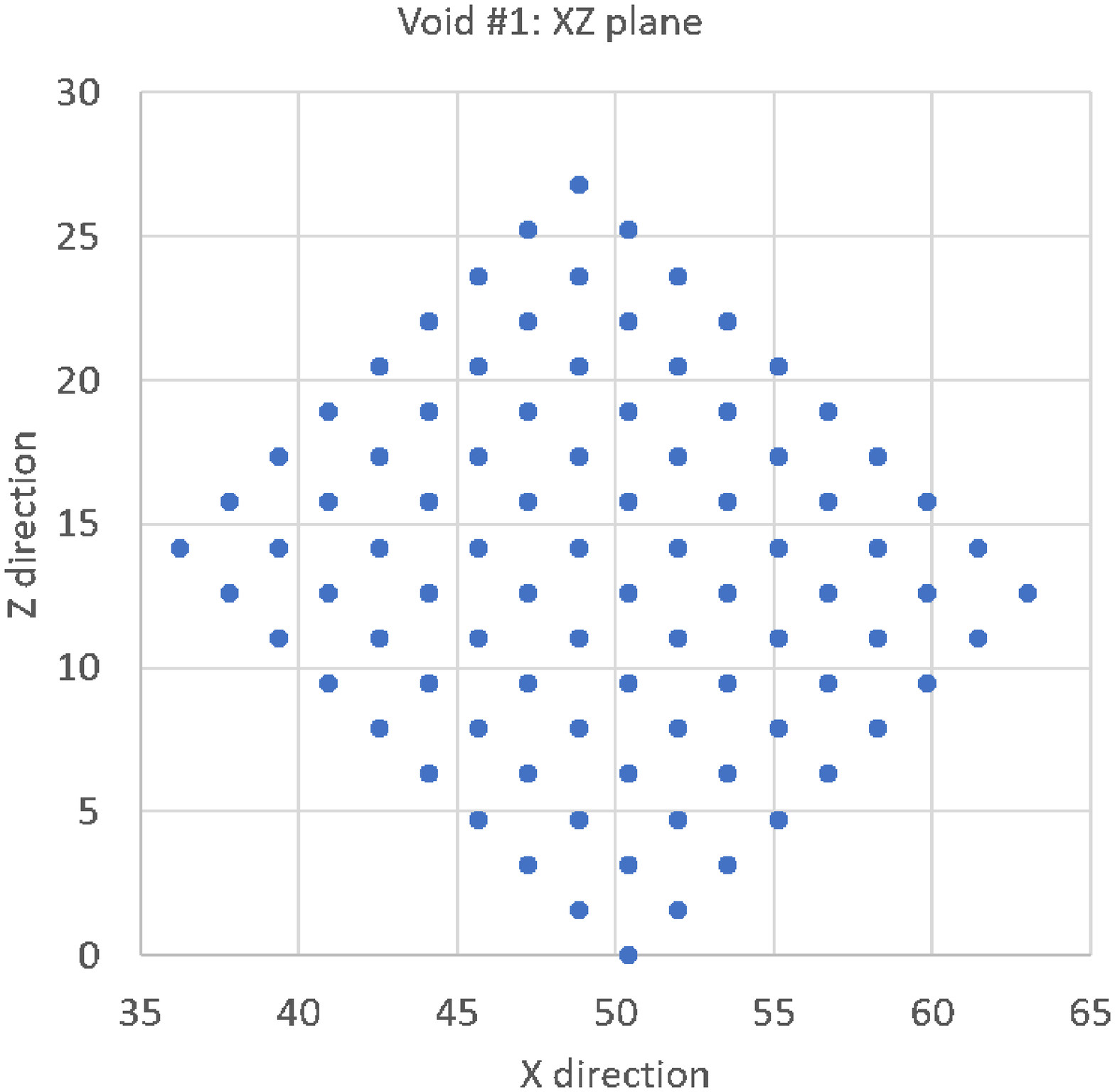} \\
     \hline
     Void 2 & \includegraphics[width=.3\linewidth]{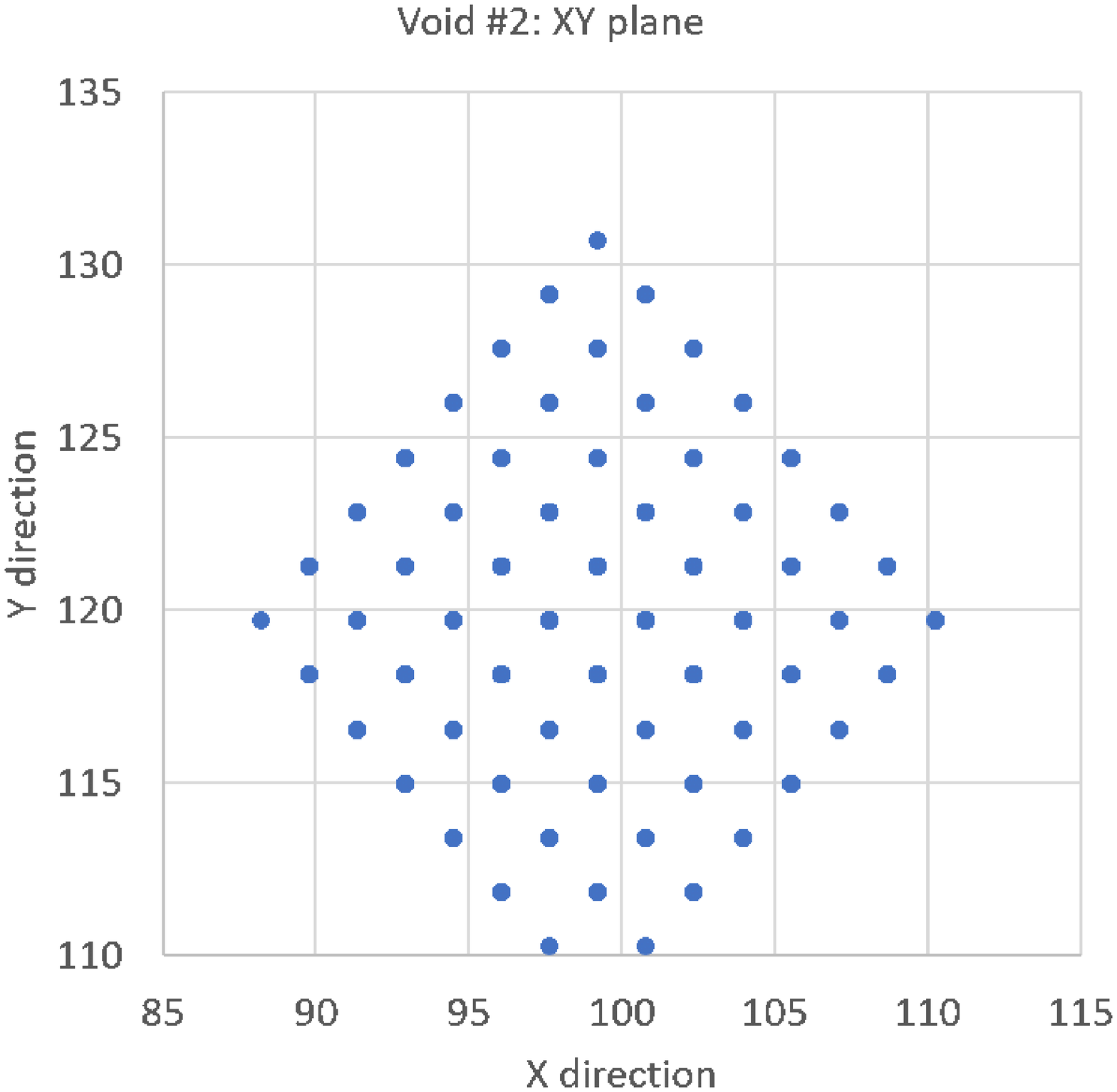} & \includegraphics[width=.3\linewidth]{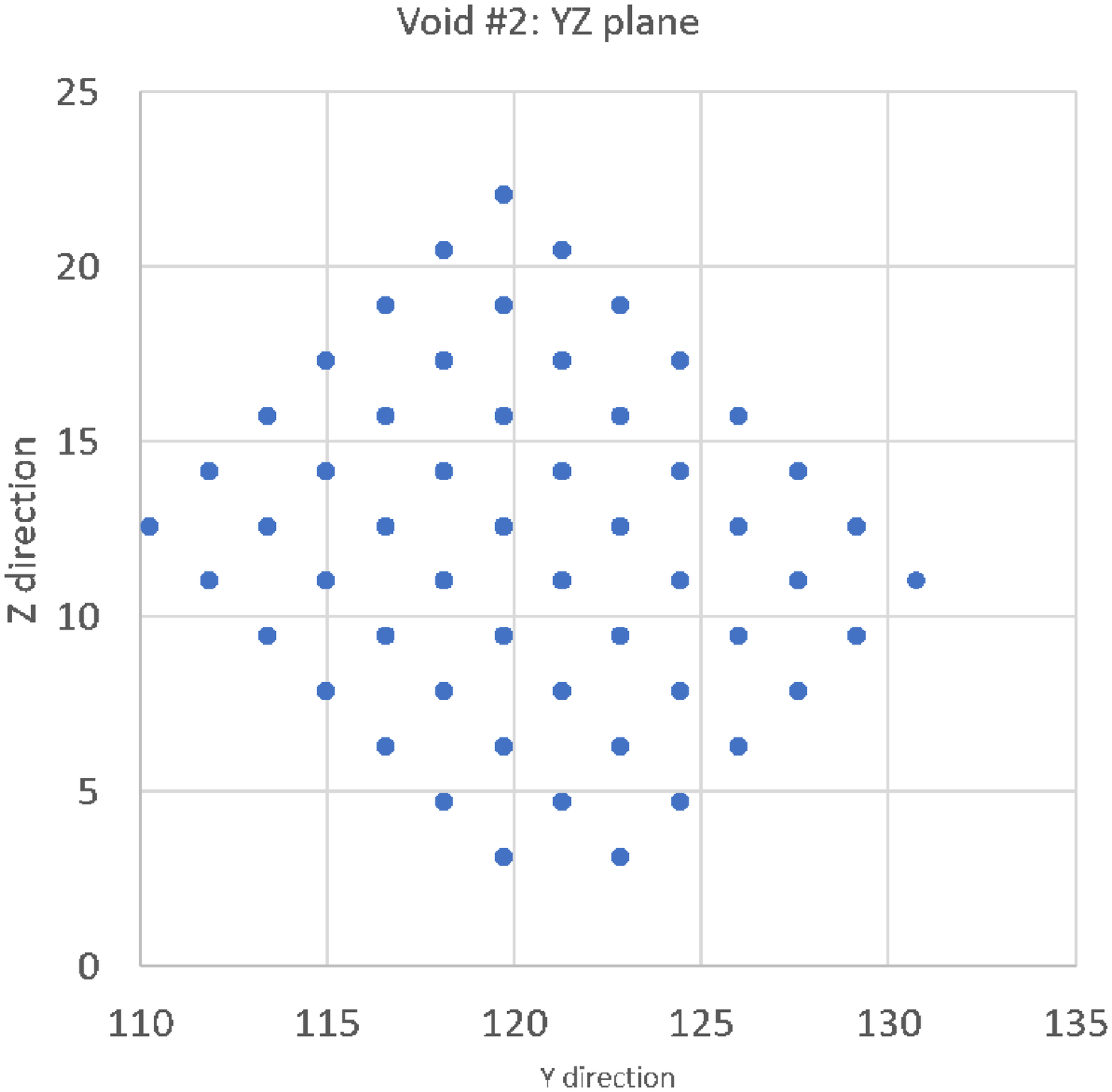} &  \includegraphics[width=.3\linewidth]{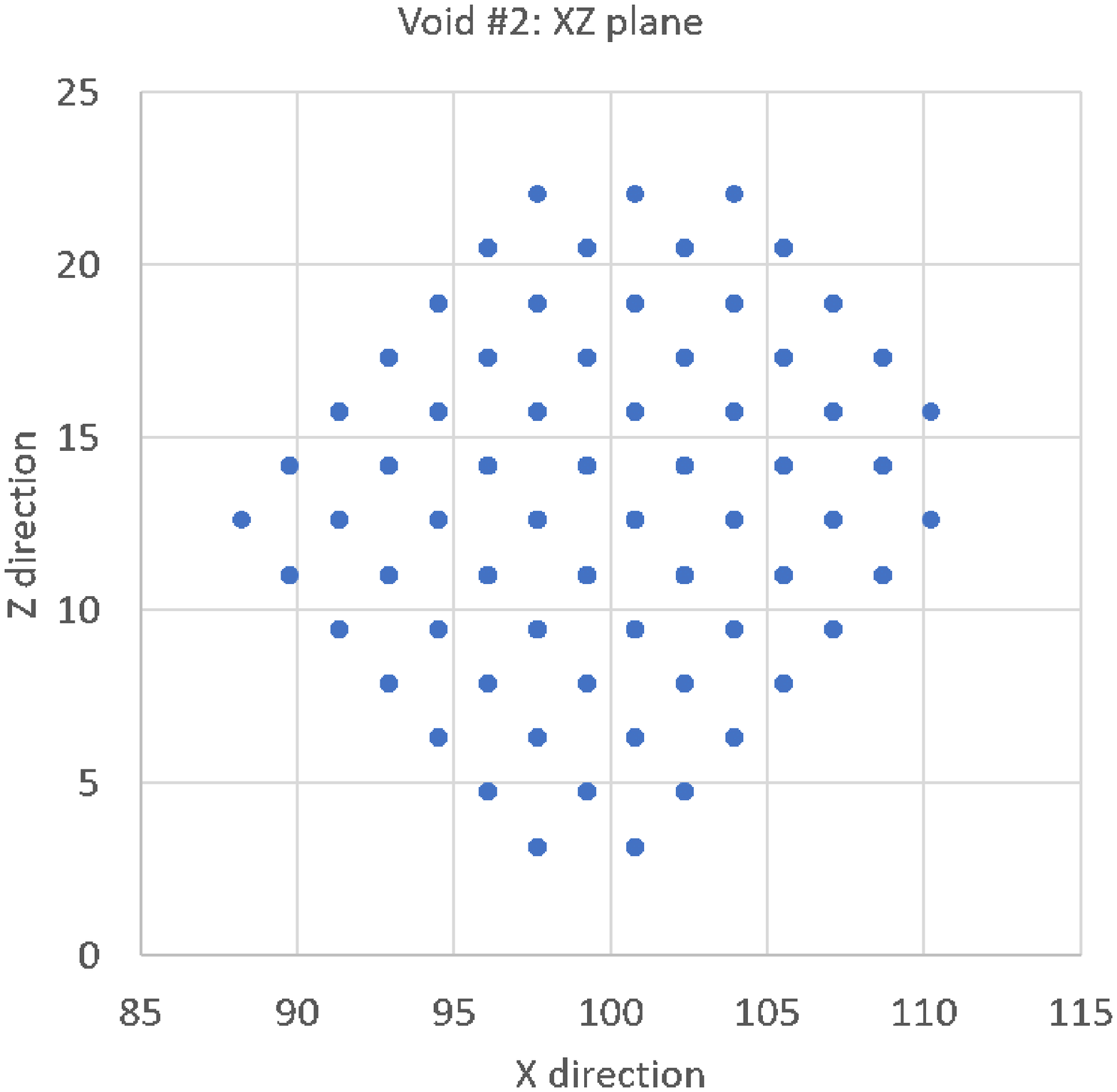} \\
     \hline
      Void 3 & \includegraphics[width=.3\linewidth]{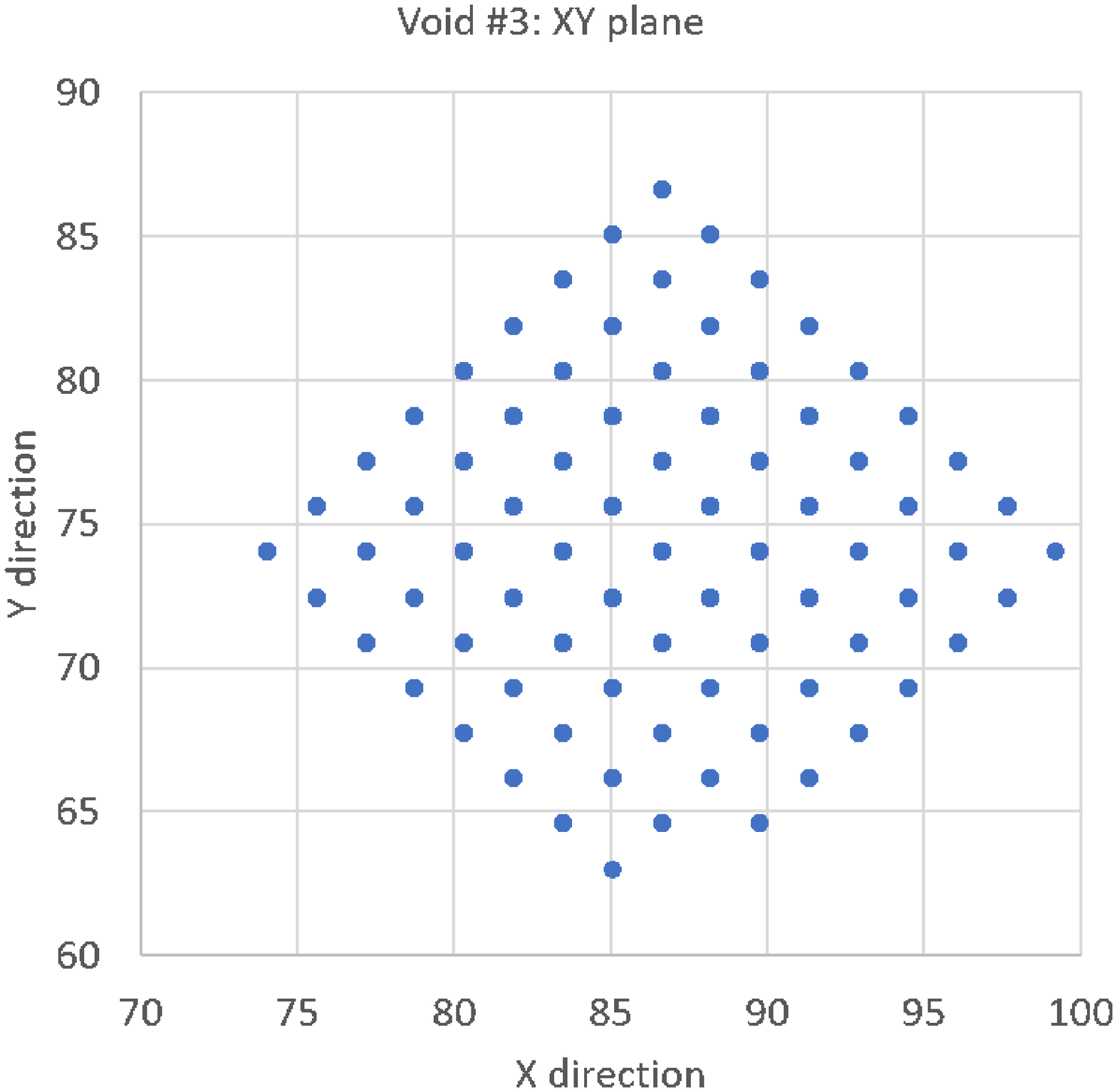} & \includegraphics[width=.3\linewidth]{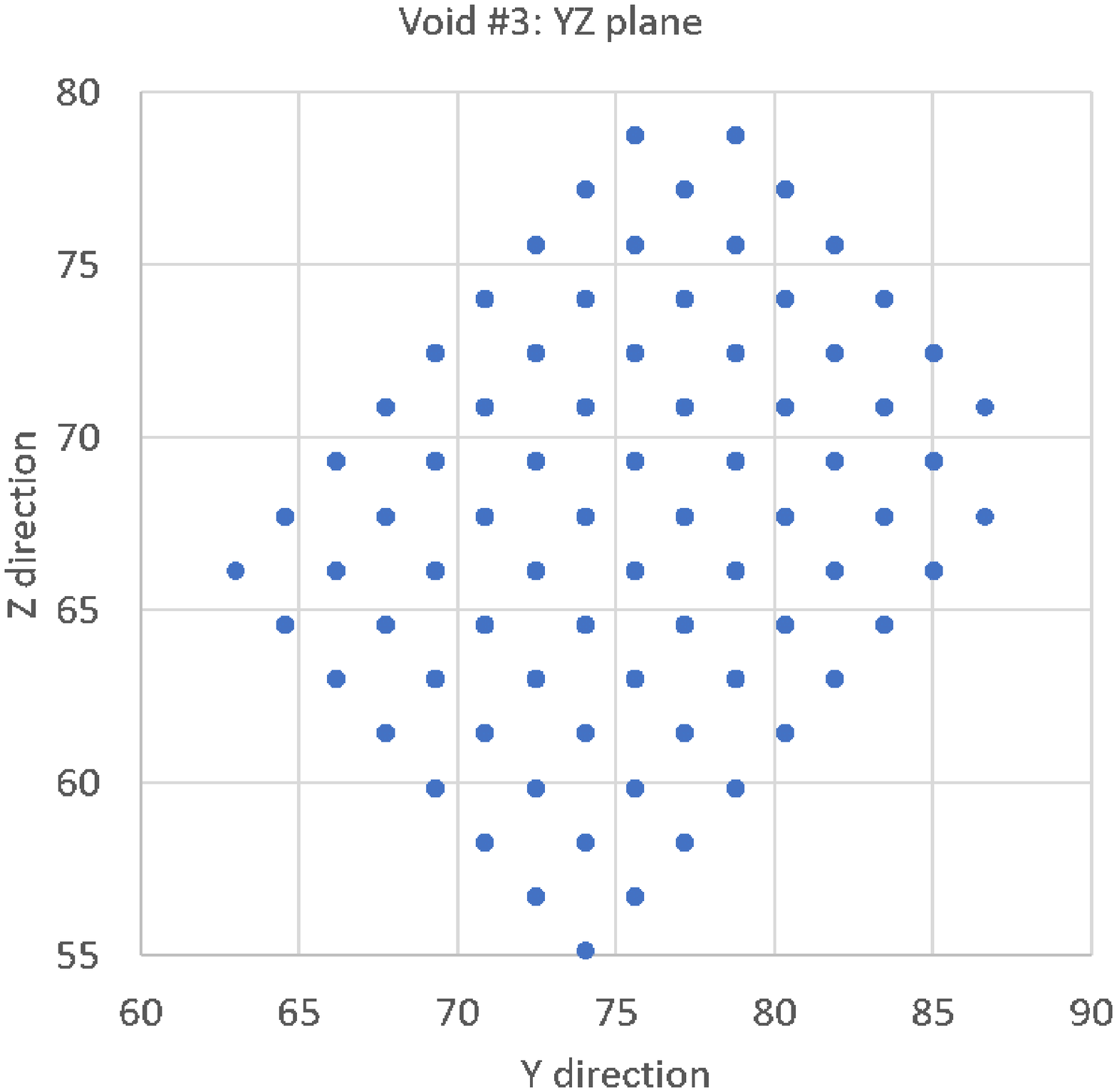} &  \includegraphics[width=.3\linewidth]{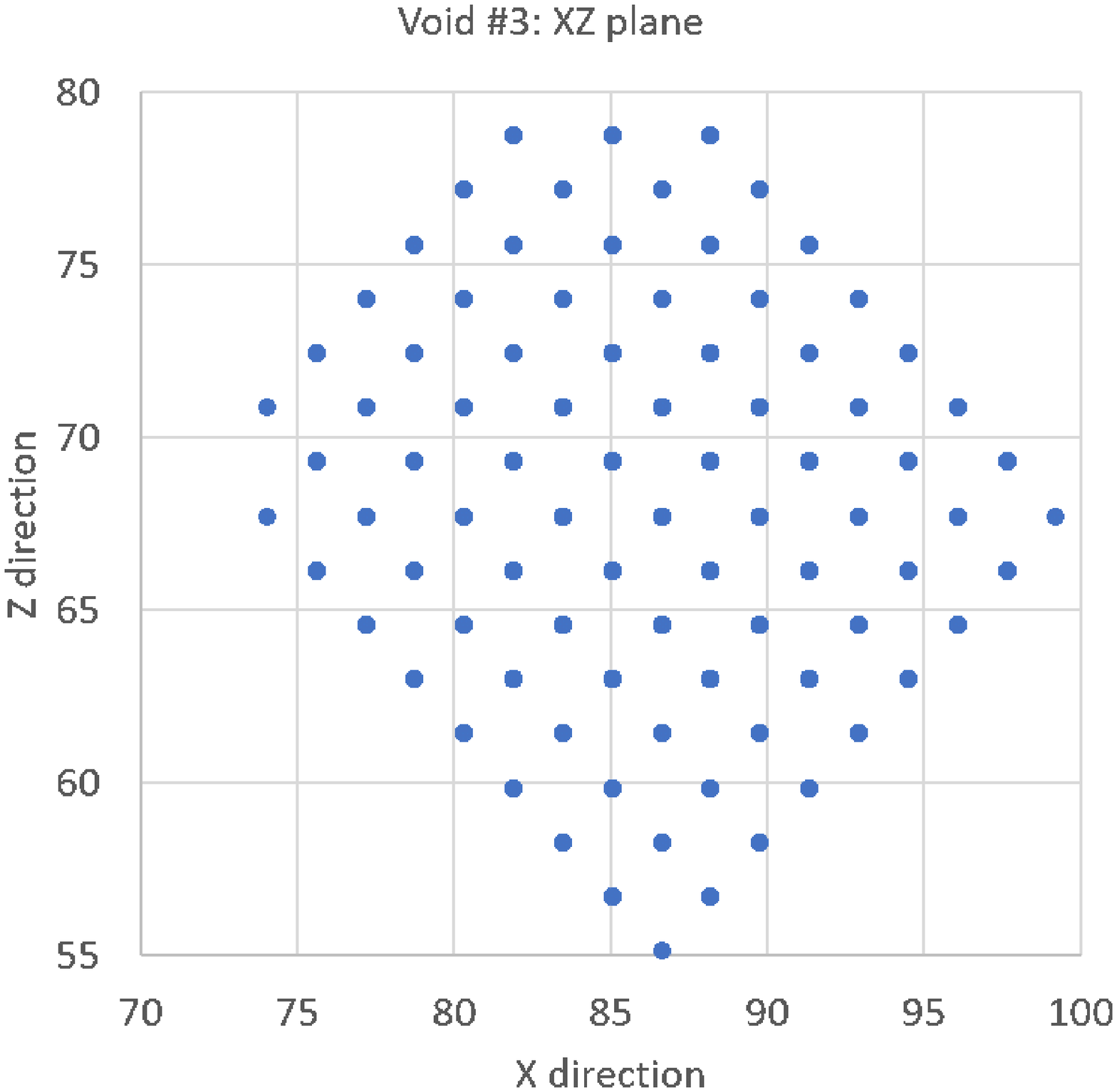} \\
      \end{tabular}%
\end{ruledtabular}
\label{fig:voids}
\end{table*}
     
  \subsection{Transmission Electron Microscopy results}

  The samples used in this study to interpret the modellings results were single and poly-crystalline tungsten neutron irradiated in the High Flux Reactor (Petten, Netherlands) for a total of 208 days at 900 $^{\circ}$C to an average damage level of 1.6 dpa \cite{Klimenkov2016}. FISPACT-II simulations by Gilbert et al. using a displacement threshold energy of 55eV predicted that a damage of 1.67dpa was produced at a dose rate of  9 × $10^{-8}$ dpa/s. The post-irradiation composition was predicted composition of W - 1.4 at.$\%$Re - 0.1 at.$\%$Os - 0.02 at.$\%$Ta from FISPACT-II \cite{Gilbert2011,Gilbert2015}. Thin foils for Transmission Electron Microscopy (TEM) investigations were prepared using the Focus Ion Beam (FIB) technique. Microstructural characterization was performed using the Talos F200X equipped with a scanning unit for STEM with a HAADF ("High Angular Dark Field" ) detector and a super EDX (energy dispersive X-ray) system for elemental analysis.
  
The TEM images revealed that voids are homogeneously distributed in mono-crystalline material and inside grains in poly-crystalline material. The size distribution of voids in both materials showed that they varies from 3 to 14 nm with a distribution maximum at 5 nm for both materials. The experimental finding supports our modelling results displaying void formation at T=1200 K from either fixed temperature or quenched MC simulations shown in the previous section. High-resolution transmission electron microscopy (HRTEM) showed that the cavities had a faceted shape, where the facets are formed in {110} planes. These observations are again very consistent with our simulations shown in Table \ref{fig:voids}. 

\begin{figure*}
\raisebox{12\height}{(a)}\includegraphics[width=0.3\linewidth]{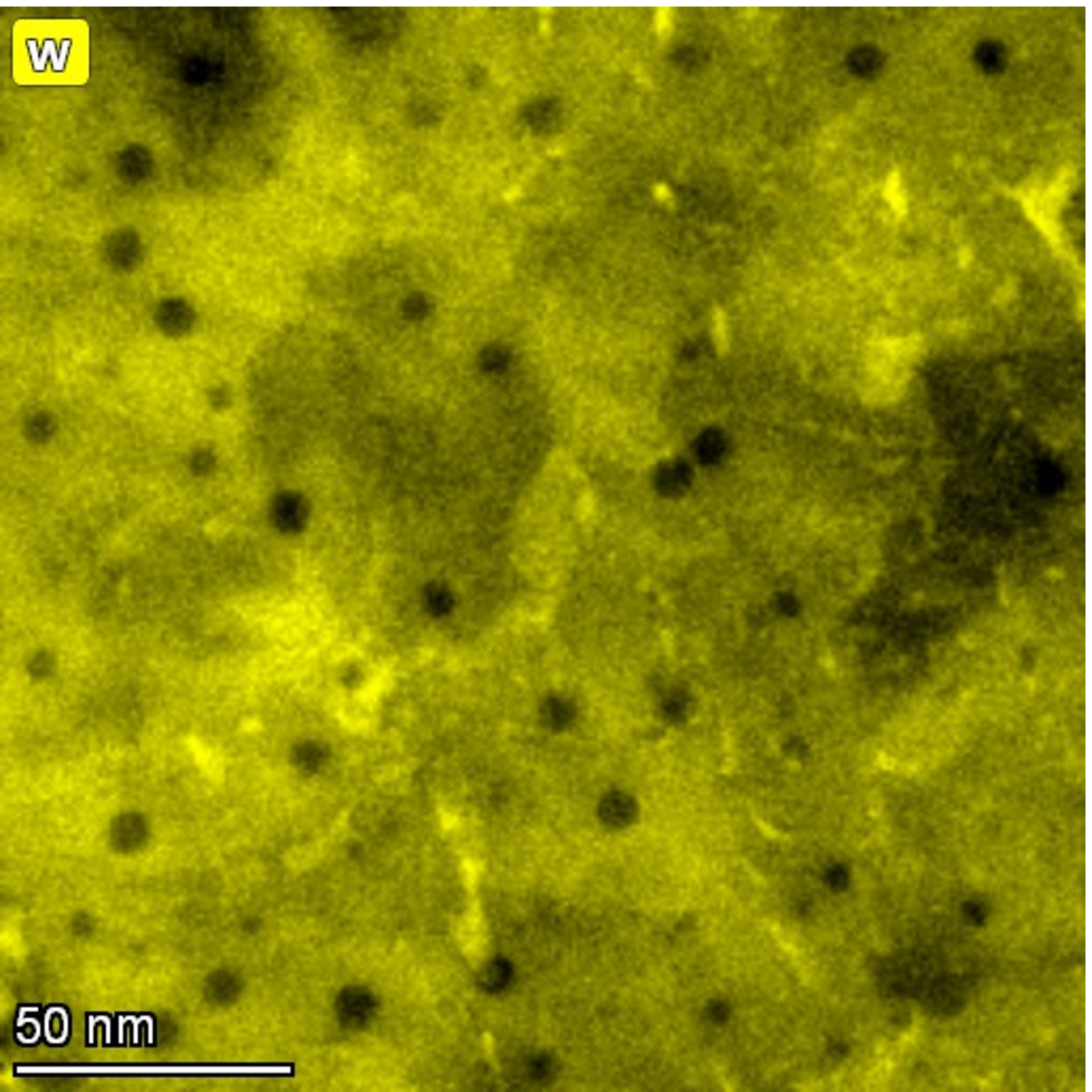}
\raisebox{12\height}{(b)}\includegraphics[width=0.3\linewidth]{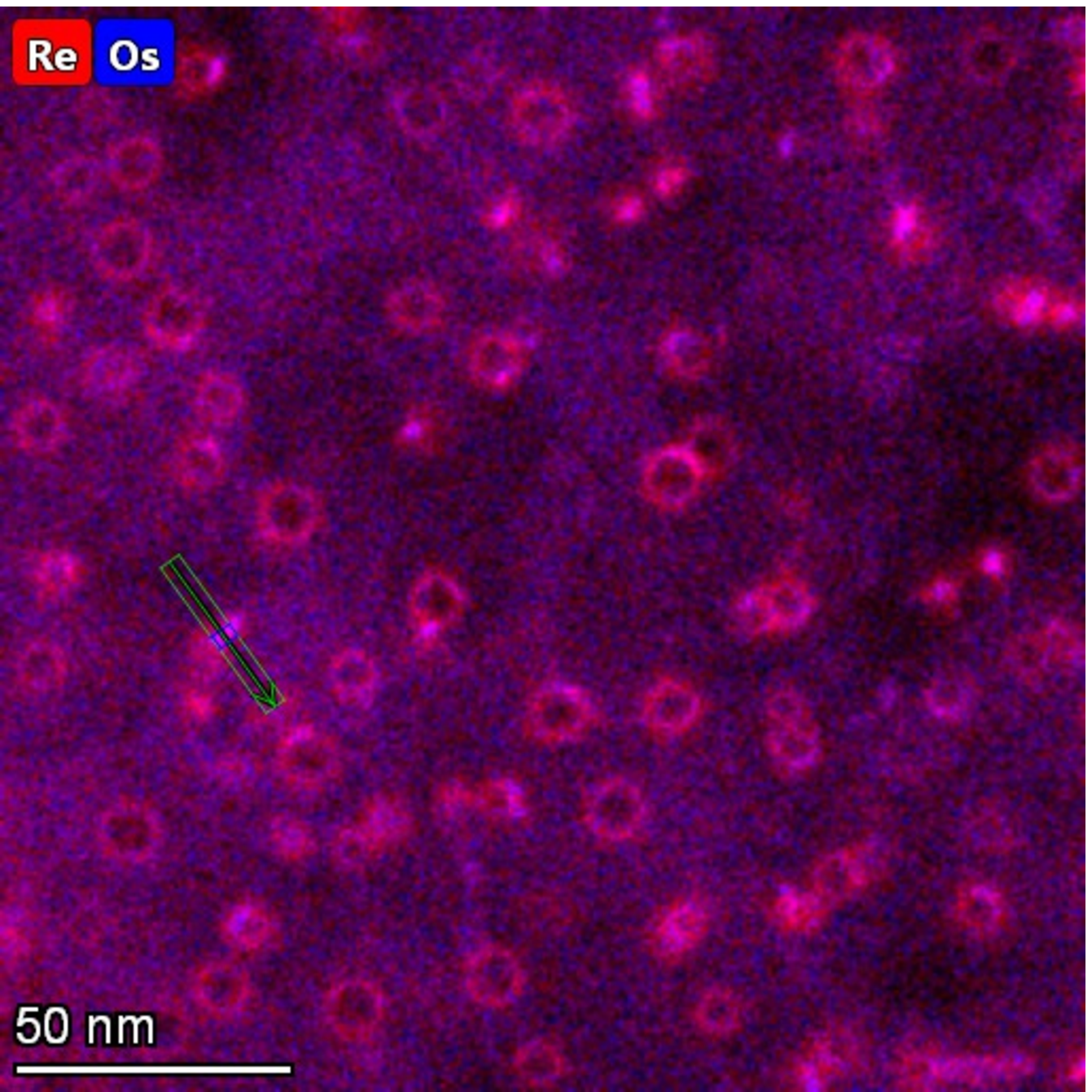}
\raisebox{12\height}{(c)}\includegraphics[width=0.3\linewidth]{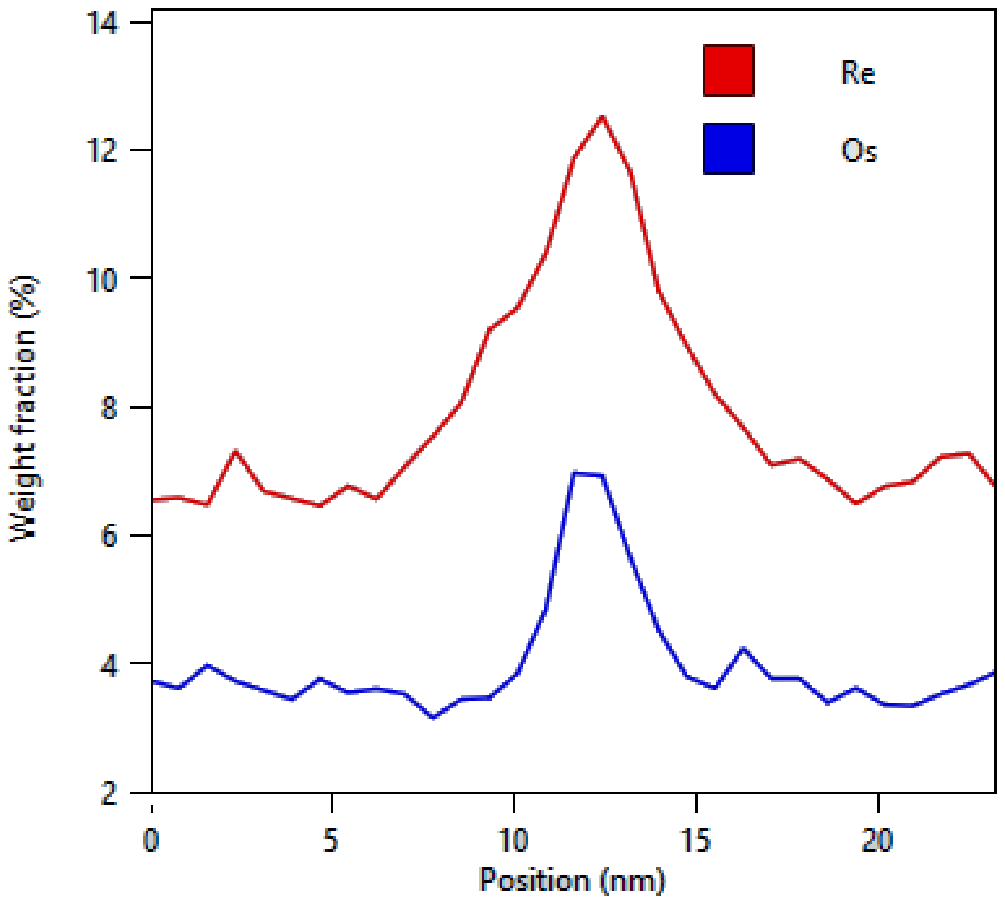}
\caption{\label{haadf} HAADF images of voids in neutron irradiated: (a) W (b) decorated Re, Os and (c) their distribution profiles across the void indicated in (b) and averaged over the sample thickness}
\label{fig:haadf}
\end{figure*}

In addition to the formation of voids, the production of transmutation elements was observed after neutron irradiation of W from the EDX measurements. Fig.\ref{fig:haadf} shows the two-dimensional elemental mapping of single crystalline tungsten using HAADF images for W (a), Re and Os (b) and the average distribution of transmutation element (c) across a chosen void with closely located Os-rich rod. Re forms a 15 nm cloud around the cavity, while Os forms rods within the cloud of 3 nm - 5 nm size (Fig.\ref{fig:haadf}c). It can be estimated that approximately 10-15$\%$ of all voids have similar, closely located, Os-rich rod. The results clearly demonstrated that both of Re and Os elements are correlated well with the void in neutron irradiated tungsten at 900 $^{\circ}$C. The decoration of voids by both Re and Os was further experimentally confirmed in \cite{Lloyd2019}. These experimental observations underpin the present theoretical approach by looking at finite-temperature modelling of the vacancy cluster interaction with Re and Os in solid solution limit.        

\subsection{Atom Probe Tomography results} 

APT has been performed on neutron and ion irradiated W samples from several different reactors. Studies of neutron irradiated W in the 1980s focussed on W-10Re and W-25Re alloys, and observed the formation of $\chi$ phase precipitates, as well as voids in field-ion microscopy experiments \cite{herschitz1984,herschitz1984_2}. More recent investigation have focused primarily on initially pure W samples, so as to simulate component evolution in a fusion reactor. Katoh et. al performed APT on single crystal W samples, neutron irradiated to 0.11dpa at 963K with a high transmutation rate, and observed coherent, Re rich spherical precipitates \cite{katoh2019}. Hwang et. al analysed pure W and W-10Re samples irradiated with a lower transmutation rate from Joyo (0.96dpa, 773K) and similarly observed spherical Re rich precipitates. Cluster analysis performed using a maximum separation method indicated that the clusters had compositions within the $\sigma$+bcc region of the ternary W-Re-Os phase diagram \cite{hwang2018}. Furthermore, comparison with TEM imaging correlated the spacing of these clusters to voids, suggesting that void decoration with Re had occurred. Lloyd et. al also saw strong evidence for void decoration with both Re and Os in APT analyses of low transmutation rate, neutron irradiated W samples from HFR (1.67dpa, 1173K) \cite{Lloyd2019,Lloyd2021}. High density artefacts are introduced into APT reconstructions due to local curvature induced trajectory aberrations at the periphery of the void, as outlined by Wang et. al \cite{wang2020}. In all of these analyses using maximum separation methods, there was no indication that either $\sigma$ or $\chi$ phase formation had occurred.

\begin{figure*}
	\includegraphics[width=0.8\linewidth]{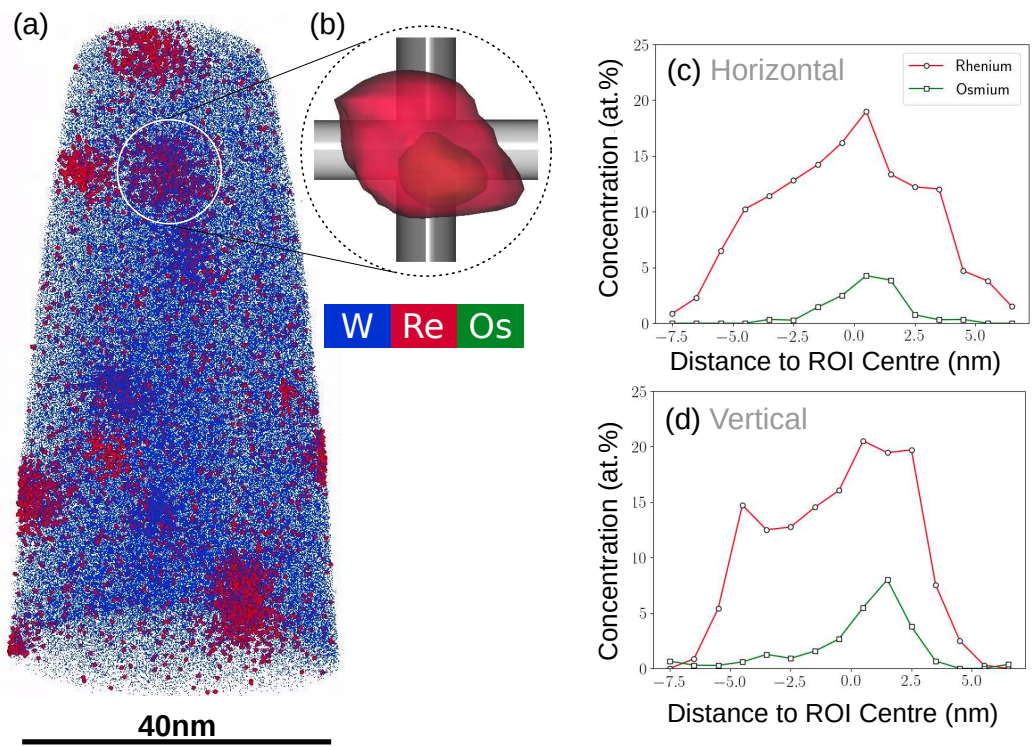}
	\caption{\label{apt} Reconstruction and analysis of APT dataset taken from single crystal W sample, neutron irradiated to at 900 $^{\circ}$C. (a) the analysis volume, with 100$\%$ of Re atoms plotted as red spheres, with around 10$\%$ of W atoms shown in blue. (b) shows a higher magnification region with the W atoms removed, and Re and Os iso-concentration surface at 5at.$\%Re$ and 1.5at.$\%$Os respectively in red and green, showing a core of Os surrounded by Re. The plots shown in (c) and (d) are line profiles taken in the horizontal and vertical directions respectively, as shown in (b), with the binned Re and Os concentration plotted as a function of distance across the ROI (Region Of Interest).}
\label{fig:apt}
\end{figure*}

Fig.\ref{fig:apt} shows reconstruction and analysis of an APT dataset taken from the single crystal W sample, neutron irradiated to 1.67dpa at 900 $^{\circ}$C at the High Flux Reactor in Petten \cite{Klimenkov2016,abernethy2019,Lloyd2019}. The predicted post-irradiation composition simulated using FISPACT-II was W-1.4Re-0.1Os, and was measured using APT and EDX analyses to be W-1.2Re-0.1Os \cite{Lloyd2019,Lloyd2021}. For visual clarity, Fig.\ref{fig:apt}a shows only around 10$\%$ of W atoms in blue and all defected Re atoms are shown in red displaying a strong tendency for Re to form precipitates. Fig.\ref{fig:apt}b shows a higher magnification image of a precipitate with a core of Os surrounded by Re atoms, for which all W atoms are removed whereas Re and Os are displayed using iso-concentration surface at 5$\%$ and 1.5$\%$, respectively.Fig. \ref{fig:apt}c and \ref{fig:apt}d show line profile analyses of the cluster shown in Fig. \ref{fig:apt}b, with the concentration of Re (red) and Os (green) plotted. Previous analysis provided strong evidence for void decoration with both Re and Os, in STEM and APT results \cite{Lloyd2019}. Strong segregation of Os to both voids and precipitates is observed, despite the low nominal concentration of Os (0.1at.$\%$) compared to Re. Statistical analysis of voids and precipitates in APT data taken from these samples demonstrate that in centres of some precipitates concentration has reached 25at.$\%$Re and 10at.$\%$Os or even higher, as highlighted by Fig. \ref{fig:apt}c and \ref{fig:apt}d. These APT findings again support the modelling results of this work.          

\section{Conclusions}

In this paper, a first-principles approach has been developed to model the RIS of solutes in multi-component systems within the framework of thermodynamics under irradiation. The matrix formation of DFT-based cluster expansion Hamiltonian combined  with a thermodynamic integration method is presented in a systematic way so as to result in  a general expression of the short-range order (SRO) parameter in K-component alloys containing defective elements. It allows us to to investigate not only configuration free energy but also the SRO between different elements at an arbitrary composition and temperature. The model is applied to studying anomalous segregation of transmutation products of neutron irradiated tungsten at high temperatures, with formation of vacancy clusters and/or voids. The effective cluster interactions of  first-principles Hamiltonians have been obtained from mapping DFT calculations, and show that the cross validation errors are small and around 10 meV for the three considered W-Ta-Vac, W-Os-Vac and W-Re-Os-Vac systems. In particular for the quaternary W-Re-Os-Vac, which is the main focus of the present work, the predictive power of the CE model has been cross-checked with an independent set of DFT calculations to confirm accuracy of the  of the set of ECIs used in our simulations.

When comparing the results of exchange MC simulations for the three systems, W-Re-Vac, W-Os-Vac and W-Ta-Vac, it is found that there is  no segregation of Ta atoms to the vacancy clusters. This finding is in a very good agreement with the finding from Atom Probe Tomography experimental observation that there are stronger effects of radiation induced segregation of Os in a comparison with Re, whereas there is no effect of Ta precipitation in the binaries under self-ion irradiation. The distinct behaviour between the 3 solute cases can be explained by analysing the short-range order parameters evaluated from MC simulations. For the Ta-Vac pair, the SRO remains positive whereas for Re-Vac and Os-Vac pairs they are negative implying the attractive interactions with vacancy clusters at high temperatures. 

Extending our investigation to the competitive interaction between Re and Os in the defective W-Re-Os-Vac system, it is found that the presence of Os has decreased  the disorder-order transition temperature in the void formation in W, is comparable to similar results from free-energy calculations for W-Re-Vac system which have been carefully investigated previously. Importantly, our results reveal that the SRO parameter between Re and Os is very sensitive to the Os concentration in presence of vacancy defects. The Re-Os SRO parameter changes the sign from positive to negative when Os concentration decreases from 1$\%$ to lower than 0.25$\%$ leading to the conclusion that both of these solute atoms can be simultaneously segregated to the precipitation with vacancy clusters at low Os concentration range.    

Focusing on the defective quaternary W-Re-Os-Vac system with Re1.5$\%$ and an Os0.1$\%$, it is shown that the SRO parameter between Re and Os is strongly negative at 1200K, resulting in very high concentrations of both Re and Os inside the formed precipitate, up to 25$\%$ and 18$\%$, respectively. Comparing with available experimental data from Transmission Electron Microscopy (TEM) and APT, performed for neutron irradiated W samples within the High Flux Reactor (HFR) with a low Re transmutation rate and with the same pre-irradiation composition and irradiation temperature, it is found that the developed model would be able to provide an accurate account on nano-scale origin of anomalous segregation of transmutation products (Re,Os) to the vacancy clusters and voids in tungsten at high irradiation temperatures. Importantly, the modelling results clearly demonstrate that the voids formed under neutron irradiation in tungsten have a faceted shape, where the facets are formed in {110} planes in excellent agreement with HR-TEM imaging. Further investigation of interstitial cluster defects within the present approach would help to understand the transmutation induced segregation effects under neutron irradiation consistently from both a thermodynamic and defect diffusion standpoints.


\section{Acknowledgements}
This work has been carried out within the framework of the EUROfusion Consortium and has received funding from the Euratom research and training programmes 2014-2018 and 2019-2020 under Grant Agreement No. 633053 and from the RCUK Energy Programme
[grant number EP/T012250/1]. We also acknowledge
funding from the European Research Council (ERC) under the European Union’s Horizon 2020 research and innovation programme (grant agreement no. 714697). The
views and opinions expressed herein do not necessarily reflect those of the European Commission.
DNM, JSW and LM acknowledge the support from high-performing computing facility MARCONI (Bologna, Italy) provided by EUROfusion. 
The work at Warsaw University of Technology has been carried out as a part of an international project co-financed from the funds of the program of the Polish Minister of Science and Higher Education entitled "PMW" in 2019; Agreement No. 5018 / H2020-Euratom / 2019/2. The simulations were also carried out with the support of the Interdisciplinary Centre for Mathematical and Computational Modelling (ICM), University of Warsaw, under grant No. GB79-6. MJL's work is also supported by the UK Engineering and Physical Sciences Research Council [EP/N509711/1] and the Culham Centre for Fusion Energy, United Kingdom Atomic Energy Authority through an Industrial CASE scholarship, [Project Reference Number 1802461]. 


\bibliography{WReOsVac}

\end{document}